\documentclass[acmsmall,screen,nonacm]{acmart}

\usepackage{tikz}
\usetikzlibrary{arrows.meta,positioning,shapes.geometric,shapes.symbols,calc,decorations.pathreplacing,patterns}
\usepackage{subfig}
\usepackage[capitalize,noabbrev]{cleveref}
\hypersetup{hidelinks}

\definecolor{tapconavy}{RGB}{82,114,146}
\definecolor{tapcoteal}{RGB}{78,128,108}
\definecolor{tapcoamber}{RGB}{176,128,72}
\definecolor{tapcored}{RGB}{172,92,86}
\definecolor{tapcoink}{RGB}{48,54,62}
\definecolor{tapcomuted}{RGB}{112,120,128}
\definecolor{tapcorule}{RGB}{196,201,207}
\definecolor{tapconavyfill}{RGB}{239,243,247}
\definecolor{tapcotealfill}{RGB}{238,245,241}
\definecolor{tapcoamberfill}{RGB}{249,244,234}
\definecolor{tapcoredfill}{RGB}{248,239,238}
\definecolor{tapcograyfill}{RGB}{245,246,248}
\colorlet{tapcoblue}{tapconavy}
\colorlet{tapcolightblue}{tapconavyfill}
\colorlet{tapcoorange}{tapcoamber}
\colorlet{tapcolightorange}{tapcoamberfill}
\colorlet{tapcogreen}{tapcoteal}
\colorlet{tapcolightgreen}{tapcotealfill}
\colorlet{tapcogray}{tapcoink}
\colorlet{tapcolightgray}{tapcograyfill}

\newcommand{\tool}[1]{\textsf{#1}}
\newcommand{\tapco}{\tool{TAPCO-G}}
\newcommand{\openroad}{\tool{OpenROAD}}

\newcommand{\dpower}{P_{\mathrm{dyn}}}
\newcommand{\lpower}{P_{\mathrm{leak}}}

\crefname{figure}{Fig.}{Figs.}
\crefname{table}{Table}{Tables}
\crefname{section}{Section}{Sections}
\crefname{equation}{Eq.}{Eqs.}

\setcopyright{none}
\copyrightyear{}
\acmDOI{}
\acmISBN{}

\author{Wuqian Tang}
\email{wqtang@cs.nthu.edu.tw}
\orcid{0009-0008-5042-5062}
\affiliation{%
  \institution{National Tsing Hua University}
  \city{Hsinchu}
  \country{Taiwan}}

\author{Zixiao Wang}
\authornote{Corresponding author.}
\email{zixiaowang97@gmail.com}
\orcid{0009-0000-8179-0996}
\affiliation{%
  \institution{The Hong Kong University of Science and Technology (Guangzhou)}
  \city{Guangzhou}
  \country{China}}

\author{Che-Chien Lin}
\email{wensen514@gmail.com}
\orcid{0009-0009-9790-966X}
\affiliation{%
  \institution{National Tsing Hua University}
  \city{Hsinchu}
  \country{Taiwan}}

\author{Fong-Hua Fu}
\email{a81869033@gmail.com}
\orcid{0009-0009-6547-2600}
\affiliation{%
  \institution{National Tsing Hua University}
  \city{Hsinchu}
  \country{Taiwan}}

\author{Laura Angélica Shion-Korsak}
\email{laura@gapp.nthu.edu.tw}
\orcid{0009-0009-8389-0961}
\affiliation{%
  \institution{National Tsing Hua University}
  \city{Hsinchu}
  \country{Taiwan}}

\author{Xinyun Zhang}
\email{zhangxy12@shanghaitech.edu.cn}
\orcid{0000-0002-7763-7507}
\affiliation{%
  \institution{ShanghaiTech University}
  \city{Shanghai}
  \country{China}}

\author{Bei Yu}
\email{byu@cse.cuhk.edu.hk}
\orcid{0000-0001-6406-4810}
\affiliation{%
  \institution{The Chinese University of Hong Kong}
  \city{Shatin}
  \country{Hong Kong}}

\author{Chun-Yao Wang}
\email{wcyao@cs.nthu.edu.tw}
\orcid{0000-0002-4536-214X}
\affiliation{%
  \institution{National Tsing Hua University}
  \city{Hsinchu}
  \country{Taiwan}}

\renewcommand{\shortauthors}{Tang et al.}

\newcommand{\PaperAcknowledgment}{%
  This work was supported in part by the National Science and Technology
  Council (NSTC) under Grants NSTC 112-2218-E-007-014,
  NSTC 112-2221-E-007-106-MY2, NSTC 112-2221-E-007-108,
  NSTC 113-2640-E-011-003, NSTC 113-2221-E-007-082-MY3,
  NSTC 114-2221-E-007-125-MY3, NSTC 114-2218-E-007-002,
  and NSTC 114-2221-E-007-126-MY3.}

\begin{document}

\title[TAPCO-G: Graph-Guided Timing and Power Co-Optimization]{TAPCO-G: LLM-Guided Design-Adaptive
  Timing and Power Co-Optimization with Graph-Based Physical Priors}

\begin{abstract}
Post-placement timing and power co-optimization is critical to meeting setup timing targets
and reducing power within placement constraints. A cell replacement can improve timing
along one path while increasing power or reducing slack on others. Conventional optimizers
follow fixed rules for scheduling and accepting repairs, limiting their ability to address
design-specific bottlenecks. A large language model (LLM) can propose changes to these
rules, but on its own it cannot reliably estimate how a cell replacement affects endpoint
slack and power. \tapco{} combines LLM-guided optimizer adaptation with graph-based
physical priors to evaluate candidate cell replacements. The outer loop tunes tool settings
using physical-design feedback and revises executable repair policies when tuning stalls.
The inner loop anchors delay estimates to static timing analysis snapshots, accounts for
upstream driver-load changes, and propagates delay shifts to timing endpoints. It combines
these estimates with coarse-to-fine power screening to prioritize candidates for physical
trials. \tapco{} produced placement-legal, logic-equivalent implementations and outscored
\tool{CERBERUS} on all eight ASAP7 benchmarks from the MLCAD 2026 Contest, with a mean
quality score gain of $5.516$ points, and scored $21.648$ points higher than \tool{TAPCO}
on average. These results demonstrate the effectiveness of the proposed co-optimization
framework.
\end{abstract}


\makeatletter
\let\tapco@originalfnsymbol\@fnsymbol
\def\@fnsymbol#1{%
  \ifnum#1=1\relax
    \tapco@originalfnsymbol{2}%
  \else
    \tapco@originalfnsymbol{#1}%
  \fi}
\maketitle
\let\@fnsymbol\tapco@originalfnsymbol
\makeatother

\section{Introduction}
\label{sec:intro}

Post-placement timing and power co-optimization is a decisive stage in modern very-large-scale integration (VLSI) physical synthesis~\cite{alpert2007physical}. Once standard cells and hard macros are placed and interconnect parasitics are extracted, the optimizer must simultaneously close setup timing violations and minimize dynamic and leakage power~\cite{luo2008totalpower} within strict placement-density and routability headroom~\cite{do2019legalization}. Discrete netlist transformations, including gate sizing~\cite{chen1998gatewire,liu2026hypas}, multi-threshold-voltage (VT) assignment, buffering~\cite{alpert2004buffer,sze2005path}, cloning, and pin swapping, exhibit intense electrical and physical coupling~\cite{stefanidis2020autonomous}. Upsizing a gate accelerates its intrinsic transition but increases input pin capacitance, imposing heavier capacitive loading on upstream drivers and elevating dynamic switching power; similarly, substituting a regular-VT (RVT) gate with a low-VT (LVT) or super-low-VT (SLVT) cell recovers path slack at the expense of exponential leakage growth. Achieving superior power, performance, and area (PPA) quality of results (QoR) therefore requires answering two intertwined questions: \emph{macroscopically}, how to schedule, budget, and commit repair passes for a specific circuit's bottleneck regime; and \emph{microscopically}, how to accurately quantify the true endpoint-level timing and power value of candidate cell replacements before committing expensive physical trials.

Existing methodologies typically address only one side of this bi-level challenge. At the microscopic operator level, classical sensitivity-based sizers~\cite{agarwal2005statistical} and Lagrangian relaxation formulations~\cite{chen1998gatewire,stefanidis2020autonomous} rely on simplified local delay models that struggle to capture discrete library quantization and multi-corner path interactions. Recent machine learning (ML)~\cite{huang2021mlsurvey,lu2024ganplace} and differentiable optimization frameworks~\cite{lin2019dreamplace,guo2022differentiable} have significantly advanced physical design modeling~\cite{liao2023dreamplace4,chhabria2023timing}. Graph neural networks (GNNs) have been deployed to predict pre-routing timing slack~\cite{guo2022timinggnn}, guide discrete gate sizing over directed acyclic graphs (DAGs, e.g., \tool{DAGSizer})~\cite{cheng2023dagsizer}, and estimate signoff power via subgraph approximation (\tool{ECO-GNN})~\cite{lu2023ecognn}. Concurrently, analytical gradient descent (\tool{AGD})~\cite{pham2025agd} and differentiable sizing frameworks~\cite{ye2024learning} relax discrete cell choices into continuous selections~\cite{du2024fusion,du2025continuity} to co-optimize timing and power~\cite{pan2026quadgradient}. Despite their modeling power, learned surrogates often suffer from distribution shift when applied to unseen netlist topologies or macro-dominated layouts, whereas myopic rule-based heuristics in industrial and open-source engines~\cite{ajayi2019openroad,openroadrsz} evaluate cell swaps locally without propagating upstream driver penalties or downstream reconvergence across the static timing analysis (STA) graph.

At the macroscopic strategy level, the repair policy governs the effectiveness of each physical operator. It determines which violating endpoints to prioritize, how to balance timing recovery against power reduction, and when to retain or roll back intermediate states. Because circuit characteristics vary dramatically from logic-dense arithmetic cores to SRAM-dominated systems-on-chip (SoCs), no single static repair schedule is universally optimal. Automated flow tuning frameworks such as \tool{PTPT}~\cite{geng2023ptpt}, \tool{REMOTune}~\cite{zheng2023flowtuning}, reinforced placement autotuning~\cite{murali2023autotuning}, and microarchitectural design space explorers~\cite{bai2024boom} cast tool customization as black-box hyperparameter optimization via Bayesian optimization~\cite{jones1998ego,snoek2012bo} or reinforcement learning. However, conventional parameter tuning operates strictly within a predefined interface of exposed scalar controls (e.g., slack margins, pass budgets, and utilization caps). An \emph{expressivity gap} can arise when a static endpoint traversal order repeatedly exhausts whitespace on shallow violations while starving deep critical cones, or when an all-or-nothing commit rule discards valuable partial progress. In either case, tuning scalar parameters merely repeats the flawed behavior more aggressively. Escaping such stagnation requires modifying the executable repair policy itself.

\begin{figure}[!t]
  \centering
  \input{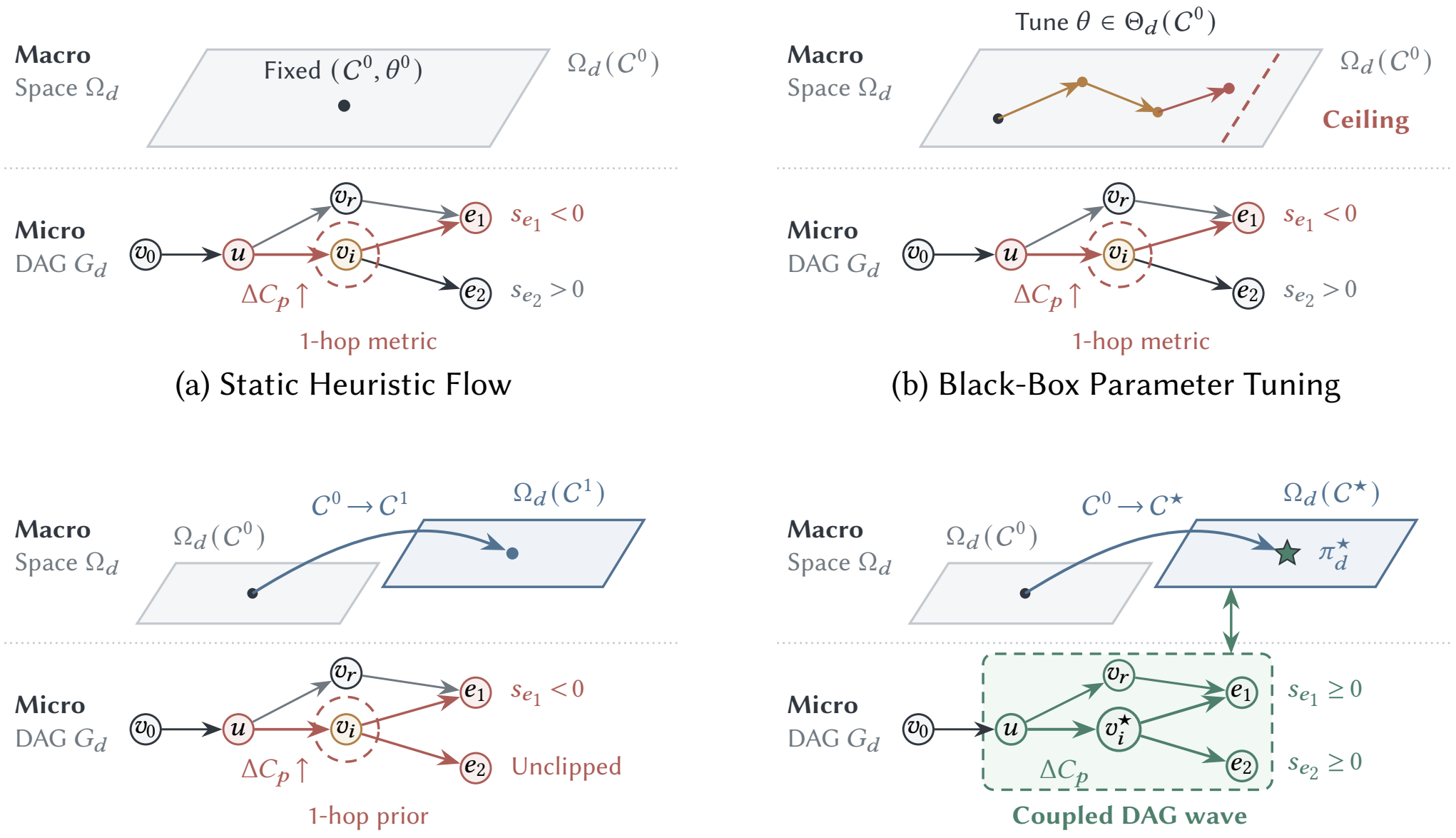}
  \caption{Evolution of post-placement timing and power optimization paradigms across macroscopic policy spaces ($\Omega_d$) and microscopic timing--power graphs ($G_d = (V_d, E_d)$). (a) Static heuristic flows fix $(\mathcal{C}^0,\theta^0)$ and evaluate isolated single-vertex moves on $G_d$, missing fanin edge load penalties and multi-hop cone conflicts. (b) Black-box parameter tuning explores $\theta \in \Theta_d(\mathcal{C}^0)$ within a single fixed policy plane and stalls at an expressivity ceiling. (c) Uncoupled outer-loop LLM policy adaptation~\cite{he2023chateda,he2025llmeda} expands macroscopic policies ($\mathcal{C}^0\!\to\!\mathcal{C}^1$) but still relies on myopic 1-hop vertex metrics on $G_d$. (d) Bi-level \tapco{} jointly evolves outer source policies to $\pi_d^\star$ and propagates inner replacements across coupled multi-hop DAG cones on $G_d$.}
  \Description{Four-panel schematic comparing static heuristic flow, black-box parameter tuning, outer-only LLM adaptation, and bi-level TAPCO-G across policy planes and timing-power DAGs.}
  \label{fig:paradigms}
\end{figure}

Large language models (LLMs)~\cite{chen2021codex} and autonomous coding agents~\cite{jimenez2024swebench,yang2024sweagent} provide a transformative paradigm for bridging this expressivity gap~\cite{yao2023react,shinn2023reflexion}. In electronic design automation (EDA)~\cite{he2025llmeda}, domain-adapted LLMs~\cite{liu2023chipnemo} and agentic workflows have demonstrated strong capabilities in script generation, design rule checking~\cite{zhu2023drcsg}, and tool orchestration~\cite{he2023chateda}. Because LLMs encode rich structural priors over source code, algorithmic control flow, and optimization heuristics, they can diagnose bottleneck symptoms from execution traces and synthesize targeted modifications to both tool configurations and source-level repair policies. Yet, language models alone lack fine-grained numerical precision on circuit netlists: an LLM cannot directly compute picosecond-level nonlinear delay model (NLDM) shifts or microwatt-level instance power tradeoffs across tens of thousands of gates.

We propose \tapco{}, a bi-level framework for post-placement timing and power
co-optimization (\cref{fig:paradigms}). In the outer loop, an LLM proposes
design-specific changes to the optimizer. To compare these proposals,
\tapco{} runs each candidate optimizer from the same initial placement.
During each run, the inner loop ranks cell replacements using graph-based
estimates of their timing and power effects, then tests selected replacements
on the placed design. Signoff results and execution diagnostics from the
completed run guide the next outer-loop proposal.

We validate \tapco{} on all eight ASAP7 7-nm benchmarks~\cite{clark2016asap7} from the MLCAD 2026 Contest~\cite{mlcad26contest,mlcad26results,ispd26contest}. Every emitted implementation undergoes strict placement legality checks and formal logic equivalence checking (LEC).

Our main contributions are summarized below.

\begin{enumerate}
  \item We develop an LLM-guided adaptation loop that uses physical-design
        feedback to tune tool settings and revise executable repair policies
        when the current policy limits further improvement.
  \item We propose a graph-based timing model anchored to STA snapshots that
        accounts for changes in upstream driver load and propagates delay shifts
        through reconvergent paths to estimate how cell replacements affect
        endpoint slack.
  \item We introduce coarse-to-fine power screening from cell families to
        individual instances and use batched physical trials with bisection to
        recover beneficial replacements from rejected batches.
  \item We evaluate \tapco{} through benchmark comparisons, physical replacement
        probes, and ablation studies, and show that it outperforms
        \tool{CERBERUS}~\cite{zhou2026cerberus} on all eight designs, with a mean
        quality score gain of $5.516$ points.
\end{enumerate}

The remainder of this article is organized as follows. \Cref{sec:background} introduces the post-placement timing and power graph, capacitive coupling models, and the constrained bi-level optimization formulation. \Cref{sec:method} presents the outer-loop LLM-guided hierarchical optimizer adaptation framework, including policy-space expansion, bottleneck-conditioned proposal generation, and valid-incumbent trajectory gating. \Cref{sec:graph} details the inner-loop graph-based physical priors, covering snapshot-anchored driver-load coupling, incremental reconvergent DAG propagation, sequential boundary handling, and coarse-to-fine instance power screening. \Cref{sec:experiments} reports comprehensive experimental evaluations, physical probe verifications, and ablation studies across the MLCAD 2026 Contest benchmark suite, and \cref{sec:conclusion} concludes the article.

\section{Preliminaries and Problem Formulation}
\label{sec:background}

We first formalize the electrical and physical models governing post-placement discrete cell transformations, and then cast design-adaptive timing and power co-optimization as a constrained bi-level optimization problem.

\subsection{Post-Placement Timing Graph and Capacitive Coupling}

Given a placed gate-level netlist $d$ with standard-cell and macro coordinates, Liberty timing/power views, and extracted interconnect parasitics, post-placement optimization applies gate sizing, multi-VT substitution, buffering, cloning, pin swapping, and other transformations that preserve functional equivalence, followed by placement legalization~\cite{chen1998gatewire,alpert2004buffer,sze2005path,do2019legalization,openroadrsz}. For each editable cell instance $i \in \mathcal{I}_d$ currently mapped to library master $c$, the technology library defines an admissible equivalent master set $\mathcal{V}(c)$ spanning alternative drive strengths and VT flavors (e.g., RVT, LVT, and SLVT).

\begin{figure}[!t]
  \centering
\begingroup
\let\tiny\scriptsize
\captionsetup[subfloat]{font=small,labelfont=normalfont,
  justification=centering,singlelinecheck=false,captionskip=4pt}
\subfloat[NLDM Surface $D_c(\sigma_{\mathrm{in}}, C_{\mathrm{load}})$ \& Load Coupling\label{fig:context-timing}]{%
\begin{minipage}[b]{0.49\linewidth}\centering
\begin{tikzpicture}[x=1cm,y=1cm,font=\sffamily\tiny,
  >=Stealth,
  gate/.style={draw=tapcoink,line width=0.65pt,fill=white}]
\path[use as bounding box] (0,0) rectangle (6.45,3.25);

\node[anchor=south,font=\sffamily\bfseries\tiny,text=tapcoink] at (1.42,2.78) {Driver--Sink Coupling};

\filldraw[gate,fill=tapconavyfill,draw=tapconavy] (0.28,1.52) -- (0.28,2.24) -- (0.94,1.88) -- cycle;
\node[font=\sffamily\bfseries\tiny,text=tapconavy] at (0.52,1.88) {$u$};
\node[anchor=north,text=tapconavy] at (0.56,1.46) {Driver};

\filldraw[gate,fill=tapcoamberfill,draw=tapcoamber,line width=0.7pt] (1.92,1.46) -- (1.92,2.30) -- (2.64,1.88) -- cycle;
\draw[tapcoamber,densely dashed,line width=0.55pt] (1.86,1.38) -- (1.86,2.38) -- (2.72,1.88) -- cycle;
\node[font=\sffamily\bfseries\tiny,text=tapcoink] at (2.18,1.88) {$i$};
\node[anchor=north,font=\sffamily\bfseries\tiny,text=tapcoamber] at (2.28,1.34) {$c \!\to\! c'$};

\draw[->,tapcoink,line width=0.65pt] (0.04,1.88) -- (0.28,1.88);
\draw[->,tapcoink,line width=0.7pt] (0.94,1.88) -- (1.92,1.88);
\draw[->,tapcoink,line width=0.65pt] (2.64,1.88) -- (2.94,1.88);

\fill[tapcoink] (1.36,1.88) circle (1.1pt);
\draw[tapcoink,line width=0.6pt] (1.36,1.88) -- (1.36,1.22);
\draw[tapcored,line width=0.8pt] (1.18,1.22) -- (1.54,1.22);
\draw[tapcored,line width=0.8pt] (1.18,1.08) -- (1.54,1.08);
\draw[tapcoink,line width=0.6pt] (1.36,1.08) -- (1.36,0.80);
\draw[tapcoink,line width=0.6pt] (1.20,0.80) -- (1.52,0.80);
\draw[tapcoink,line width=0.6pt] (1.26,0.71) -- (1.46,0.71);
\draw[tapcoink,line width=0.6pt] (1.32,0.62) -- (1.40,0.62);
\node[anchor=north,font=\sffamily\bfseries\tiny,text=tapcored] at (1.36,0.56) {$C_p^{\mathrm{pin}}\!\uparrow \;\Rightarrow\; C_u^{\mathrm{load}}\!\uparrow$};

\node[anchor=south,font=\sffamily\bfseries\tiny,text=tapcoink] at (4.82,2.78) {NLDM $D_c(\sigma_{\mathrm{in}}, C_{\mathrm{load}})$};

\filldraw[fill=tapconavyfill,draw=tapcorule,line width=0.5pt] (3.42,0.68) rectangle (6.16,2.62);
\fill[tapcotealfill] (3.95,0.68) .. controls (3.95,1.45) and (4.45,2.05) .. (6.16,2.05) -- (6.16,2.62) -- (3.42,2.62) -- (3.42,1.38) .. controls (3.72,1.38) and (3.95,1.05) .. (3.95,0.68) -- cycle;
\fill[tapcoamberfill] (4.58,0.68) .. controls (4.58,1.35) and (5.05,1.82) .. (6.16,1.82) -- (6.16,2.62) -- (3.42,2.62) -- (3.42,1.92) .. controls (4.05,1.92) and (4.58,1.35) .. (4.58,0.68) -- cycle;
\fill[tapcoredfill] (5.22,0.68) .. controls (5.22,1.25) and (5.58,1.62) .. (6.16,1.62) -- (6.16,2.62) -- (4.18,2.62) .. controls (4.72,2.15) and (5.22,1.42) .. (5.22,0.68) -- cycle;

\draw[tapconavy,densely dashed,line width=0.55pt] (3.42,1.38) .. controls (3.75,1.35) and (3.95,1.05) .. (3.95,0.68);
\draw[tapcoteal,densely dashed,line width=0.55pt] (3.42,1.92) .. controls (4.05,1.85) and (4.58,1.35) .. (4.58,0.68);
\draw[tapcored,densely dashed,line width=0.55pt] (4.18,2.62) .. controls (4.72,2.15) and (5.22,1.42) .. (5.22,0.68);

\draw[->,tapcoink,line width=0.65pt] (3.42,0.68) -- (6.32,0.68);
\draw[->,tapcoink,line width=0.65pt] (3.42,0.68) -- (3.42,2.76);
\node[anchor=north,text=tapcoink,font=\sffamily\bfseries\tiny] at (4.85,0.62) {Output load $C_{\mathrm{load}}$};
\node[anchor=south,rotate=90,text=tapcoink,font=\sffamily\bfseries\tiny] at (3.32,1.68) {Input slew $\sigma_{\mathrm{in}}$};

\fill[tapconavy] (4.02,1.18) circle (1.6pt);
\node[anchor=north,text=tapconavy,font=\sffamily\bfseries\tiny] at (4.02,1.12) {$c$};
\fill[tapcored] (5.38,2.12) circle (1.7pt);
\node[anchor=south,text=tapcored,font=\sffamily\bfseries\tiny] at (5.35,2.16) {$c': \sigma\!\uparrow, D\!\uparrow$};
\draw[->,tapcored,line width=0.8pt] (4.08,1.22) .. controls (4.62,1.42) and (4.95,1.72) .. (5.30,2.06);
\node[anchor=west,text=tapcored,font=\sffamily\bfseries\tiny] at (5.16,1.15) {$+\Delta C_p$};
\end{tikzpicture}
\end{minipage}}\hfill
\subfloat[Activity-Dependent Multi-VT Power\label{fig:context-power}]{%
\begin{minipage}[b]{0.49\linewidth}\centering
\begin{tikzpicture}[x=1cm,y=1cm,font=\sffamily\tiny,
  >=Stealth,
  gate/.style={draw=tapcoink,line width=0.65pt,fill=white}]
\path[use as bounding box] (0,0) rectangle (6.45,3.25);

\node[anchor=west,font=\sffamily\bfseries\tiny,text=tapconavy] at (0.08,2.92) {High $\rho_p$: $\dpower$ dominates};
\draw[tapconavy,line width=0.7pt]
  (0.12,2.18) -- (0.26,2.18) -- (0.26,2.58) -- (0.42,2.58) -- (0.42,2.18) --
  (0.58,2.18) -- (0.58,2.58) -- (0.74,2.58) -- (0.74,2.18) --
  (0.90,2.18) -- (0.90,2.58) -- (1.06,2.58) -- (1.06,2.18) -- (1.22,2.18);
\draw[->,tapcoink,line width=0.65pt] (1.26,2.38) -- (1.68,2.38);
\filldraw[gate,fill=tapconavyfill,draw=tapconavy] (1.68,2.04) -- (1.68,2.72) -- (2.32,2.38) -- cycle;
\node[font=\sffamily\bfseries\tiny,text=tapconavy] at (1.92,2.38) {$i_1$};
\draw[->,tapcoink,line width=0.65pt] (2.32,2.38) -- (2.78,2.38);
\node[anchor=north,text=tapcomuted] at (2.00,2.00) {Master $c$};

\node[anchor=west,font=\sffamily\bfseries\tiny,text=tapcoamber] at (0.08,1.42) {Low $\rho_p$: $\lpower$ dominates};
\draw[tapcoamber,line width=0.7pt]
  (0.12,0.68) -- (0.58,0.68) -- (0.58,1.08) -- (0.74,1.08) -- (0.74,0.68) -- (1.22,0.68);
\draw[->,tapcoink,line width=0.65pt] (1.26,0.88) -- (1.68,0.88);
\filldraw[gate,fill=tapcoamberfill,draw=tapcoamber] (1.68,0.54) -- (1.68,1.22) -- (2.32,0.88) -- cycle;
\node[font=\sffamily\bfseries\tiny,text=tapcoamber] at (1.92,0.88) {$i_2$};
\draw[->,tapcoink,line width=0.65pt] (2.32,0.88) -- (2.78,0.88);
\node[anchor=north,text=tapcomuted] at (2.00,0.50) {Master $c$};

\draw[->,tapcorule,densely dashed,line width=0.6pt] (2.82,2.38) -- (3.32,2.08);
\draw[->,tapcorule,densely dashed,line width=0.6pt] (2.82,0.88) -- (3.32,1.22);

\draw[tapcoink,line width=0.65pt] (3.42,0.78) -- (6.28,0.78);
\draw[->,tapcoink,line width=0.65pt] (3.42,0.78) -- (3.42,2.82);
\node[anchor=south,font=\sffamily\bfseries\tiny,text=tapcoink] at (4.85,2.85) {Multi-VT Power vs.\ Speed};

\filldraw[fill=tapcoamber,draw=tapcoink,line width=0.45pt] (3.72,0.78) rectangle (4.22,0.96);
\filldraw[fill=tapconavy,draw=tapcoink,line width=0.45pt] (3.72,0.96) rectangle (4.22,1.56);
\node[anchor=north,font=\sffamily\bfseries\tiny,text=tapcoink] at (3.97,0.74) {RVT};

\filldraw[fill=tapcoamber,draw=tapcoink,line width=0.45pt] (4.60,0.78) rectangle (5.10,1.22);
\filldraw[fill=tapconavy,draw=tapcoink,line width=0.45pt] (4.60,1.22) rectangle (5.10,1.88);
\node[anchor=north,font=\sffamily\bfseries\tiny,text=tapcoink] at (4.85,0.74) {LVT};

\filldraw[fill=tapcoamber,draw=tapcoink,line width=0.45pt] (5.48,0.78) rectangle (5.98,1.92);
\filldraw[fill=tapconavy,draw=tapcoink,line width=0.45pt] (5.48,1.92) rectangle (5.98,2.62);
\node[anchor=north,font=\sffamily\bfseries\tiny,text=tapcoink] at (5.73,0.74) {SLVT};

\draw[->,tapcoteal,line width=0.75pt] (3.85,1.72) -- node[above,sloped,font=\sffamily\bfseries\tiny,text=tapcoteal]{Speed $\uparrow$, $\lpower \uparrow\uparrow$} (5.58,2.72);

\fill[tapconavy] (3.55,0.06) rectangle (3.75,0.20);
\node[anchor=west,text=tapconavy,font=\sffamily\bfseries\tiny] at (3.79,0.13) {$\dpower(\rho_p)$};
\fill[tapcoamber] (4.98,0.06) rectangle (5.18,0.20);
\node[anchor=west,text=tapcoamber,font=\sffamily\bfseries\tiny] at (5.22,0.13) {$\lpower(\mathrm{VT})$};
\end{tikzpicture}
\end{minipage}}
\endgroup
  \caption{Electrical and physical coupling fundamentals in post-placement co-optimization. (a) NLDM lookup surface $D_c(\sigma_{\mathrm{in}}, C_{\mathrm{load}})$ and physical driver--sink capacitive coupling: upsizing sink $i$ ($c \!\to\! c'$) increases pin capacitance $C_p^{\mathrm{pin}}$, shifting the upstream driver operating point into a steeper delay/slew regime ($\sigma\!\uparrow$). (b) Instance-level switching activity density $\rho_p$ governs the balance between dynamic switching power $\dpower(\rho_p)$ and multi-VT static leakage $\lpower(\mathrm{VT})$ across RVT, LVT, and SLVT flavors.}
  \Description{Two-panel physical foundation showing the nonlinear NLDM delay-slew-load surface and driver-sink coupling on the left, and activity-dependent dynamic versus multi-VT leakage power tradeoff on the right.}
  \label{fig:physical-context}
\end{figure}
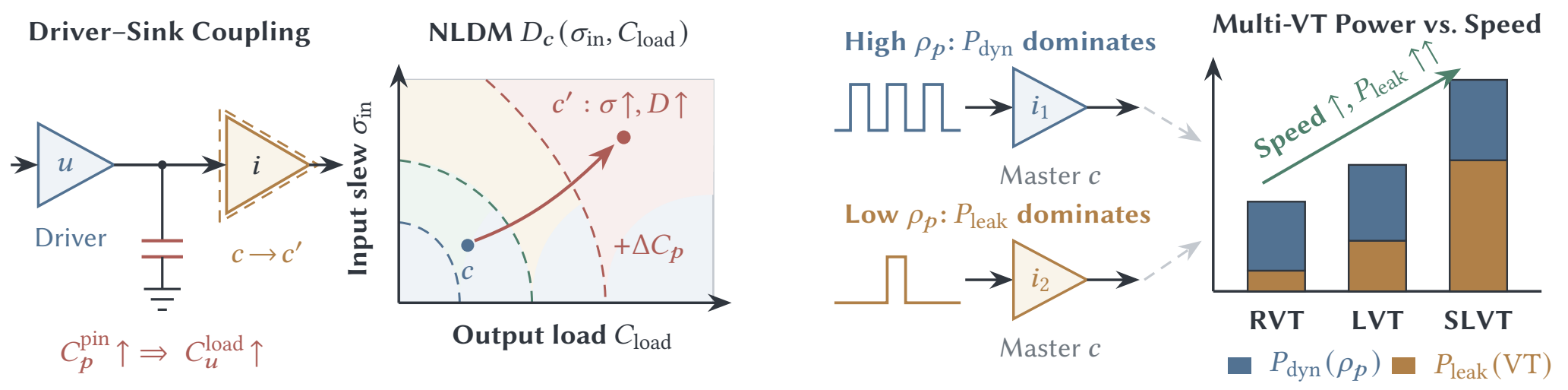

STA models signal propagation across combinational logic cones between sequential launch and capture boundaries as a timing and power DAG $G_d = (V_d, E_d)$. Each vertex $v \in V_d$ represents a pin transition (rise or fall), and each directed edge $(u,v) \in E_d$ represents a cell timing arc or interconnect wire arc with delay $\tau_{uv}$. Given launch arrival boundary conditions, internal arrival times propagate recursively via
\begin{equation}
  a_v = \max_{u \in \operatorname{pred}(v)} \bigl(a_u + \tau_{uv}\bigr).
  \label{eq:arrival}
\end{equation}
Let $\mathcal{E}_d$ denote the set of constrained setup endpoints, and let $V_e \subset V_d$ denote the analyzed transition vertices associated with endpoint $e \in \mathcal{E}_d$. With required arrival time $r_v$, the endpoint slack $s_e$ and circuit-wide setup TNS $T_d \le 0$ are defined as
\begin{equation}
  s_e = \min_{v \in V_e} \bigl(r_v - a_v\bigr), \qquad
  T_d = \sum_{e \in \mathcal{E}_d} \min(0, s_e).
  \label{eq:tns}
\end{equation}
Under standard NLDM lookup tables, the delay $D_c(\sigma_u, C_v^{\mathrm{load}})$ and output slew of a cell arc $(u,v)$ mapped to master $c$ are nonlinear functions of the input transition slew $\sigma_u$ and the effective output capacitive load $C_v^{\mathrm{load}}$. For a driver output pin $u$ driving sink pin set $\mathcal{F}(u)$ over interconnect wire capacitance $C_u^{\mathrm{wire}}$, the effective load is
\begin{equation}
  C_u^{\mathrm{load}} = C_u^{\mathrm{wire}} + \sum_{p \in \mathcal{F}(u)} C_p^{\mathrm{pin}}.
  \label{eq:load}
\end{equation}
As depicted in \cref{fig:context-timing}, substituting master $c \to c'$ at instance $i$ alters both its own arc delays $D_{c'}(\cdot)$ and the input pin capacitances $C_p^{\mathrm{pin}}(c')$, which directly perturbs the load $C_u^{\mathrm{load}}$ and delay of its upstream drivers~\cite{pan2026quadgradient,cheng2023dagsizer}. Furthermore, because $T_d$ sums clipped endpoint slacks $\min(0, s_e)$ across reconvergent cones, local delay shifts only improve $T_d$ to the extent that they reduce the dominant arrival at violating endpoints ($s_e < 0$).

\subsection{Dynamic and Leakage Power Decomposition}

Total circuit power consists of dynamic power $P_{d,\mathrm{dyn}}$ and static cell leakage power $P_{d,\mathrm{leak}}$~\cite{luo2008totalpower,lu2023ecognn}. Dynamic power includes net switching power and internal cell power due to short-circuit current and parasitics. We write these components as
\begin{equation}
  P_{d,\mathrm{dyn}} = \sum_{i \in \mathcal{I}_d} \bigl(P_{i,\mathrm{switch}} + P_{i,\mathrm{internal}}\bigr), \qquad
  P_{d,\mathrm{leak}} = \sum_{i \in \mathcal{I}_d} P_{i,\mathrm{leak}}.
  \label{eq:power}
\end{equation}
While leakage power $P_{i,\mathrm{leak}}$ is primarily determined by the cell master's VT and transistor width, dynamic power depends strongly on instance-level operating conditions: $P_{i,\mathrm{switch}}$ scales with net toggle activity and load capacitance, whereas $P_{i,\mathrm{internal}}$ depends on pin toggle rates, input slew, and output capacitance. Consequently, as shown in \cref{fig:context-power}, two instances $i_1$ and $i_2$ instantiated from the exact same master $c$ can exhibit vastly different dynamic-to-leakage ratios and timing slack headroom. Effective co-optimization must therefore account for instance-specific electrical context rather than relying solely on static library cell rankings.

\subsection{Bi-Level Optimizer Adaptation Problem Formulation}

Let $x_d^0$ denote the initial placed physical state of design $d$. An inner post-placement optimizer is parameterized by an executable repair policy $\mathcal{C}$ (governing endpoint traversal order, operator sequencing, effort allocation, and commit/rollback logic) and an exposed configuration vector $\theta$ (specifying setup/hold margins, pass budgets, utilization caps, and operator switches). Executing the optimizer on $x_d^0$ yields the optimized implementation
\begin{equation}
  x_d = A_{\mathcal{C},\theta}(x_d^0).
  \label{eq:inner}
\end{equation}
Relative to the initial state $x_d^0$ with setup TNS $T_d^0 < 0$ and baseline power components $P_{d,\mathrm{dyn}}^0, P_{d,\mathrm{leak}}^0 > 0$, the normalized timing and power improvements of implementation $x_d$ are defined as
\begin{align}
  \Delta T_d &= \frac{T_d - T_d^0}{|T_d^0|}, \label{eq:normalizedtiming}\\
  \Delta P_{d,x} &= \frac{P_{d,x}^0 - P_{d,x}}{P_{d,x}^0}, \qquad x \in \{\mathrm{dyn}, \mathrm{leak}\}.
  \label{eq:normalized}
\end{align}
A signoff domain evaluator measures the composite quality score $S_d(x_d)$ by combining weighted PPA improvements with normalized physical-design penalties $p_{d,j}$ (capturing electrical slew and capacitance violations, cell displacement, and routing congestion overflow):
\begin{equation}
  S_d(x_d) = w_T \Delta T_d + w_D \Delta P_{d,\mathrm{dyn}} + w_L \Delta P_{d,\mathrm{leak}} - \sum_j \lambda_j p_{d,j},
  \label{eq:score}
\end{equation}
where $(w_T, w_D, w_L)$ are nonnegative tradeoff weights and $\lambda_j \ge 0$ are penalty coefficients. Crucially, every emitted design must satisfy a strict binary feasibility predicate $\operatorname{valid}(x_d) \in \{0,1\}$, which enforces placement legality (zero cell overlap and valid site alignment) and formal LEC against $x_d^0$.

We formulate design-adaptive timing and power co-optimization as searching for a design-specialized policy $\mathcal{C}_d^\star$ and configuration $\theta_d^\star$ that maximize the evaluated signoff objective:
\begin{equation}
  \begin{aligned}
    (\mathcal{C}_d^\star, \theta_d^\star) &= \arg\max_{\mathcal{C} \in \mathfrak{C}_d,\, \theta \in \Theta_d(\mathcal{C})} S_d\!\left(A_{\mathcal{C},\theta}(x_d^0)\right), \\
    \text{s.t.}\qquad &\operatorname{valid}\!\left(A_{\mathcal{C},\theta}(x_d^0)\right) = 1.
  \end{aligned}
  \label{eq:outer}
\end{equation}
\begin{figure}[!t]
  \centering
\begingroup
\let\scriptsize\footnotesize
\captionsetup[subfloat]{font=small,labelfont=normalfont,justification=centering,singlelinecheck=false,captionskip=3pt}
\subfloat[Outer-Loop: Evidence-grounded LLM optimizer adaptation (\cref{sec:method})\label{fig:workflow-a}]{%
\begin{minipage}[b]{\linewidth}\centering
\begin{tikzpicture}[x=1cm,y=1cm,>=Latex]
\path[use as bounding box] (0,0) rectangle (13.2,2.25);

\draw[draw=tapcoink,line width=0.9pt,rounded corners=3.5pt,fill=white] (0.10,0.52) rectangle (3.90,2.16);
\node[align=center,font=\bfseries\scriptsize,text=tapcoink] at (2.00,1.76) {Structure Search Space\\[-1pt]Parameter \& policy slices};
\fill[fill=tapconavyfill,rounded corners=2.5pt] (0.28,0.64) rectangle (3.72,1.34);
\node[align=center,font=\scriptsize,text=tapconavy] at (2.00,0.99) {Coupled $(\mathcal{C}, \theta)$ hierarchy\\[-1pt]Policy slice $\Omega_d(\mathcal{C})$};
\node[align=center,font=\bfseries\scriptsize,text=tapcoink] at (2.00,0.22) {Sec.~\ref{sec:method-space}};

\draw[->,draw=tapcoink,line width=0.9pt] (3.90,1.34) -- (4.70,1.34);

\draw[draw=tapcoink,line width=0.9pt,rounded corners=3.5pt,fill=white] (4.70,0.52) rectangle (8.50,2.16);
\node[align=center,font=\bfseries\scriptsize,text=tapcoink] at (6.60,1.76) {Diagnose \& Replay\\[-1pt]Anchor-reset evaluation};
\fill[fill=tapconavyfill,rounded corners=2.5pt] (4.88,0.64) rectangle (8.32,1.34);
\node[align=center,font=\scriptsize,text=tapconavy] at (6.60,0.99) {Evidence-guided edit $\delta_d^k$\\[-1pt]Replay from anchor $x_d^0$};
\node[align=center,font=\bfseries\scriptsize,text=tapcoink] at (6.60,0.22) {Secs.~\ref{sec:method-proposal} and \ref{sec:method-update}};

\draw[->,draw=tapcoink,line width=0.9pt] (8.50,1.34) -- (9.30,1.34);

\draw[draw=tapcoink,line width=0.9pt,rounded corners=3.5pt,fill=white] (9.24,0.52) rectangle (13.16,2.16);
\node[align=center,font=\bfseries\scriptsize,text=tapcoink] at (11.20,1.76) {Expand Source Policy\\[-1pt]Escape stagnation ceiling};
\fill[fill=tapconavyfill,rounded corners=2.5pt] (9.38,0.64) rectangle (13.02,1.34);
\node[align=center,font=\scriptsize,text=tapconavy] at (11.20,0.99) {Stagnation-triggered $\mathcal{C}^k \!\to\! \mathcal{C}^{k+1}$\\[-1pt]Schedule \& commit evolution};
\node[align=center,font=\bfseries\scriptsize,text=tapcoink] at (11.20,0.22) {Sec.~\ref{sec:method-expansion}};
\end{tikzpicture}
\end{minipage}}\\[4pt]
\subfloat[Inner-Loop: Graph-guided physical replacement priors (\cref{sec:graph})\label{fig:workflow-b}]{%
\begin{minipage}[b]{\linewidth}\centering
\begin{tikzpicture}[x=1cm,y=1cm,>=Latex]
\path[use as bounding box] (0,0) rectangle (13.2,2.25);

\draw[draw=tapcoink,line width=0.9pt,rounded corners=3.5pt,fill=white] (0.10,0.52) rectangle (3.90,2.16);
\node[align=center,font=\bfseries\scriptsize,text=tapcoink] at (2.00,1.76) {Couple Driver Load\\[-1pt]Snapshot NLDM differential};
\fill[fill=tapcotealfill,rounded corners=2.5pt] (0.26,0.64) rectangle (3.74,1.34);
\filldraw[fill=white,draw=tapcoteal,line width=0.6pt] (0.48,0.99) circle (0.10cm);
\filldraw[fill=tapcoamberfill,draw=tapcoamber,line width=0.6pt] (0.92,0.99) circle (0.10cm);
\draw[->,>=Stealth,draw=tapcoteal,line width=0.65pt] (0.58,0.99) -- (0.82,0.99);
\node[align=center,font=\scriptsize,text=tapcoteal] at (2.38,0.99) {Anchor to live STA $\tau_a^s$\\[-1pt]Upstream $\Delta C_p$ edge};
\node[align=center,font=\bfseries\scriptsize,text=tapcoink] at (2.00,0.22) {Sec.~\ref{sec:graph-snapshot}};

\draw[->,draw=tapcoink,line width=0.9pt] (3.90,1.34) -- (4.70,1.34);

\draw[draw=tapcoink,line width=0.9pt,rounded corners=3.5pt,fill=white] (4.70,0.52) rectangle (8.50,2.16);
\node[align=center,font=\bfseries\scriptsize,text=tapcoink] at (6.60,1.76) {Propagate \& Clip Slack\\[-1pt]DAG cone \& sequential roles};
\fill[fill=tapcotealfill,rounded corners=2.5pt] (4.86,0.64) rectangle (8.34,1.34);
\filldraw[fill=white,draw=tapcoteal,line width=0.55pt] (5.04,0.99) circle (0.075cm);
\filldraw[fill=white,draw=tapcoteal,line width=0.55pt] (5.32,1.15) circle (0.075cm);
\filldraw[fill=white,draw=tapcomuted,line width=0.5pt] (5.32,0.83) circle (0.075cm);
\filldraw[fill=tapcoteal,draw=tapcoink,line width=0.5pt] (5.60,0.99) circle (0.075cm);
\draw[->,>=Stealth,draw=tapcoteal,line width=0.55pt] (5.11,1.03) -- (5.25,1.12);
\draw[->,>=Stealth,draw=tapcomuted,densely dashed,line width=0.5pt] (5.11,0.95) -- (5.25,0.86);
\draw[->,>=Stealth,draw=tapcoteal,line width=0.55pt] (5.39,1.12) -- (5.53,1.03);
\node[align=center,font=\scriptsize,text=tapcoteal] at (7.00,0.99) {Event-driven $\delta a_v$ wave\\[-1pt]Clipped endpoint $\widehat{g}_T(q)$};
\node[align=center,font=\bfseries\scriptsize,text=tapcoink] at (6.60,0.22) {Secs.~\ref{sec:graph-prop} and \ref{sec:graph-seq}};

\draw[->,draw=tapcoink,line width=0.9pt] (8.50,1.34) -- (9.30,1.34);

\draw[draw=tapcoink,line width=0.9pt,rounded corners=3.5pt,fill=white] (9.20,0.52) rectangle (13.18,2.16);
\node[align=center,font=\bfseries\scriptsize,text=tapcoink] at (11.19,1.76) {Screen Power \& Commit\\[-1pt]Coarse-to-fine \& bisection};
\fill[fill=tapcotealfill,rounded corners=2.5pt] (9.32,0.64) rectangle (13.06,1.34);
\filldraw[fill=white,draw=tapcoteal,line width=0.55pt] (9.62,1.15) circle (0.075cm);
\filldraw[fill=tapcoteal,draw=tapcoink,line width=0.5pt] (9.46,0.84) circle (0.075cm);
\filldraw[fill=tapcoredfill,draw=tapcored,line width=0.5pt] (9.78,0.84) circle (0.075cm);
\draw[->,>=Stealth,draw=tapcoteal,line width=0.55pt] (9.58,1.08) -- (9.49,0.92);
\draw[->,>=Stealth,draw=tapcored,line width=0.55pt] (9.66,1.08) -- (9.75,0.92);
\node[align=center,font=\fontsize{7.5pt}{8.5pt}\selectfont,text=tapcoteal] at (11.44,0.99) {Activity-weighted $\rho_p$ funnel\\[-1pt]Recursive batch bisection};
\node[align=center,font=\bfseries\scriptsize,text=tapcoink] at (11.19,0.22) {Secs.~\ref{sec:graph-power} and \ref{sec:graph-trials}};
\end{tikzpicture}
\end{minipage}}
\endgroup
  \caption{Bi-level methodological overview of \tapco{}. (a) The outer-loop LLM optimizer adaptation engine (\cref{sec:method}) structures coupled $(\mathcal{C},\theta)$ slices, replays evidence-conditioned edits from the clean placement anchor $x_d^0$, and expands source policies when parameter sweeps stagnate. (b) The inner-loop graph-guided replacement engine (\cref{sec:graph}) couples snapshot NLDM differentials with upstream driver loads, propagates clipped arrival shifts across reconvergent cones, and screens candidates via an activity-weighted power funnel and recursive batch bisection.}
  \Description{Two-tier three-card methodological overview of TAPCO-G showing the outer-loop LLM adaptation stages on top and the inner-loop graph-guided physical prior stages on bottom.}
  \label{fig:workflow}
\end{figure}
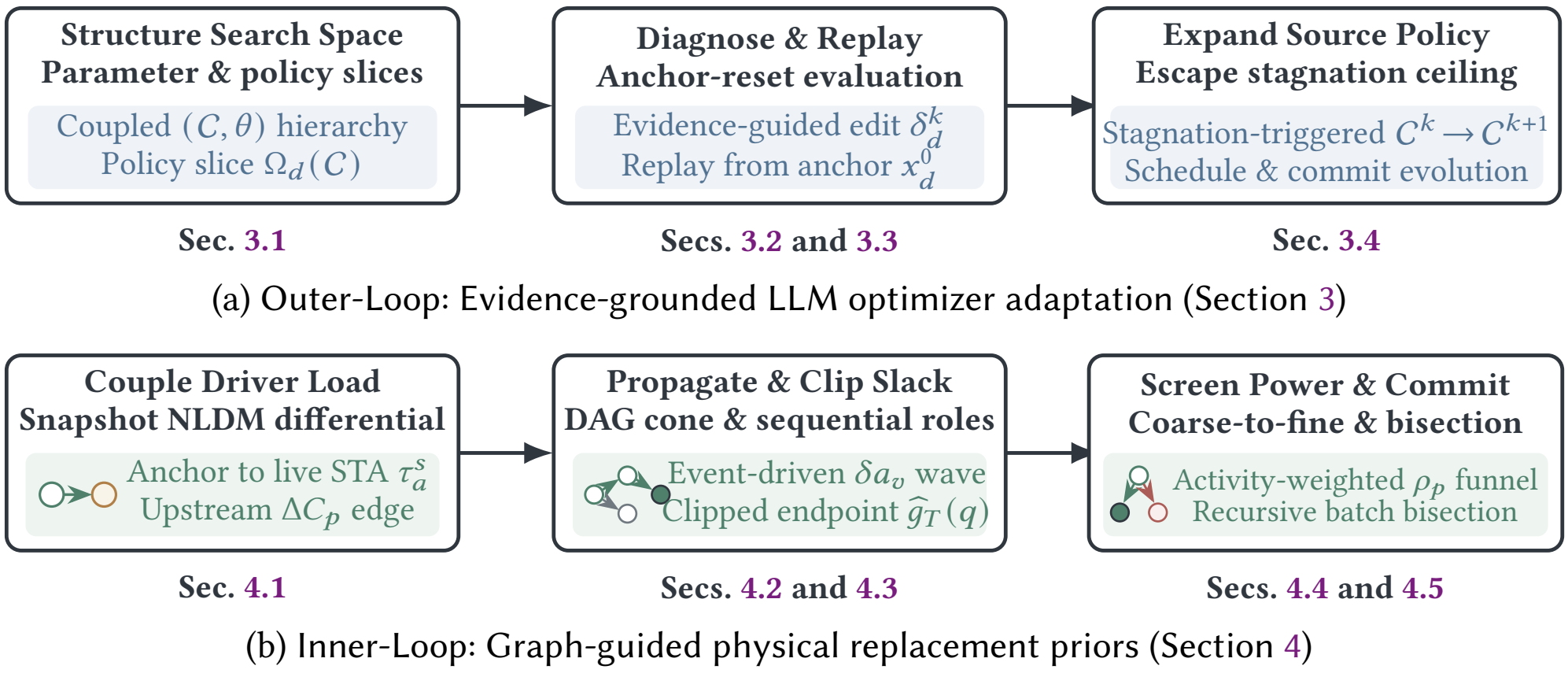

In \tapco{}, the outer loop (\cref{sec:method}) solves \cref{eq:outer} by leveraging an LLM program prior and evaluator feedback to jointly evolve $\mathcal{C}$ and $\theta$, while the inner loop (\cref{sec:graph}) embeds graph-based physical priors inside $A_{\mathcal{C},\theta}$ to maximize the accuracy and efficiency of candidate cell replacements.

\section{Outer-Loop: LLM-Guided Hierarchical Optimizer Adaptation}
\label{sec:method}

Solving the bi-level co-optimization problem in \cref{eq:outer} requires more than sampling numeric configurations within a static parameter box: the attainable quality frontier itself depends on how the optimizer schedules repair effort and commits physical transformations. \tapco{} realizes outer-loop optimizer adaptation as an iterative, evidence-conditioned process (\cref{fig:workflow}) that bridges macroscopic program reasoning and microscopic physical execution. At each iteration, the LLM leverages its structural prior over source code and optimization heuristics to diagnose the limiting bottleneck and synthesize a bounded modification at either the parameter or policy level. The physical-design pipeline then runs the candidate optimizer on the placed circuit, invoking the inner graph-guided replacement engine developed in \cref{sec:graph}. The pipeline returns objective PPA scores, legality/equivalence verdicts, and fine-grained execution diagnostics to ground subsequent adaptation.

\subsection{Hierarchical Adaptation Space}
\label{sec:method-space}

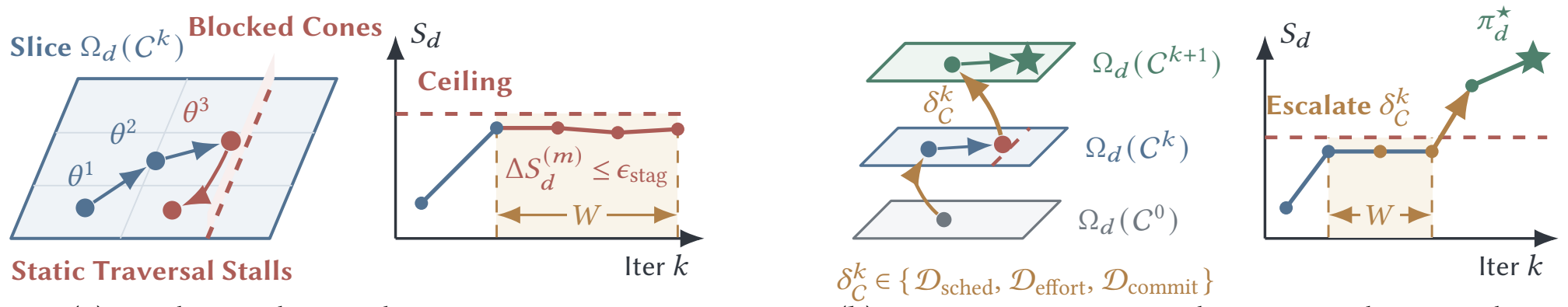
\begin{figure}[!t]
  \centering
\begingroup
\let\tiny\scriptsize
\captionsetup[subfloat]{font=small,labelfont=normalfont,justification=centering,singlelinecheck=false,captionskip=3pt}
\subfloat[Within-policy ceiling \& stagnation ($\Delta S_d^{(m)} \le \epsilon_{\mathrm{stag}}$)\label{fig:mot-3-1-a}]{%
\begin{minipage}[b]{0.48\linewidth}\centering
\begin{tikzpicture}[x=1cm,y=1cm,font=\sffamily\scriptsize]
\path[use as bounding box] (0,0) rectangle (6.2,2.55);

\filldraw[fill=tapconavyfill,draw=tapconavy,line width=0.75pt,line join=round]
  (0.15,0.42) -- (2.35,0.42) -- (2.92,1.78) -- (0.72,1.78) -- cycle;
\draw[tapconavy!25,line width=0.35pt] (0.34,0.87) -- (2.54,0.87);
\draw[tapconavy!25,line width=0.35pt] (0.53,1.32) -- (2.73,1.32);
\draw[tapconavy!25,line width=0.35pt] (1.08,0.42) -- (1.65,1.78);
\node[anchor=south west,font=\sffamily\bfseries\tiny,text=tapconavy] at (0.02,1.80) {Slice $\Omega_d(\mathcal{C}^k)$};

\fill[tapcoredfill] (1.82,0.42) -- (2.39,1.78) -- (2.39,2.00) -- (1.82,0.64) -- cycle;
\draw[tapcored,dashed,line width=1.05pt] (1.82,0.42) -- (2.39,1.78);
\node[anchor=south,font=\sffamily\bfseries\tiny,text=tapcored] at (2.48,2.02) {Blocked Cones};

\node[circle,fill=tapconavy,inner sep=1.6pt] (t1) at (0.78,0.68) {};
\node[circle,fill=tapconavy,inner sep=1.6pt] (t2) at (1.38,1.08) {};
\node[circle,fill=tapcored,inner sep=1.7pt] (t3) at (2.02,1.25) {};
\node[circle,fill=tapcored,inner sep=1.6pt] (t4) at (1.52,0.66) {};
\node[anchor=south,font=\sffamily\tiny,text=tapconavy] at (0.75,0.74) {$\theta^1$};
\node[anchor=south east,font=\sffamily\tiny,text=tapconavy] at (1.36,1.10) {$\theta^2$};
\node[anchor=south east,font=\sffamily\bfseries\tiny,text=tapcored] at (1.98,1.28) {$\theta^3$};
\draw[->,>=Latex,tapconavy,line width=0.8pt] (t1) -- (t2);
\draw[->,>=Latex,tapconavy,line width=0.8pt] (t2) -- (t3);
\draw[->,>=Latex,tapcored,line width=0.85pt] (t3) .. controls (1.88,0.92) and (1.72,0.75) .. (t4);
\node[anchor=north,font=\sffamily\bfseries\tiny,text=tapcored] at (1.35,0.36) {Static Traversal Stalls};

\draw[->,>=Latex,tapcoink,line width=0.65pt] (3.42,0.42) -- (6.05,0.42) node[below left,font=\sffamily\tiny] {Iter $k$};
\draw[->,>=Latex,tapcoink,line width=0.65pt] (3.42,0.42) -- (3.42,2.15) node[right,font=\sffamily\bfseries\scriptsize] {$S_d$};

\fill[tapcoamberfill] (4.28,0.43) rectangle (5.82,1.48);
\draw[tapcoamber,dashed,line width=0.55pt] (4.28,0.43) -- (4.28,1.48);
\draw[tapcoamber,dashed,line width=0.55pt] (5.82,0.43) -- (5.82,1.48);
\draw[<->,>=Latex,tapcoamber,line width=0.6pt] (4.28,0.62) -- (5.82,0.62);
\node[fill=tapcoamberfill,inner sep=1pt,font=\sffamily\bfseries\tiny,text=tapcoamber] at (5.05,0.62) {$W$};
\node[font=\sffamily\bfseries\tiny,text=tapcored] at (5.05,1.02) {$\Delta S_d^{(m)} \!\le\! \epsilon_{\mathrm{stag}}$};

\draw[tapcored,dashed,line width=0.95pt] (3.42,1.48) -- (5.95,1.48);
\node[anchor=south west,font=\sffamily\bfseries\tiny,text=tapcored] at (3.48,1.50) {Ceiling};

\draw[tapconavy,line width=1.0pt] (3.64,0.72) -- (4.28,1.36);
\draw[tapcored,line width=1.0pt] (4.28,1.36) -- (4.80,1.36) -- (5.31,1.32) -- (5.82,1.35);
\fill[tapconavy] (3.64,0.72) circle (1.6pt);
\fill[tapconavy] (4.28,1.36) circle (1.6pt);
\fill[tapcored] (4.80,1.36) circle (1.6pt);
\fill[tapcored] (5.31,1.32) circle (1.6pt);
\fill[tapcored] (5.82,1.35) circle (1.6pt);
\end{tikzpicture}
\end{minipage}}\hfill
\subfloat[Stagnation-triggered source-policy escalation\label{fig:mot-3-1-b}]{%
\begin{minipage}[b]{0.48\linewidth}\centering
\begin{tikzpicture}[x=1cm,y=1cm,font=\sffamily\scriptsize]
\path[use as bounding box] (0,0) rectangle (6.2,2.55);

\filldraw[fill=tapcograyfill,draw=tapcomuted,line width=0.6pt,line join=round]
  (0.08,0.42) -- (1.52,0.42) -- (1.84,0.74) -- (0.40,0.74) -- cycle;
\node[anchor=west,font=\sffamily\bfseries\tiny,text=tapcomuted] at (1.86,0.58) {$\Omega_d(\mathcal{C}^0)$};
\node[circle,fill=tapcomuted,inner sep=1.3pt] (s0) at (0.85,0.58) {};

\filldraw[fill=tapconavyfill,draw=tapconavy,line width=0.65pt,line join=round]
  (0.14,1.04) -- (1.58,1.04) -- (1.90,1.36) -- (0.46,1.36) -- cycle;
\draw[tapcored,dashed,line width=0.85pt] (1.26,1.04) -- (1.58,1.36);
\node[anchor=west,font=\sffamily\bfseries\tiny,text=tapconavy] at (1.92,1.20) {$\Omega_d(\mathcal{C}^k)$};
\node[circle,fill=tapconavy,inner sep=1.4pt] (sk1) at (0.72,1.18) {};
\node[circle,fill=tapcored,inner sep=1.5pt] (sk2) at (1.35,1.22) {};
\draw[->,>=Latex,tapconavy,line width=0.7pt] (sk1) -- (sk2);

\filldraw[fill=tapcotealfill,draw=tapcoteal,line width=0.75pt,line join=round]
  (0.20,1.76) -- (1.64,1.76) -- (1.96,2.08) -- (0.52,2.08) -- cycle;
\node[anchor=west,font=\sffamily\bfseries\tiny,text=tapcoteal] at (1.98,1.92) {$\Omega_d(\mathcal{C}^{k+1})$};
\node[circle,fill=tapcoteal,inner sep=1.4pt] (sk3) at (0.92,1.90) {};
\node[star,star points=5,star point ratio=2.2,fill=tapcoteal,draw=tapcoteal,inner sep=1.3pt] (sk4) at (1.56,1.94) {};
\draw[->,>=Latex,tapcoteal,line width=0.8pt] (sk3) -- (sk4);

\draw[->,>=Latex,tapcoamber,line width=0.85pt] (s0) .. controls (0.62,0.78) and (0.60,0.98) .. (sk1);
\draw[->,>=Latex,tapcoamber,line width=1.05pt] (sk2) .. controls (1.28,1.54) and (1.10,1.72) .. (sk3);
\node[anchor=east,font=\sffamily\bfseries\tiny,text=tapcoamber] at (1.08,1.56) {$\delta_{\mathcal{C}}^k$};

\node[anchor=north,font=\sffamily\bfseries\tiny,text=tapcoamber] at (1.55,0.34) {$\delta_{\mathcal{C}}^k \!\in\! \{\mathcal{D}_{\mathrm{sched}},\mathcal{D}_{\mathrm{effort}},\mathcal{D}_{\mathrm{commit}}\}$};

\draw[->,>=Latex,tapcoink,line width=0.65pt] (3.58,0.42) -- (6.08,0.42) node[below left,font=\sffamily\tiny] {Iter $k$};
\draw[->,>=Latex,tapcoink,line width=0.65pt] (3.58,0.42) -- (3.58,2.15) node[right,font=\sffamily\bfseries\scriptsize] {$S_d$};

\fill[tapcoamberfill] (4.12,0.43) rectangle (5.00,1.28);
\draw[tapcoamber,dashed,line width=0.55pt] (4.12,0.43) -- (4.12,1.28);
\draw[tapcoamber,dashed,line width=0.55pt] (5.00,0.43) -- (5.00,1.28);
\draw[<->,>=Latex,tapcoamber,line width=0.6pt] (4.12,0.62) -- (5.00,0.62);
\node[fill=tapcoamberfill,inner sep=1pt,font=\sffamily\bfseries\tiny,text=tapcoamber] at (4.56,0.62) {$W$};

\draw[tapcored,dashed,line width=0.9pt] (3.58,1.28) -- (5.98,1.28);

\draw[tapconavy,line width=1.0pt] (3.76,0.68) -- (4.12,1.16) -- (5.00,1.16);
\fill[tapconavy] (3.76,0.68) circle (1.6pt);
\fill[tapconavy] (4.12,1.16) circle (1.6pt);
\fill[tapcoamber] (4.56,1.16) circle (1.6pt);
\fill[tapcoamber] (5.00,1.16) circle (1.6pt);

\draw[->,>=Latex,tapcoamber,line width=1.1pt] (5.00,1.16) -- (5.34,1.72);
\node[anchor=east,font=\sffamily\bfseries\tiny,text=tapcoamber] at (4.96,1.54) {Escalate $\delta_{\mathcal{C}}^k$};

\draw[tapcoteal,line width=1.1pt] (5.34,1.72) -- (5.86,1.95);
\fill[tapcoteal] (5.34,1.72) circle (1.6pt);
\node[star,star points=5,star point ratio=2.2,fill=tapcoteal,draw=tapcoteal,inner sep=1.3pt] at (5.86,1.95) {};
\node[anchor=south east,font=\sffamily\bfseries\scriptsize,text=tapcoteal] at (5.82,1.98) {$\pi_d^\star$};
\end{tikzpicture}
\end{minipage}}
\endgroup
  \caption{Hierarchical adaptation space and stagnation-triggered policy escalation. (a) Within a fixed policy slice $\Omega_d(\mathcal{C}^k)$, scalar parameter tuning stalls at an expressivity ceiling ($\Delta S_d^{(m)} \le \epsilon_{\mathrm{stag}}$ over window $W$) as static traversal repeatedly hits blocked cones. (b) Detecting stagnation triggers a source-policy edit $\delta_{\mathcal{C}}^k$ (\cref{sec:method-expansion}), transitioning from $\Omega_d(\mathcal{C}^k)$ to $\Omega_d(\mathcal{C}^{k+1})$ to unlock higher QoR frontiers.}
  \Description{Comparison between parameter tuning stalling at a policy slice ceiling over window W versus stagnation-triggered source-policy escalation jumping to an expanded policy slice.}
  \label{fig:mot-3-1}
\end{figure}

Conventional automated flow tuning treats an EDA tool as an immutable black box and restricts search to its exposed scalar hyperparameters~\cite{geng2023ptpt,zheng2023flowtuning,murali2023autotuning}. However, when the underlying repair schedule or commit logic mismatches a circuit's topological structure, varying scalar margins or pass budgets rapidly hits an expressivity ceiling (\cref{fig:mot-3-1-a}). Conversely, unconstrained code synthesis risks destabilizing core physical operators. To bridge this gap, our hierarchical adaptation space organizes optimizer customization into coupled parameter slices and executable policy planes, allowing \tapco{} to exploit fast parameter specialization when sufficient and seamlessly expand into new policy slices when structural stagnation occurs (\cref{fig:mot-3-1-b}).

Formally, let $\pi = (\mathcal{C}, \theta)$ denote a complete candidate optimizer comprising an executable source policy $\mathcal{C}$ and its exposed configuration vector $\theta \in \Theta_d(\mathcal{C})$. Holding the executable policy $\mathcal{C}$ fixed induces a policy-conditioned parameter slice $\Omega_d(\mathcal{C})$, while varying $\mathcal{C}$ across the family of reachable policies $\mathfrak{C}_d$ defines the full hierarchical search space:
\begin{equation}
  \Omega_d(\mathcal{C}) = \bigl\{(\mathcal{C}, \theta) \;\big|\; \theta \in \Theta_d(\mathcal{C})\bigr\}, \qquad
  \Omega_d^{\mathrm{TAPCO-G}} = \bigcup_{\mathcal{C} \in \mathfrak{C}_d} \Omega_d(\mathcal{C}).
  \label{eq:hierarchicalspace}
\end{equation}
This hierarchy spans three tightly coupled architectural layers:
\begin{enumerate}
  \item \emph{Physical Kernel Layer:} Provides primitive netlist and layout transformations, including gate sizing, multi-VT substitution, buffer insertion/removal, gate cloning, pin swapping, and detailed placement legalization, alongside the inner-loop graph-guided replacement operator (\cref{sec:graph}).
  \item \emph{Parameter Envelope Layer ($\theta$):} Exposes continuous and categorical controls that bound the repair envelope, including setup/hold slack margins, electrical slew and capacitance tolerances, maximum wire-length limits, endpoint TNS coverage ratios, pass iteration caps, and discrete operator enablement switches.
  \item \emph{Executable Source-Policy Layer ($\mathcal{C}$):} Defines how the optimizer uses the parameter envelope to prioritize and re-sort violating endpoints across passes, scale repair effort with violation depth, filter candidate pools for timing and power, and decide when to commit or roll back intermediate netlist states.
\end{enumerate}
Rather than treating hyperparameter tuning and code synthesis as disjoint paradigms, \tapco{} navigates $\Omega_d^{\mathrm{TAPCO-G}}$ from the inside out: it prioritizes lightweight parameter edits within the active slice $\Omega_d(\mathcal{C}^k)$ whenever bottlenecks map to exposed controls, and invokes a policy transition $\mathcal{C}^k \to \mathcal{C}^{k+1}$ when empirical evidence reveals a structural limitation in $\mathcal{C}^k$.

With \cref{eq:hierarchicalspace}, \tapco{} strictly generalizes conventional hyperparameter tuning ($\Omega_d(\mathcal{C}^0) \subset \Omega_d^{\mathrm{TAPCO-G}}$): it preserves the sample efficiency and stability of parameter optimization within well-behaved slices while empowering the optimizer to reshape its own decision logic and configuration interface $\Theta_d(\mathcal{C})$ on complex corner cases.

\subsection{Evidence-Conditioned LLM Proposal}
\label{sec:method-proposal}

\begin{figure}[!t]
  \centering
\begingroup
\let\tiny\scriptsize
\captionsetup[subfloat]{font=small,labelfont=normalfont,justification=centering,singlelinecheck=false,captionskip=3pt}
\subfloat[Scalar blindness \& chained netlist drift\label{fig:mot-3-2-a}]{%
\begin{minipage}[b]{0.48\linewidth}\centering
\begin{tikzpicture}[x=1cm,y=1cm,font=\sffamily\scriptsize]
\path[use as bounding box] (0,0) rectangle (6.2,2.80);

\node[anchor=west,font=\sffamily\bfseries\tiny,text=tapcoamber] at (0.12,2.62) {Shallow Wide};
\draw[tapcoink,line width=0.5pt] (0.15,2.44) -- (1.42,2.44);
\foreach \x/\h in {0.20/0.20, 0.42/0.23, 0.64/0.18, 0.86/0.22, 1.08/0.20, 1.26/0.19}
  \fill[tapcoamber] (\x,2.44) rectangle +(0.14,-\h);

\node[anchor=west,font=\sffamily\bfseries\tiny,text=tapcored] at (0.12,1.98) {Deep Cone};
\draw[tapcoink,line width=0.5pt] (0.15,1.80) -- (1.42,1.80);
\fill[tapcored!35] (0.22,1.80) rectangle (0.38,1.72);
\fill[tapcored]    (0.48,1.80) rectangle (0.68,1.44);
\fill[tapcored]    (0.78,1.80) rectangle (0.98,1.48);
\fill[tapcored!35] (1.08,1.80) rectangle (1.24,1.73);

\draw[->,>=Latex,tapcoink,line width=0.7pt] (1.48,2.30) -- (2.12,2.14);
\draw[->,>=Latex,tapcoink,line width=0.7pt] (1.48,1.66) -- (2.12,1.90);
\filldraw[fill=tapcograyfill,draw=tapcoink,line width=0.65pt,line join=round]
  (2.15,2.34) -- (3.08,2.16) -- (3.08,1.88) -- (2.15,1.70) -- cycle;
\node[font=\sffamily\bfseries\tiny,text=tapcoink] at (2.60,2.02) {Scalar};

\draw[->,>=Latex,tapcoink,line width=0.75pt] (3.10,2.02) -- (3.48,2.02);
\filldraw[fill=tapcoredfill,draw=tapcored,rounded corners=1.5pt,line width=0.65pt]
  (3.50,1.78) rectangle (5.25,2.26);
\node[font=\sffamily\bfseries\scriptsize,text=tapcored] at (4.38,2.02) {$S_d = -48\text{ ns}$};
\draw[->,>=Latex,tapcored,line width=0.75pt] (5.27,2.02) -- (5.62,2.02);
\node[circle,fill=tapcoredfill,draw=tapcored,line width=0.7pt,inner sep=1.1pt,font=\sffamily\bfseries\scriptsize,text=tapcored] at (5.86,2.02) {?};

\draw[tapcorule,dashed,line width=0.6pt] (0.15,1.32) -- (6.05,1.32);

\fill[tapcoredfill,rounded corners=3pt] (2.85,0.08) rectangle (6.02,1.16);
\draw[tapcored!65,dashed,line width=0.6pt,rounded corners=3pt] (2.85,0.08) rectangle (6.02,1.16);
\node[anchor=north east,font=\sffamily\bfseries\tiny,text=tapcored] at (5.94,1.12) {Chained Drift / DRC!};

\node[circle,draw=tapcoink,fill=white,line width=0.8pt,inner sep=1.9pt] (x0) at (0.55,0.68) {};
\node[circle,draw=tapconavy,fill=tapconavy,line width=0.8pt,inner sep=1.9pt] (x1) at (1.85,0.82) {};
\node[circle,draw=tapcored,fill=tapcoredfill,line width=0.85pt,inner sep=2.1pt] (x2) at (3.48,0.68) {};
\node[circle,draw=tapcored,fill=tapcored,line width=0.8pt,inner sep=1.9pt] (x3) at (4.95,0.34) {};

\node[anchor=south,font=\sffamily\bfseries\scriptsize,text=tapcoink] at (0.55,0.78) {$x_d^0$};
\node[anchor=south,font=\sffamily\bfseries\scriptsize,text=tapconavy] at (1.85,0.92) {$x_d^1$};
\node[anchor=north east,font=\sffamily\bfseries\scriptsize,text=tapcored] at (3.42,0.58) {$x_d^2$};
\node[anchor=west,font=\sffamily\bfseries\scriptsize,text=tapcored] at (5.08,0.34) {$x_d^3$};

\draw[->,>=Latex,tapconavy,line width=0.9pt] (x0) .. controls (1.10,0.78) and (1.45,0.82) .. (x1);
\draw[->,>=Latex,tapcored,line width=0.9pt] (x1) .. controls (2.45,0.82) and (2.95,0.74) .. (x2);
\draw[->,>=Latex,tapcored,dashed,line width=0.9pt] (x2) .. controls (4.05,0.54) and (4.48,0.40) .. (x3);
\draw[tapcored,line width=1.0pt] (3.38,0.58) -- (3.58,0.78) (3.38,0.78) -- (3.58,0.58);
\end{tikzpicture}
\end{minipage}}\hfill
\subfloat[Diagnostic-guided radial replay from anchor $x_d^0$\label{fig:mot-3-2-b}]{%
\begin{minipage}[b]{0.48\linewidth}\centering
\begin{tikzpicture}[x=1cm,y=1cm,font=\sffamily\scriptsize]
\path[use as bounding box] (0,0) rectangle (6.2,2.80);

\filldraw[fill=tapconavyfill,draw=tapconavy,rounded corners=2pt,line width=0.7pt]
  (0.10,1.38) rectangle (2.28,2.68);
\node[anchor=north west,font=\sffamily\bfseries\tiny,text=tapconavy] at (0.18,2.64) {Diagnostics $g_d^k$};

\draw[tapcoink,line width=0.45pt] (0.22,2.04) -- (1.12,2.04);
\fill[tapconavy!55] (0.28,2.04) rectangle (0.41,1.88);
\fill[tapcored]     (0.47,2.04) rectangle (0.60,1.70);
\fill[tapcored]     (0.66,2.04) rectangle (0.79,1.74);
\fill[tapconavy!55] (0.85,2.04) rectangle (0.98,1.90);
\node[anchor=north,font=\sffamily\tiny,text=tapcoink] at (0.67,1.68) {Slack};

\filldraw[fill=tapcoamberfill,draw=tapcoamber,line width=0.55pt,line join=round]
  (1.32,1.72) -- (2.08,1.72) -- (1.70,2.04) -- cycle;
\fill[tapcored] (1.70,1.84) circle (1.7pt);
\node[anchor=north,font=\sffamily\tiny,text=tapcoink] at (1.70,1.68) {Cone};

\draw[->,>=Latex,tapconavy,line width=0.8pt] (1.19,1.36) -- (1.19,1.04);
\filldraw[fill=tapcotealfill,draw=tapcoteal,rounded corners=2pt,line width=0.7pt]
  (0.10,0.32) rectangle (2.28,1.02);
\node[align=center,font=\sffamily\bfseries\scriptsize,text=tapcoteal] at (1.19,0.67) {LLM Proposal\\[-1pt]\tiny $(h_d^k,\, \delta_d^k)$};

\node[circle,draw=tapcoink,fill=white,line width=0.9pt,inner sep=2.3pt] (anc) at (2.88,1.48) {};
\node[anchor=south east,font=\sffamily\bfseries\scriptsize,text=tapcoink] at (2.84,1.56) {$x_d^0$};
\node[anchor=north east,font=\sffamily\tiny,text=tapcomuted] at (2.82,1.40) {Anchor};

\draw[->,>=Latex,tapcoteal,line width=0.8pt] (2.30,0.67) .. controls (2.68,0.67) and (2.88,1.02) .. (anc.south);

\node[circle,fill=tapcoteal,inner sep=1.7pt] (r1) at (3.92,2.22) {};
\node[circle,fill=tapcoredfill,draw=tapcored,line width=0.7pt,inner sep=1.7pt] (r2) at (3.92,1.48) {};
\node[star,star points=5,star point ratio=2.2,fill=tapcoteal,draw=tapcoteal,inner sep=1.4pt] (r3) at (3.92,0.82) {};

\draw[->,>=Latex,tapcoteal,line width=0.85pt] (anc) -- (r1);
\draw[->,>=Latex,tapcored,line width=0.85pt] (anc) -- (r2);
\draw[->,>=Latex,tapcoteal,line width=0.85pt] (anc) -- (r3);

\draw[tapcored,line width=0.95pt] (3.83,1.39) -- (4.01,1.57) (3.83,1.57) -- (4.01,1.39);

\node[anchor=south,font=\sffamily\bfseries\tiny,text=tapcoteal] at (3.92,2.28) {$\widetilde{x}_d^1$};
\node[anchor=south,font=\sffamily\bfseries\tiny,text=tapcored] at (3.92,1.58) {$\widetilde{x}_d^2$};
\node[anchor=south,font=\sffamily\bfseries\tiny,text=tapcoteal] at (3.92,0.90) {$\widetilde{x}_d^3$};

\filldraw[fill=tapcoamberfill,draw=tapcoamber,rounded corners=1.5pt,line width=0.65pt]
  (3.15,0.08) rectangle (4.85,0.48);
\node[font=\sffamily\bfseries\tiny,text=tapcoamber] at (4.00,0.28) {History $H_d^k$};
\draw[->,>=Latex,tapcoamber,dashed,line width=0.8pt] (4.05,1.38) .. controls (4.42,1.12) and (4.42,0.75) .. (4.28,0.50);
\draw[->,>=Latex,tapcoamber,dashed,line width=0.75pt] (3.13,0.28) .. controls (2.68,0.28) and (2.52,0.45) .. (2.30,0.52);

\filldraw[fill=tapcotealfill,draw=tapcoteal,rounded corners=2pt,line width=0.75pt]
  (4.58,1.18) rectangle (6.18,2.48);
\node[anchor=north,font=\sffamily\bfseries\tiny,text=tapcoteal] at (5.38,2.44) {Incumbent $I_d^k$};
\draw[tapcoteal,line width=1.05pt] (4.82,1.32) -- (5.18,1.32) -- (5.18,1.60) -- (5.58,1.60) -- (5.58,1.90) -- (5.98,1.90);
\fill[tapcoteal] (5.18,1.60) circle (1.4pt);
\fill[tapcoteal] (5.58,1.90) circle (1.4pt);

\draw[->,>=Latex,tapcoteal,line width=0.85pt] (r1) -- (4.63,2.08);
\draw[->,>=Latex,tapcoteal,line width=0.85pt] (r3) .. controls (4.32,0.88) and (4.45,1.32) .. (4.63,1.48);
\end{tikzpicture}
\end{minipage}}
\endgroup
  \caption{Diagnostic-guided LLM proposal and common-anchor execution protocol. (a) Scalar black-box scores mask distinct physical slack distributions while chained netlist executions accumulate placement and electrical drift. (b) \tapco{} conditions proposals on multi-channel diagnostics $g_d^k$ and replays every candidate radially from the clean anchor $x_d^0$ (\cref{sec:method-update}), logging all trials in $H_d^k$ while monotonically updating the valid incumbent $I_d^k$.}
  \Description{Comparison showing scalar ambiguity and chained netlist drift on the left, versus multi-channel diagnostic conditioning and radial common-anchor replay with monotonic incumbent tracking on the right.}
  \label{fig:mot-3-2}
\end{figure}
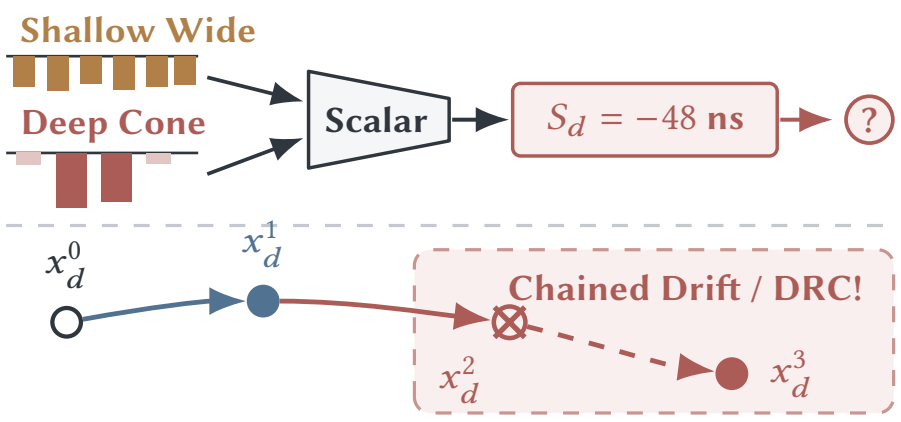
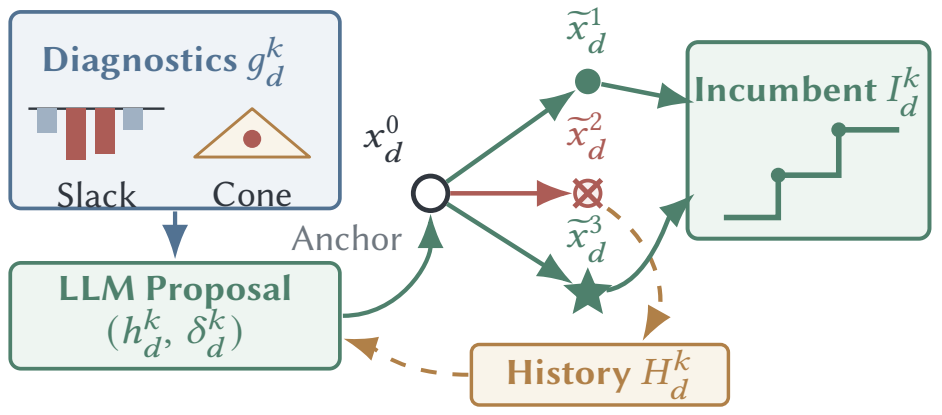

Black-box optimization surrogates compress an entire physical-design execution into a single scalar objective value $S_d$~\cite{jones1998ego,snoek2012bo}. In post-placement co-optimization, however, identical scalar scores can mask diametrically opposed physical bottlenecks. For instance, a $-48\text{ ns}$ TNS caused by thousands of shallow path violations requires broad multi-VT recovery, whereas the same $-48\text{ ns}$ TNS concentrated in two deeply congested logic cones demands targeted buffering and cloning (\cref{fig:mot-3-2-a}). Our proposal mechanism resolves this ambiguity by conditioning the LLM's code and strategy prior on rich, multi-channel physical diagnostics alongside the full historical trajectory of accepted and rejected trials, transforming black-box guessing into hypothesis-driven engineering (\cref{fig:mot-3-2-b}).

Specifically, at outer iteration $k$, \tapco{} maintains both the active working optimizer $\pi_d^k$ and the best valid incumbent tuple $I_d^k$:
\begin{equation}
  I_d^k = \bigl(\pi_{d,b}^k,\, x_{d,b}^k,\, S_d(x_{d,b}^k)\bigr), \qquad x_{d,b}^k = A_{\pi_{d,b}^k}(x_d^0).
  \label{eq:incumbent}
\end{equation}
Each evaluated candidate optimizer produces a decomposed evaluation vector $z_d^k$ paired with a structured diagnostic trace $g_d^k$:
\begin{equation}
  z_d^k = \bigl[\Delta T_d,\, \Delta P_{d,\mathrm{dyn}},\, \Delta P_{d,\mathrm{leak}},\, \{p_{d,j}\},\, v_d\bigr]^k, \qquad
  H_d^k = \bigl\{\bigl(\pi_d^t, z_d^t, g_d^t\bigr)\bigr\}_{t=0}^k,
  \label{eq:state}
\end{equation}
where $v_d = \operatorname{valid}(x_d) \in \{0,1\}$ records placement legality and formal LEC status, and $g_d^k$ captures the endpoint slack distribution, worst violating cones, breakdown of applied move classes (sizing, VT swap, buffering, cloning), residual electrical design rule violation (DRV) counts, and any compilation or legalization failure logs.

A structured prompt constructor $\Phi$ compiles the design profile, current working optimizer $\pi_d^k$, incumbent reference $I_d^k$, cumulative trial history $H_d^k$, and writable code/parameter scope $\mathcal{B}_d^k$ into a decision context. Conditioned on this context, the LLM agent $\mathcal{L}_\phi$ generates a structured triplet:
\begin{equation}
  (h_d^k,\, \ell_d^k,\, \delta_d^k) = \mathcal{L}_\phi\!\left(\Phi\bigl(d, \pi_d^k, I_d^k, H_d^k, \mathcal{B}_d^k\bigr)\right), \qquad
  \ell_d^k \in \{\mathrm{parameter}, \mathrm{policy}\},
  \label{eq:proposal}
\end{equation}
where $h_d^k$ is an explicit root-cause bottleneck hypothesis linking observed symptoms in $g_d^k$ to specific optimizer mechanisms, $\ell_d^k$ selects the adaptation level, and $\delta_d^k$ is a single bounded diff applied within $\mathcal{B}_d^k$. Restricting $\delta_d^k$ to one coherent intervention per iteration ensures unambiguous credit assignment when the next measurement $z_d^{k+1}$ arrives.

With \cref{eq:incumbent,eq:state,eq:proposal}, \tapco{} decouples hypothesis generation from objective evaluation: the LLM exploits fine-grained diagnostic semantics $g_d^k$ and historical contrast $H_d^k$ to propose high-leverage edits $\delta_d^k$, eliminating blind random perturbations while maintaining clear causal attribution across iterations.

\subsection{Physical-Design Execution and Evaluator-Grounded Update}
\label{sec:method-update}

When adapting an optimizer across sequential trials, two pitfalls commonly arise: first, chaining candidate executions on top of previously modified netlists causes cumulative state drift, where a single aggressive pass that introduces subtle placement congestion or electrical degradation permanently corrupts downstream trials (\cref{fig:mot-3-2-a}); second, greedy hill-climbing discards failed or regressing trials from memory, causing the agent to repeatedly revisit invalid parameter regions. To prevent state drift while preserving negative evidence, our execution and update protocol anchors every candidate trial to the identical initial placed state $x_d^0$ and separates non-monotonic diagnostic memory accumulation from strict monotonic incumbent retention (\cref{fig:mot-3-2-b}).

Given the proposed edit $(\ell_d^k, \delta_d^k)$, an instantiation operator $U_{\ell_d^k}$ constructs the candidate optimizer $\widetilde{\pi}_d^{k+1}$ and executes it from the clean initial placement $x_d^0$:
\begin{equation}
  \widetilde{\pi}_d^{k+1} = U_{\ell_d^k}\bigl(\pi_d^k, \delta_d^k\bigr), \qquad
  \widetilde{x}_d^{k+1} = A_{\widetilde{\pi}_d^{k+1}}(x_d^0).
  \label{eq:candidate}
\end{equation}
When $\ell_d^k = \mathrm{parameter}$, $U_{\ell_d^k}$ updates the Tcl/JSON configuration $\theta$; when $\ell_d^k = \mathrm{policy}$, $U_{\ell_d^k}$ patches the source policy $\mathcal{C}$, compiles the updated binary, and verifies syntactic and structural integrity prior to netlist execution. During $A_{\widetilde{\pi}_d^{k+1}}(x_d^0)$, the pipeline performs initial parasitic extraction and STA, executes electrical DRV repair (fixing slew, capacitance, fanout, and wire-length violations), conducts setup timing and multi-VT power co-optimization integrated with the inner-loop graph-guided replacement module (\cref{sec:graph}), legalizes modified cell placements, and runs signoff STA, power analysis, and independent LEC verification.

Upon completion, the incumbent optimizer $\pi_{d,b}^{k+1}$ is updated strictly under verified improvement:
\begin{equation}
  \pi_{d,b}^{k+1} =
  \begin{cases}
    \widetilde{\pi}_d^{k+1}, & \text{if } \operatorname{valid}(\widetilde{x}_d^{k+1}) = 1 \;\wedge\; S_d(\widetilde{x}_d^{k+1}) > S_d(x_{d,b}^k), \\
    \pi_{d,b}^k, & \text{otherwise}.
  \end{cases}
  \label{eq:accept}
\end{equation}
Crucially, whether or not $\widetilde{\pi}_d^{k+1}$ updates $I_d^{k+1}$, the observation record $(\widetilde{\pi}_d^{k+1}, z_d^{k+1}, g_d^{k+1})$ is appended to the history $H_d^{k+1}$. A trial that triggers legalization overflow or trades excessive leakage for marginal slack serves as informative negative evidence, guiding $\mathcal{L}_\phi$ in iteration $k+1$ to refine margins or bound operator aggressiveness.

With \cref{eq:candidate,eq:accept}, \tapco{} guarantees monotonically nondecreasing signoff quality $S_d(x_{d,b}^k)$ and $100\%$ legality/LEC compliance on the retained solution, while harvesting informative diagnostic signals from both successful and unsuccessful experiments without state contamination.

\subsection{Adaptive Policy-Space Expansion}
\label{sec:method-expansion}

In complex post-placement regimes, scaling up scalar budgets (such as pass counts or endpoint coverage percentages) under a fixed control policy often amplifies inefficiency rather than resolving violations. For example, if a static schedule visits endpoints in a fixed initial order and aborts whenever a pass fails to improve global score, increasing the pass budget simply re-executes expensive buffering on blocked paths and discards valid local timing recoveries (\cref{fig:mot-3-1-a}). Adaptive policy-space expansion overcomes this bottleneck by detecting expressivity saturation from historical traces $H_d^k$ and synthesizing targeted source-policy interventions that restructure endpoint ordering, effort allocation, and commit granularity (\cref{fig:mot-3-1-b}).

To formalize this transition, let $\Delta S_d^{(m)} = S_d(x_{d,b}^k) - S_d(x_{d,b}^{k-m})$ denote the incumbent improvement over a trailing window of $m$ parameter-tuning trials within slice $\Omega_d(\mathcal{C}^k)$. When $\Delta S_d^{(m)} \le \epsilon_{\mathrm{stag}}$ and diagnostic traces $g_d^{k-m:k}$ reveal recurring structural bottlenecks, the prompt constructor $\Phi$ escalates $\ell_d^k \gets \mathrm{policy}$. The traces may show the same critical endpoints left untouched across parameter sweeps, high move-rejection rates, or power-recovery rollbacks. Formally, a policy edit $\delta_{\mathcal{C}}^k$ maps the current policy $\mathcal{C}^k$ to a new executable policy $\mathcal{C}^{k+1} = U_{\mathrm{policy}}(\mathcal{C}^k, \delta_{\mathcal{C}}^k)$ across four primary structural dimensions:
\begin{equation}
  \mathcal{C}^{k+1} = \mathcal{C}^k \oplus \delta_{\mathcal{C}}^k, \qquad
  \delta_{\mathcal{C}}^k \in \bigl\{\mathcal{D}_{\mathrm{sched}},\, \mathcal{D}_{\mathrm{effort}},\, \mathcal{D}_{\mathrm{commit}},\, \mathcal{D}_{\mathrm{cand}}\bigr\},
  \label{eq:policyexpansion}
\end{equation}
where $\mathcal{D}_{\mathrm{sched}}$ modifies dynamic endpoint prioritization (recomputing and re-sorting the violating frontier after intermediate netlist updates), $\mathcal{D}_{\mathrm{effort}}$ implements violation-severity-aware effort allocation (scaling transformation budgets proportionally to path slack depth and skipping repeatedly unrepairable cones), $\mathcal{D}_{\mathrm{commit}}$ refines state retention (preserving beneficial partial sub-passes and enabling bounded rollback recovery), and $\mathcal{D}_{\mathrm{cand}}$ customizes candidate pre-filtering and inner-loop graph-prior invocation schedules. Once $\mathcal{C}^{k+1}$ is instantiated, \tapco{} resumes parameter tuning over the newly induced configuration domain $\Theta_d(\mathcal{C}^{k+1})$.

With \cref{eq:policyexpansion}, \tapco{} breaks free from the static expressive limits of off-the-shelf EDA engines: it automatically evolves custom scheduling, effort-allocation, and commit algorithms tailored to each circuit's topology, and then fine-tunes their numeric parameters to reach operating points unattainable by hyperparameter tuning alone.

\section{Inner-Loop: Graph-Guided Physical Priors for Cell Replacement}
\label{sec:graph}

While the outer loop (\cref{sec:method}) adapts the macroscopic scheduling and commit policy $\pi = (\mathcal{C}, \theta)$, the inner loop must evaluate and execute thousands of discrete cell replacements (sizing and multi-VT substitutions) during timing and power repair. Evaluating every candidate replacement via full incremental STA and signoff power analysis is computationally prohibitive, whereas relying on isolated library cell lookups leads to severe estimation errors due to upstream capacitive loading, reconvergent path switching, and instance-level activity variations. To achieve both high physical fidelity and fast candidate screening, \tapco{} embeds a five-stage graph-guided physical prior (\cref{fig:workflow-b}) directly inside the inner optimization loop.

\subsection{Snapshot-Anchored Replacement and Driver-Load Coupling}
\label{sec:graph-snapshot}

\begin{figure}[!t]
  \centering
\begingroup
\let\tiny\scriptsize
\captionsetup[subfloat]{font=small,labelfont=normalfont,justification=centering,singlelinecheck=false,captionskip=3pt}
\subfloat[Physical netlist to annotated DAG $G_d=(V_d,E_d)$\label{fig:mot-4-1-a}]{%
\begin{minipage}[b]{0.485\linewidth}\centering
\begin{tikzpicture}[x=1cm,y=1cm,font=\sffamily\scriptsize,>=Stealth]
\path[use as bounding box] (0,0.05) rectangle (6.2,2.75);

\node[text=tapcomuted,font=\sffamily\bfseries\tiny,anchor=west] at (0.12,2.58) {Physical Netlist};

\filldraw[fill=tapcograyfill,draw=tapcoink,line width=0.7pt] (0.28,1.86) rectangle (0.76,2.42);
\node[font=\sffamily\tiny,text=tapcoink] at (0.52,2.22) {FF};
\draw[line width=0.55pt,draw=tapcoink] (0.28,1.92) -- (0.40,1.98) -- (0.28,2.04);

\draw[line width=0.65pt,draw=tapcoink,->] (0.76,2.14) -- (1.35,2.14);

\filldraw[fill=tapconavyfill,draw=tapconavy,line width=0.75pt]
  (1.35,1.90) -- (1.35,2.38) -- (1.88,2.14) -- cycle;
\node[font=\sffamily\bfseries\tiny,text=tapconavy] at (1.52,2.14) {$u$};

\draw[line width=0.65pt,draw=tapcoink] (1.88,2.14) -- (2.32,2.14);
\fill[tapcoink] (2.32,2.14) circle (1.1pt);
\draw[line width=0.65pt,draw=tapcoink,->] (2.32,2.14) |- (2.78,2.32);
\draw[line width=0.6pt,draw=tapcomuted,->] (2.32,2.14) |- (2.78,1.88);

\filldraw[fill=tapcotealfill,draw=tapcoteal,line width=0.8pt]
  (2.78,2.08) -- (2.78,2.56) -- (3.34,2.32) -- cycle;
\node[font=\sffamily\bfseries\tiny,text=tapcoteal] at (2.96,2.32) {$i$};

\filldraw[fill=tapcograyfill,draw=tapcomuted,line width=0.65pt]
  (2.78,1.70) -- (2.78,2.04) -- (3.18,1.87) -- cycle;

\draw[line width=0.65pt,draw=tapcoink,->] (3.34,2.32) -- (4.15,2.32);

\filldraw[fill=tapcograyfill,draw=tapcoink,line width=0.7pt] (4.15,1.86) rectangle (4.63,2.42);
\node[font=\sffamily\tiny,text=tapcoink] at (4.39,2.22) {FF};
\draw[line width=0.55pt,draw=tapcoink] (4.15,1.92) -- (4.27,1.98) -- (4.15,2.04);

\foreach \x in {0.76, 1.62, 3.06, 4.15} {
  \draw[line width=0.6pt,draw=tapconavy,dashed,->] (\x,1.80) -- (\x,1.42);
}
\node[fill=white,inner sep=1.2pt,text=tapconavy,font=\sffamily\bfseries\tiny] at (2.45,1.61) {Pin-Transition DAG Mapping};

\node[text=tapcoink,font=\sffamily\bfseries\tiny,anchor=west] at (0.12,1.26) {$G_d=(V_d,E_d)$};

\node[circle,draw=tapcoink,fill=tapcograyfill,line width=0.7pt,minimum size=0.30cm,inner sep=0pt,font=\sffamily\tiny] (vq) at (0.76,0.92) {$q_0$};
\node[circle,draw=tapconavy,fill=tapconavyfill,line width=0.75pt,minimum size=0.30cm,inner sep=0pt,font=\sffamily\tiny] (pu) at (1.50,0.92) {$p_u$};
\node[circle,draw=tapconavy,fill=tapconavyfill,line width=0.75pt,minimum size=0.30cm,inner sep=0pt,font=\sffamily\tiny] (yu) at (2.26,0.92) {$u$};
\node[circle,draw=tapcoteal,fill=tapcotealfill,line width=0.8pt,minimum size=0.32cm,inner sep=0pt,font=\sffamily\bfseries\tiny,text=tapcoteal] (pi) at (3.06,1.12) {$p_i$};
\node[circle,draw=tapcomuted,fill=tapcograyfill,line width=0.65pt,minimum size=0.28cm,inner sep=0pt,font=\sffamily\tiny,text=tapcomuted] (pj) at (3.06,0.58) {$p_j$};
\node[circle,draw=tapcoteal,fill=tapcotealfill,line width=0.8pt,minimum size=0.32cm,inner sep=0pt,font=\sffamily\bfseries\tiny,text=tapcoteal] (yi) at (3.86,1.12) {$y_i$};
\node[circle,draw=tapcoink,fill=tapcograyfill,line width=0.7pt,minimum size=0.30cm,inner sep=0pt,font=\sffamily\tiny] (e1) at (4.60,1.12) {$e_1$};

\draw[line width=0.7pt,draw=tapcoamber,dashed,->] (vq) -- (pu);
\draw[line width=0.7pt,draw=tapcoamber,dashed,->] (yu) -- (pi);
\draw[line width=0.65pt,draw=tapcoamber,dashed,->] (yu) -- (pj);
\draw[line width=0.7pt,draw=tapcoamber,dashed,->] (yi) -- (e1);
\draw[line width=0.9pt,draw=tapconavy,->] (pu) -- (yu);
\draw[line width=0.95pt,draw=tapcoteal,->] (pi) -- (yi);

\draw[rounded corners=2pt,draw=tapconavy,fill=tapconavyfill,line width=0.65pt] (4.72,1.50) rectangle (6.18,2.44);
\node[text=tapconavy,font=\sffamily\bfseries\tiny] at (5.45,2.24) {Vertex $v \in V_d$};
\node[text=tapcoink,font=\sffamily\tiny] at (5.45,1.96) {$(a_v^s,\, s_v^s,\, \rho_p)$};
\node[text=tapcomuted,font=\sffamily\tiny] at (5.45,1.68) {STA + Activity};
\draw[line width=0.55pt,draw=tapconavy,->] (4.72,1.82) -- (3.98,1.24);

\draw[line width=0.9pt,draw=tapconavy,->] (0.04,0.22) -- (0.42,0.22);
\node[text=tapcoink,font=\sffamily\tiny,anchor=west] at (0.45,0.22) {Cell arc $\tau_{uv}^{\mathrm{cell}}\!=\!D_{c,a}(\sigma_a^s,C_a^s)$};
\draw[line width=0.7pt,draw=tapcoamber,dashed,->] (3.76,0.22) -- (4.14,0.22);
\node[text=tapcoink,font=\sffamily\tiny,anchor=west] at (4.17,0.22) {Net arc $\tau_{uv}^{\mathrm{net}}(C_u^s)$};
\end{tikzpicture}
\end{minipage}}\hfill
\subfloat[Snapshot-anchored fanin--target edge coupling\label{fig:mot-4-1-b}]{%
\begin{minipage}[b]{0.485\linewidth}\centering
\begin{tikzpicture}[x=1cm,y=1cm,font=\sffamily\scriptsize,>=Stealth]
\path[use as bounding box] (0,0.05) rectangle (6.2,2.75);

\draw[rounded corners=4pt,draw=tapcoteal,dashed,line width=0.8pt,fill=tapcotealfill,fill opacity=0.35]
  (0.15,0.48) rectangle (4.18,2.42);
\node[text=tapcoteal,font=\sffamily\bfseries\tiny,anchor=south] at (2.16,2.43) {Coupled 2-Hop Subgraph on $G_d$};

\node[circle,draw=tapconavy,fill=tapconavyfill,line width=0.75pt,minimum size=0.34cm,inner sep=0pt,font=\sffamily\tiny] (pdrv) at (0.52,1.42) {$p_d$};
\node[circle,draw=tapcoamber,fill=tapcoamberfill,line width=0.85pt,minimum size=0.36cm,inner sep=0pt,font=\sffamily\bfseries\tiny,text=tapcoamber] (udrv) at (1.68,1.42) {$u_{\mathrm{drv}}$};
\node[circle,draw=tapcoteal,fill=tapcotealfill,line width=0.85pt,minimum size=0.36cm,inner sep=0pt,font=\sffamily\bfseries\tiny,text=tapcoteal] (pi) at (2.78,1.76) {$p_i$};
\node[circle,draw=tapcoteal,fill=tapcotealfill,line width=0.85pt,minimum size=0.36cm,inner sep=0pt,font=\sffamily\bfseries\tiny,text=tapcoteal] (yi) at (3.82,1.76) {$y_i$};
\node[circle,draw=tapcomuted,fill=tapcograyfill,line width=0.7pt,minimum size=0.32cm,inner sep=0pt,font=\sffamily\tiny,text=tapcomuted] (pj) at (2.78,0.84) {$p_j$};

\draw[line width=1.05pt,draw=tapcoamber,->] (pdrv) -- (udrv);
\node[text=tapcoamber,font=\sffamily\bfseries\tiny,anchor=south] at (1.10,1.52) {$\Delta\tau_{\mathrm{drv}}\!=\!+3\text{ ps}$};
\node[text=tapcoink,font=\sffamily\tiny,anchor=north] at (1.10,1.30) {$\widehat{C}_u\!=\!C_u^s\!+\!\Delta C_p$};

\draw[line width=0.75pt,draw=tapconavy,dashed,->] (udrv) -- (pi);
\draw[line width=0.65pt,draw=tapcomuted,dashed,->] (udrv) -- (pj);
\node[text=tapcomuted,font=\sffamily\tiny,anchor=west] at (3.02,0.84) {Sink $p_j$};

\draw[line width=0.8pt,draw=tapcoamber,->,bend right=32] (pi.north west) to (udrv.north);
\node[text=tapcoamber,font=\sffamily\bfseries\tiny,anchor=south] at (2.18,2.02) {$\Delta C_p(q)\!\uparrow$};

\draw[line width=1.05pt,draw=tapcoteal,->] (pi) -- (yi);
\node[text=tapcoteal,font=\sffamily\bfseries\tiny,anchor=south] at (3.30,1.88) {$\Delta\tau_i\!=\!-8\text{ ps}$};
\node[text=tapcoteal,font=\sffamily\tiny,anchor=north] at (3.30,1.64) {$c \to c'$};

\draw[rounded corners=2pt,draw=tapcorule,fill=tapcograyfill,line width=0.6pt] (4.32,1.26) rectangle (6.12,2.42);
\node[text=tapcoink,font=\sffamily\bfseries\tiny] at (5.22,2.26) {NLDM Anchor $\tau_a^s$};
\draw[line width=0.55pt,draw=tapcoink,->] (4.48,1.42) -- (5.98,1.42);
\draw[line width=0.55pt,draw=tapcoink,->] (4.48,1.42) -- (4.48,2.10);
\draw[line width=0.8pt,draw=tapconavy] (4.54,1.50) .. controls (4.95,1.56) and (5.45,1.82) .. (5.90,2.04);
\fill[tapcoink] (4.95,1.63) circle (1.3pt);
\fill[tapcoamber] (5.52,1.86) circle (1.3pt);
\draw[line width=0.6pt,draw=tapcoamber,->] (5.00,1.66) -- (5.46,1.84);
\node[text=tapcoink,font=\sffamily\tiny,anchor=south] at (4.84,1.64) {$C_u^s$};
\node[text=tapcoamber,font=\sffamily\bfseries\tiny,anchor=south] at (5.54,1.86) {$\widehat{C}_u$};

\draw[rounded corners=2pt,draw=tapcored,fill=tapcoredfill,line width=0.65pt] (4.24,0.46) rectangle (6.18,1.14);
\node[text=tapcored,font=\sffamily\bfseries\tiny] at (5.21,0.92) {Isolated Lookup};
\node[text=tapcored,font=\sffamily\tiny] at (5.21,0.64) {Misses $\Delta C_p$ ($+3\text{ ps}$)};

\node[text=tapcoteal,font=\sffamily\bfseries\tiny] at (2.16,0.22) {Coupled Edge Sum: $\Delta\tau_{\mathrm{drv}} + \Delta\tau_i = -5\text{ ps}$};
\end{tikzpicture}
\end{minipage}}
\endgroup
  \caption{Circuit-to-graph mapping and snapshot-anchored edge coupling on $G_d = (V_d, E_d)$. (a) Physical netlist cells, flip-flops, and interconnects map to pin-transition vertices $v \in V_d$ annotated with snapshot state $(a_v^s, s_v^s, \rho_p)$ and directed cell/net timing arcs $e \in E_d$. (b) Replacing a candidate vertex sub-graph ($c \to c'$) anchors differential NLDM lookups to $\tau_a^s$ and couples intrinsic cell-arc speedups with upstream fanin driver-edge capacitance perturbations $\Delta C_p$.}
  \Description{Two-panel graph formulation showing physical circuit mapping to annotated timing-power DAG G_d on the left, and snapshot-anchored differential cell and upstream fanin edge coupling on the right.}
  \label{fig:mot-4-1}
\end{figure}
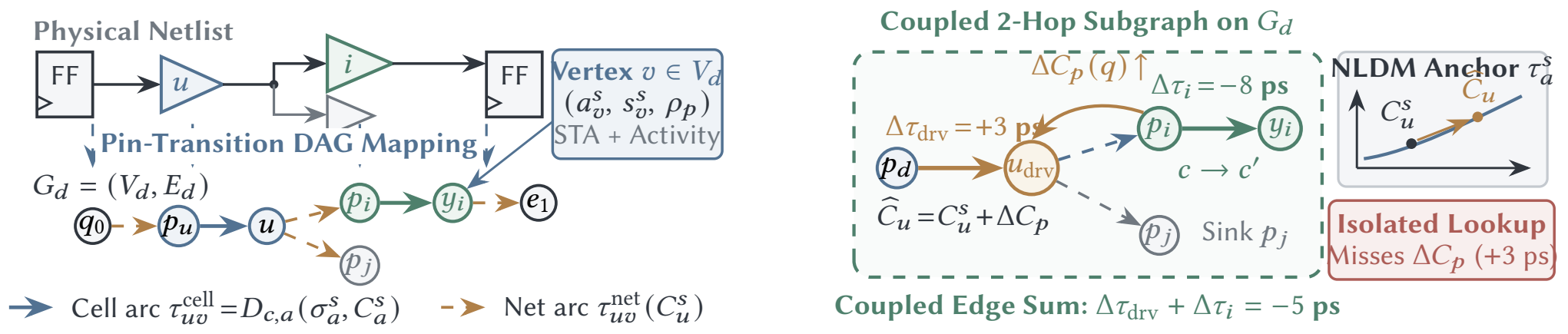

Conventional greedy gate sizers evaluate a candidate replacement $q = (i, c, c')$ by querying the isolated delay reduction of instance $i$ from Liberty lookup tables. In sub-10-nm technologies such as ASAP7, however, upsizing a cell to a higher drive strength substantially increases its input pin capacitances $C_p(c')$. When driven by a weak or heavily loaded upstream gate $u_{\mathrm{drv}}$, this extra capacitance degrades the driver's output transition and arc delay, often erasing or even reversing the local delay gain (\cref{fig:mot-4-1-a}). To capture this bidirectional electrical interaction, our snapshot-anchored replacement model anchors differential NLDM delay perturbations directly to live STA operating points while explicitly coupling each candidate's intrinsic delay shift with the induced upstream driver-load penalty (\cref{fig:mot-4-1-b}).

Formally, let superscript $s$ denote a live STA snapshot recorded on the current working netlist, capturing all enabled cell and interconnect timing arcs, rise/fall arrival times $a_v^s$, transition slews $\sigma_a^s$, effective output loads $C_a^s$, and endpoint slacks $s_e^s$. For a candidate replacement $q = (i, c, c')$ that maps instance $i$ from current master $c$ to an equivalent master $c' \in \mathcal{V}(c)$, we model the perturbed delay $\widehat{\tau}_a(q)$ of any affected cell timing arc $a$ as a snapshot-anchored first-order NLDM differential:
\begin{equation}
  \widehat{\tau}_a(q) = \tau_a^s + D_{c',a}\bigl(\sigma_a^s, \widehat{C}_a\bigr) - D_{c,a}\bigl(\sigma_a^s, C_a^s\bigr),
  \label{eq:graphdelay}
\end{equation}
where $D_{c,a}(\sigma, C)$ denotes the bilinearly interpolated Liberty NLDM delay surface of master $c$ on arc $a$ for the corresponding rise/fall transition polarity. Anchoring to $\tau_a^s$ ensures exact zero-error consistency ($\widehat{\tau}_a = \tau_a^s$) when $c' = c$ and $\widehat{C}_a = C_a^s$, while using $D_{c',a} - D_{c,a}$ captures the nonlinear sensitivity of the replacement at the exact local slew-load operating point.

Simultaneously, replacing $c \to c'$ alters the input pin capacitance of every input pin $p \in \operatorname{inputs}(i)$, which perturbs the effective capacitive load $\widehat{C}_u$ seen by each upstream driver output pin $u$ driving sink set $\mathcal{F}(u)$:
\begin{equation}
  \widehat{C}_u(q) = C_u^s + \sum_{p \in \mathcal{F}(u) \cap \operatorname{inputs}(i)} \Delta C_p(q), \qquad
  \Delta C_p(q) = C_p(c') - C_p(c).
  \label{eq:graphload}
\end{equation}
For each timing arc $a_{\mathrm{drv}}$ of the upstream driver feeding $u$, its perturbed delay $\widehat{\tau}_{a_{\mathrm{drv}}}(q)$ is evaluated via \cref{eq:graphdelay} with its fixed driver master $c_{\mathrm{drv}}$ and updated load $\widehat{C}_u(q)$ from \cref{eq:graphload}. When a batch of replacements is evaluated jointly, their pin capacitance differentials $\Delta C_p$ sum linearly at shared drivers before querying the nonlinear driver delay surface.

With \cref{eq:graphdelay,eq:graphload}, \tapco{} eliminates false-positive upsizing moves by evaluating every candidate replacement as a coupled two-stage electrical perturbation across both the target instance and its immediate fanin drivers, anchored directly to signoff-accurate STA operating conditions.

\subsection{Incremental Timing Graph Propagation}
\label{sec:graph-prop}

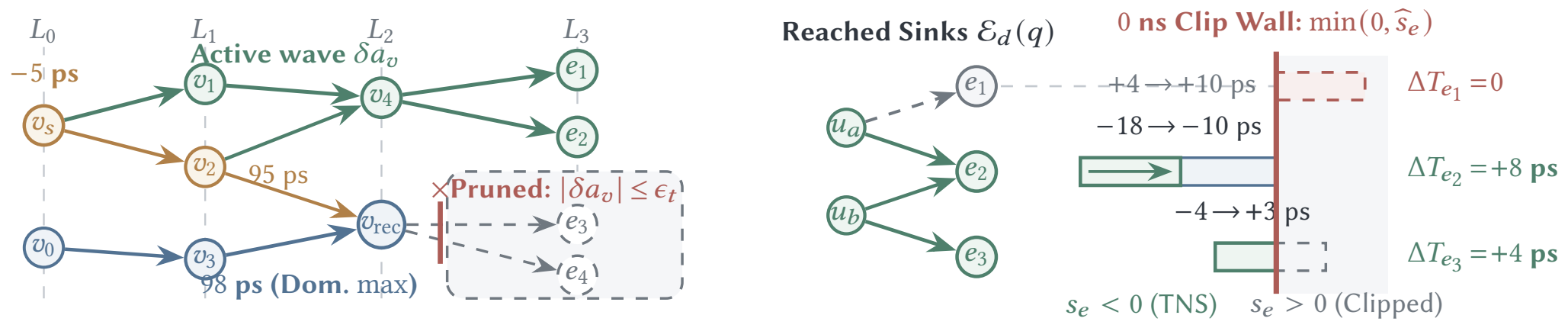
\begin{figure}[!t]
  \centering
\begingroup
\let\tiny\scriptsize
\captionsetup[subfloat]{font=small,labelfont=normalfont,justification=centering,singlelinecheck=false,captionskip=3pt}
\subfloat[Event-driven wavefront \& reconvergent pruning on $G_d$\label{fig:mot-4-2-a}]{%
\begin{minipage}[b]{0.485\linewidth}\centering
\begin{tikzpicture}[x=1cm,y=1cm,font=\sffamily\scriptsize,>=Stealth]
\path[use as bounding box] (0,0.05) rectangle (6.2,2.65);

\foreach \x/\lbl in {0.62/L_0, 2.02/L_1, 3.55/L_2, 5.25/L_3} {
  \node[text=tapcomuted,font=\sffamily\bfseries\tiny] at (\x,2.50) {$\lbl$};
  \draw[line width=0.45pt,draw=tapcorule,dashed] (\x,0.18) -- (\x,2.36);
}

\draw[rounded corners=4pt,draw=tapcomuted,dashed,line width=0.65pt,fill=tapcograyfill]
  (4.12,0.18) rectangle (6.16,1.28);
\node[text=tapcored,font=\sffamily\bfseries\tiny] at (5.14,1.12) {Pruned: $|\delta a_v|\!\le\!\epsilon_t$};

\node[circle,draw=tapcoamber,fill=tapcoamberfill,line width=0.85pt,minimum size=0.34cm,inner sep=0pt,font=\sffamily\bfseries\tiny,text=tapcoamber] (vsrc) at (0.62,1.68) {$v_s$};
\node[text=tapcoamber,font=\sffamily\bfseries\tiny,anchor=south] at (0.62,1.88) {$-5\text{ ps}$};
\node[circle,draw=tapconavy,fill=tapconavyfill,line width=0.75pt,minimum size=0.34cm,inner sep=0pt,font=\sffamily\tiny,text=tapconavy] (v0) at (0.62,0.62) {$v_0$};

\node[circle,draw=tapcoteal,fill=tapcotealfill,line width=0.85pt,minimum size=0.34cm,inner sep=0pt,font=\sffamily\bfseries\tiny,text=tapcoteal] (v1) at (2.02,2.04) {$v_1$};
\node[circle,draw=tapcoamber,fill=tapcoamberfill,line width=0.8pt,minimum size=0.34cm,inner sep=0pt,font=\sffamily\tiny,text=tapcoamber] (v2) at (2.02,1.32) {$v_2$};
\node[circle,draw=tapconavy,fill=tapconavyfill,line width=0.8pt,minimum size=0.34cm,inner sep=0pt,font=\sffamily\tiny,text=tapconavy] (v3) at (2.02,0.52) {$v_3$};

\node[circle,draw=tapcoteal,fill=tapcotealfill,line width=0.85pt,minimum size=0.34cm,inner sep=0pt,font=\sffamily\bfseries\tiny,text=tapcoteal] (v4) at (3.55,1.92) {$v_4$};
\node[circle,draw=tapconavy,fill=tapconavyfill,line width=0.85pt,minimum size=0.36cm,inner sep=0pt,font=\sffamily\bfseries\tiny,text=tapconavy] (vrec) at (3.55,0.82) {$v_{\mathrm{rec}}$};

\node[circle,draw=tapcoteal,fill=tapcotealfill,line width=0.85pt,minimum size=0.34cm,inner sep=0pt,font=\sffamily\bfseries\tiny,text=tapcoteal] (e1) at (5.25,2.16) {$e_1$};
\node[circle,draw=tapcoteal,fill=tapcotealfill,line width=0.85pt,minimum size=0.34cm,inner sep=0pt,font=\sffamily\bfseries\tiny,text=tapcoteal] (e2) at (5.25,1.58) {$e_2$};
\node[circle,draw=tapcomuted,fill=white,dashed,line width=0.65pt,minimum size=0.34cm,inner sep=0pt,font=\sffamily\tiny,text=tapcomuted] (e3) at (5.25,0.82) {$e_3$};
\node[circle,draw=tapcomuted,fill=white,dashed,line width=0.65pt,minimum size=0.34cm,inner sep=0pt,font=\sffamily\tiny,text=tapcomuted] (e4) at (5.25,0.38) {$e_4$};

\draw[line width=0.95pt,draw=tapcoteal,->] (vsrc) -- (v1);
\draw[line width=0.85pt,draw=tapcoamber,->] (vsrc) -- (v2);
\draw[line width=0.95pt,draw=tapcoteal,->] (v1) -- (v4);
\draw[line width=0.85pt,draw=tapcoteal,->] (v2) -- (v4);
\draw[line width=0.95pt,draw=tapcoteal,->] (v4) -- (e1);
\draw[line width=0.95pt,draw=tapcoteal,->] (v4) -- (e2);
\node[text=tapcoteal,font=\sffamily\bfseries\tiny,anchor=south] at (2.80,2.06) {Active wave $\delta a_v$};

\draw[line width=0.8pt,draw=tapcoamber,->] (v2) -- (vrec);
\node[text=tapcoamber,font=\sffamily\tiny,anchor=south west,inner sep=1pt] at (2.35,1.08) {$95\text{ ps}$};
\draw[line width=0.85pt,draw=tapconavy,->] (v0) -- (v3);
\draw[line width=0.9pt,draw=tapconavy,->] (v3) -- (vrec);
\node[text=tapconavy,font=\sffamily\bfseries\tiny,anchor=north] at (2.92,0.52) {$98\text{ ps}$ (Dom.\ $\max$)};

\draw[line width=0.65pt,draw=tapcomuted,dashed,->] (vrec) -- (e3);
\draw[line width=0.65pt,draw=tapcomuted,dashed,->] (vrec) -- (e4);
\draw[line width=1.25pt,draw=tapcored] (4.06,0.48) -- (4.06,1.02);
\node[text=tapcored,font=\sffamily\bfseries\scriptsize] at (4.06,1.12) {$\times$};
\end{tikzpicture}
\end{minipage}}\hfill
\subfloat[Endpoint $0\text{-ns}$ positive-slack clipping\label{fig:mot-4-2-b}]{%
\begin{minipage}[b]{0.485\linewidth}\centering
\begin{tikzpicture}[x=1cm,y=1cm,font=\sffamily\scriptsize,>=Stealth]
\path[use as bounding box] (0,0.05) rectangle (6.2,2.65);

\node[text=tapcoink,font=\sffamily\bfseries\tiny] at (1.05,2.48) {Reached Sinks $\mathcal{E}_d(q)$};

\node[circle,draw=tapcoteal,fill=tapcotealfill,line width=0.8pt,minimum size=0.32cm,inner sep=0pt,font=\sffamily\tiny,text=tapcoteal] (ua) at (0.42,1.66) {$u_a$};
\node[circle,draw=tapcoteal,fill=tapcotealfill,line width=0.8pt,minimum size=0.32cm,inner sep=0pt,font=\sffamily\tiny,text=tapcoteal] (ub) at (0.42,0.90) {$u_b$};

\node[circle,draw=tapcomuted,fill=tapcograyfill,line width=0.75pt,minimum size=0.34cm,inner sep=0pt,font=\sffamily\bfseries\tiny,text=tapcomuted] (s1) at (1.55,2.02) {$e_1$};
\node[circle,draw=tapcoteal,fill=tapcotealfill,line width=0.85pt,minimum size=0.34cm,inner sep=0pt,font=\sffamily\bfseries\tiny,text=tapcoteal] (s2) at (1.55,1.28) {$e_2$};
\node[circle,draw=tapcoteal,fill=tapcotealfill,line width=0.85pt,minimum size=0.34cm,inner sep=0pt,font=\sffamily\bfseries\tiny,text=tapcoteal] (s3) at (1.55,0.54) {$e_3$};

\draw[line width=0.75pt,draw=tapcomuted,dashed,->] (ua) -- (s1);
\draw[line width=0.9pt,draw=tapcoteal,->] (ua) -- (s2);
\draw[line width=0.85pt,draw=tapcoteal,->] (ub) -- (s2);
\draw[line width=0.9pt,draw=tapcoteal,->] (ub) -- (s3);

\fill[tapcograyfill] (4.15,0.22) rectangle (5.12,2.28);
\node[text=tapcomuted,font=\sffamily\tiny] at (4.76,0.12) {$s_e > 0$ (Clipped)};
\node[text=tapcoteal,font=\sffamily\tiny] at (2.98,0.12) {$s_e < 0$ (TNS)};

\draw[line width=0.5pt,draw=tapcorule,dashed] (1.76,2.02) -- (4.15,2.02);
\filldraw[fill=tapcoredfill,draw=tapcored,dashed,line width=0.7pt] (4.15,1.90) rectangle (4.92,2.14);
\node[text=tapcomuted,font=\sffamily\tiny,anchor=east] at (4.10,2.02) {$+4\!\to\!+10\text{ ps}$};
\node[text=tapcored,font=\sffamily\bfseries\tiny,anchor=west] at (5.16,2.02) {$\Delta T_{e_1}\!=\!0$};

\filldraw[fill=tapconavyfill,draw=tapconavy,line width=0.65pt] (2.45,1.16) rectangle (4.15,1.40);
\filldraw[fill=tapcotealfill,draw=tapcoteal,line width=0.85pt] (2.45,1.16) rectangle (3.32,1.40);
\draw[line width=0.7pt,draw=tapcoteal,->] (2.52,1.28) -- (3.25,1.28);
\node[text=tapcoink,font=\sffamily\tiny,anchor=south] at (3.30,1.42) {$-18\!\to\!-10\text{ ps}$};
\node[text=tapcoteal,font=\sffamily\bfseries\tiny,anchor=west] at (5.16,1.28) {$\Delta T_{e_2}\!=\!+8\text{ ps}$};

\filldraw[fill=tapcotealfill,draw=tapcoteal,line width=0.85pt] (3.62,0.42) rectangle (4.15,0.66);
\filldraw[fill=tapcograyfill,draw=tapcomuted,dashed,line width=0.65pt] (4.15,0.42) rectangle (4.58,0.66);
\node[text=tapcoink,font=\sffamily\tiny,anchor=south] at (3.85,0.68) {$-4\!\to\!+3\text{ ps}$};
\node[text=tapcoteal,font=\sffamily\bfseries\tiny,anchor=west] at (5.16,0.54) {$\Delta T_{e_3}\!=\!+4\text{ ps}$};

\draw[line width=1.35pt,draw=tapcored] (4.15,0.22) -- (4.15,2.32);
\node[text=tapcored,font=\sffamily\bfseries\tiny,anchor=south] at (4.15,2.32) {$0\text{ ns}$ Clip Wall: $\min(0,\widehat{s}_e)$};
\end{tikzpicture}
\end{minipage}}
\endgroup
  \caption{Event-driven topological wave propagation, reconvergent pruning, and endpoint slack clipping on $G_d = (V_d, E_d)$. (a) Arrival shifts $\delta a_v(q)$ propagate across the active fanout cone on $G_d$ while branches masked by dominant reconvergent paths ($|\delta a_v| \le \epsilon_t$) are immediately pruned. (b) At sink endpoints $\mathcal{E}_d(q)$, positive slacks are clipped at $0\text{ ns}$ via $\min(0, \widehat{s}_e)$ so non-violating sinks ($e_1$) contribute zero artificial gain while true violating sinks ($e_2$) are accurately priced.}
  \Description{DAG wave propagation showing active fanout cone and pruned reconvergent branches on the left, and 0-ns endpoint slack clipping across sink vertices on the right.}
  \label{fig:mot-4-2}
\end{figure}

Even when the net local delay shift of a cell and its driver is negative, summing local delay improvements does not reflect circuit-level timing recovery. In reconvergent logic cones, accelerating one branch by $-5\text{ ps}$ yields only a $-2\text{ ps}$ arrival improvement if a parallel side branch arriving $2\text{ ps}$ earlier becomes the new dominant critical path (\cref{fig:mot-4-2-a}); conversely, a single replacement on a high-fanout stem can simultaneously improve dozens of downstream violating endpoints. We therefore propagate local arc perturbations $\widehat{\tau}_{uv}(q)$ incrementally through the topological cone of the STA DAG $G_d$, computing exact max-arrival transitions and multi-endpoint fanout impacts without invoking full-chip STA (\cref{fig:mot-4-2-b}).

Given the perturbed arc delays $\widehat{\tau}_{uv}(q)$ on the target cell and its upstream drivers (with $\widehat{\tau}_{uv}(q) = \tau_{uv}^s$ for all unperturbed arcs), the predicted arrival time $\widehat{a}_v(q)$ and arrival shift $\delta a_v(q)$ at any downstream vertex $v \in V_d$ satisfy the recursive Bellman-Ford max-arrival relation:
\begin{equation}
  \widehat{a}_v(q) = \max_{u \in \operatorname{pred}(v)} \bigl(\widehat{a}_u(q) + \widehat{\tau}_{uv}(q)\bigr), \qquad
  \delta a_v(q) = \widehat{a}_v(q) - a_v^s.
  \label{eq:grapharrival}
\end{equation}
Rather than recomputing \cref{eq:grapharrival} over the entire circuit graph $G_d$, \tapco{} executes an event-driven, level-ordered topological wave propagation restricted to the active perturbation cone:
\begin{enumerate}
  \item \emph{Wave Initialization:} All output transition vertices of the upstream drivers $u_{\mathrm{drv}}$ and the replaced instance $i$ whose arc delays changed ($\widehat{\tau}_{uv}(q) \ne \tau_{uv}^s$) are inserted into a min-heap priority queue $\mathcal{Q}$ keyed by topological level $\operatorname{lvl}(v)$.
  \item \emph{Dominant-Arrival Re-evaluation:} Popping vertex $v$ from $\mathcal{Q}$, \tapco{} evaluates $\widehat{a}_v(q)$ over $\operatorname{pred}(v)$, reading $\widehat{a}_u(q)$ for visited predecessors and the baseline snapshot arrival $a_u^s$ for unvisited predecessors outside the perturbed cone.
  \item \emph{Zero-Shift Cone Pruning:} If $|\delta a_v(q)| \le \epsilon_t$ (i.e., the arrival shift is masked by a dominant parallel reconvergent path), outgoing edges of $v$ are pruned immediately; otherwise, all fanout neighbors $w \in \operatorname{succ}(v)$ are enqueued into $\mathcal{Q}$.
\end{enumerate}

With \cref{eq:grapharrival} and event-driven cone pruning, \tapco{} resolves both reconvergent path masking and shared-stem fanout amplification in microseconds per candidate, confining computation strictly to vertices where arrival times actually shift.

\subsection{Endpoint Value and Sequential Role Disambiguation}
\label{sec:graph-seq}

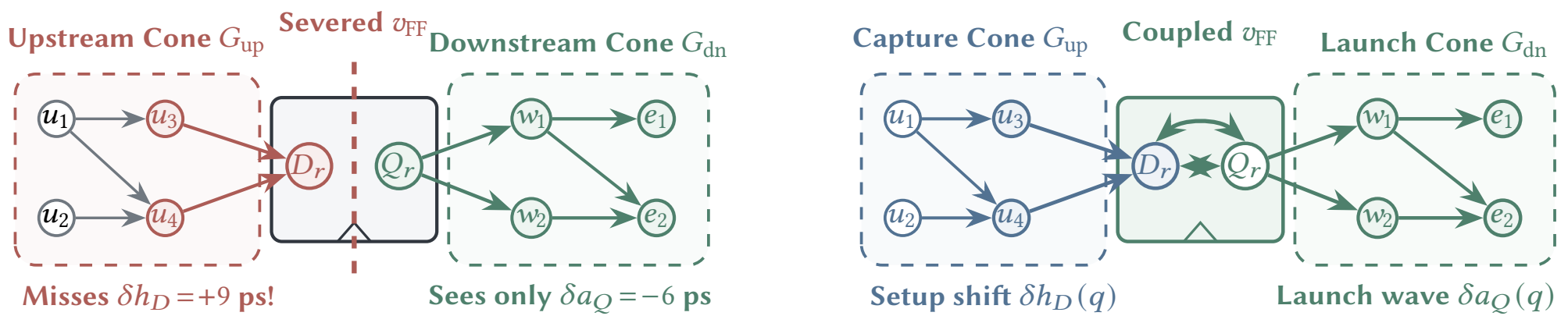
\begin{figure}[!t]
  \centering
\begingroup
\let\tiny\scriptsize
\captionsetup[subfloat]{font=small,labelfont=normalfont,justification=centering,singlelinecheck=false,captionskip=3pt}
\subfloat[Severed single-cone graph at $v_{\mathrm{FF}}$\label{fig:mot-4-3-a}]{%
\begin{minipage}[b]{0.485\linewidth}\centering
\begin{tikzpicture}[x=1cm,y=1cm,font=\sffamily\scriptsize,>=Stealth]
\path[use as bounding box] (0,0.05) rectangle (6.2,2.65);

\draw[rounded corners=5pt,draw=tapcored,dashed,line width=0.7pt,fill=tapcoredfill,fill opacity=0.35]
  (0.15,0.52) rectangle (2.22,2.14);
\node[text=tapcored,font=\sffamily\bfseries\tiny,anchor=south] at (1.18,2.16) {Upstream Cone $G_{\mathrm{up}}$};

\draw[rounded corners=5pt,draw=tapcoteal,dashed,line width=0.7pt,fill=tapcotealfill,fill opacity=0.35]
  (3.82,0.52) rectangle (6.02,2.14);
\node[text=tapcoteal,font=\sffamily\bfseries\tiny,anchor=south] at (4.92,2.16) {Downstream Cone $G_{\mathrm{dn}}$};

\node[circle,draw=tapcomuted,fill=white,line width=0.7pt,minimum size=0.30cm,inner sep=0pt,font=\sffamily\tiny] (u1) at (0.50,1.76) {$u_1$};
\node[circle,draw=tapcomuted,fill=white,line width=0.7pt,minimum size=0.30cm,inner sep=0pt,font=\sffamily\tiny] (u2) at (0.50,0.92) {$u_2$};
\node[circle,draw=tapcored,fill=tapcoredfill,line width=0.75pt,minimum size=0.30cm,inner sep=0pt,font=\sffamily\tiny,text=tapcored] (u3) at (1.42,1.76) {$u_3$};
\node[circle,draw=tapcored,fill=tapcoredfill,line width=0.75pt,minimum size=0.30cm,inner sep=0pt,font=\sffamily\tiny,text=tapcored] (u4) at (1.42,0.92) {$u_4$};

\filldraw[rounded corners=2.5pt,fill=tapcograyfill,draw=tapcoink,line width=0.8pt] (2.32,0.72) rectangle (3.72,1.94);
\draw[line width=0.6pt,draw=tapcoink] (2.88,0.72) -- (3.02,0.88) -- (3.16,0.72);
\node[circle,draw=tapcored,fill=tapcoredfill,line width=0.8pt,minimum size=0.34cm,inner sep=0pt,font=\sffamily\bfseries\tiny,text=tapcored] (Dr) at (2.64,1.36) {$D_r$};
\node[circle,draw=tapcoteal,fill=tapcotealfill,line width=0.8pt,minimum size=0.34cm,inner sep=0pt,font=\sffamily\bfseries\tiny,text=tapcoteal] (Qr) at (3.40,1.36) {$Q_r$};

\node[circle,draw=tapcoteal,fill=tapcotealfill,line width=0.75pt,minimum size=0.30cm,inner sep=0pt,font=\sffamily\tiny,text=tapcoteal] (w1) at (4.52,1.76) {$w_1$};
\node[circle,draw=tapcoteal,fill=tapcotealfill,line width=0.75pt,minimum size=0.30cm,inner sep=0pt,font=\sffamily\tiny,text=tapcoteal] (w2) at (4.52,0.92) {$w_2$};
\node[circle,draw=tapcoteal,fill=tapcotealfill,line width=0.8pt,minimum size=0.30cm,inner sep=0pt,font=\sffamily\bfseries\tiny,text=tapcoteal] (e1) at (5.58,1.76) {$e_1$};
\node[circle,draw=tapcoteal,fill=tapcotealfill,line width=0.8pt,minimum size=0.30cm,inner sep=0pt,font=\sffamily\bfseries\tiny,text=tapcoteal] (e2) at (5.58,0.92) {$e_2$};

\draw[line width=0.65pt,draw=tapcomuted,->] (u1) -- (u3);
\draw[line width=0.65pt,draw=tapcomuted,->] (u2) -- (u4);
\draw[line width=0.65pt,draw=tapcomuted,->] (u1) -- (u4);
\draw[line width=0.8pt,draw=tapcored,->] (u3) -- (Dr);
\draw[line width=0.8pt,draw=tapcored,->] (u4) -- (Dr);

\draw[line width=0.85pt,draw=tapcoteal,->] (Qr) -- (w1);
\draw[line width=0.85pt,draw=tapcoteal,->] (Qr) -- (w2);
\draw[line width=0.8pt,draw=tapcoteal,->] (w1) -- (e1);
\draw[line width=0.75pt,draw=tapcoteal,->] (w1) -- (e2);
\draw[line width=0.8pt,draw=tapcoteal,->] (w2) -- (e2);

\draw[line width=1.2pt,draw=tapcored,dashed] (3.02,0.45) -- (3.02,2.32);
\node[text=tapcored,font=\sffamily\bfseries\tiny,anchor=south] at (3.02,2.34) {Severed $v_{\mathrm{FF}}$};

\node[text=tapcored,font=\sffamily\bfseries\tiny] at (1.28,0.24) {Misses $\delta h_D\!=\!+9\text{ ps}$!};
\node[text=tapcoteal,font=\sffamily\bfseries\tiny] at (4.85,0.24) {Sees only $\delta a_Q\!=\!-6\text{ ps}$};
\end{tikzpicture}
\end{minipage}}\hfill
\subfloat[Coupled bi-stage graph across $v_{\mathrm{FF}}$\label{fig:mot-4-3-b}]{%
\begin{minipage}[b]{0.485\linewidth}\centering
\begin{tikzpicture}[x=1cm,y=1cm,font=\sffamily\scriptsize,>=Stealth]
\path[use as bounding box] (0,0.05) rectangle (6.2,2.65);

\draw[rounded corners=5pt,draw=tapconavy,dashed,line width=0.7pt,fill=tapconavyfill,fill opacity=0.38]
  (0.15,0.52) rectangle (2.22,2.14);
\node[text=tapconavy,font=\sffamily\bfseries\tiny,anchor=south] at (1.08,2.16) {Capture Cone $G_{\mathrm{up}}$};

\draw[rounded corners=5pt,draw=tapcoteal,dashed,line width=0.7pt,fill=tapcotealfill,fill opacity=0.38]
  (3.82,0.52) rectangle (6.02,2.14);
\node[text=tapcoteal,font=\sffamily\bfseries\tiny,anchor=south] at (5.00,2.16) {Launch Cone $G_{\mathrm{dn}}$};

\node[circle,draw=tapconavy,fill=white,line width=0.7pt,minimum size=0.30cm,inner sep=0pt,font=\sffamily\tiny,text=tapconavy] (u1) at (0.50,1.76) {$u_1$};
\node[circle,draw=tapconavy,fill=white,line width=0.7pt,minimum size=0.30cm,inner sep=0pt,font=\sffamily\tiny,text=tapconavy] (u2) at (0.50,0.92) {$u_2$};
\node[circle,draw=tapconavy,fill=tapconavyfill,line width=0.8pt,minimum size=0.30cm,inner sep=0pt,font=\sffamily\tiny,text=tapconavy] (u3) at (1.42,1.76) {$u_3$};
\node[circle,draw=tapconavy,fill=tapconavyfill,line width=0.8pt,minimum size=0.30cm,inner sep=0pt,font=\sffamily\tiny,text=tapconavy] (u4) at (1.42,0.92) {$u_4$};

\filldraw[rounded corners=2.5pt,fill=tapcotealfill,draw=tapcoteal,line width=0.85pt] (2.32,0.72) rectangle (3.72,1.94);
\draw[line width=0.6pt,draw=tapcoteal] (2.88,0.72) -- (3.02,0.88) -- (3.16,0.72);
\node[circle,draw=tapconavy,fill=tapconavyfill,line width=0.85pt,minimum size=0.34cm,inner sep=0pt,font=\sffamily\bfseries\tiny,text=tapconavy] (Dr) at (2.64,1.36) {$D_r$};
\node[circle,draw=tapcoteal,fill=white,line width=0.85pt,minimum size=0.34cm,inner sep=0pt,font=\sffamily\bfseries\tiny,text=tapcoteal] (Qr) at (3.40,1.36) {$Q_r$};

\draw[line width=1.05pt,draw=tapcoteal,<->] (Dr) -- (Qr);
\draw[line width=1.0pt,draw=tapcoteal,<->,bend left=42] (Dr.north) to (Qr.north);
\node[text=tapcoteal,font=\sffamily\bfseries\tiny,anchor=south] at (3.02,2.24) {Coupled $v_{\mathrm{FF}}$};

\node[circle,draw=tapcoteal,fill=tapcotealfill,line width=0.8pt,minimum size=0.30cm,inner sep=0pt,font=\sffamily\tiny,text=tapcoteal] (w1) at (4.52,1.76) {$w_1$};
\node[circle,draw=tapcoteal,fill=tapcotealfill,line width=0.8pt,minimum size=0.30cm,inner sep=0pt,font=\sffamily\tiny,text=tapcoteal] (w2) at (4.52,0.92) {$w_2$};
\node[circle,draw=tapcoteal,fill=tapcotealfill,line width=0.85pt,minimum size=0.30cm,inner sep=0pt,font=\sffamily\bfseries\tiny,text=tapcoteal] (e1) at (5.58,1.76) {$e_1$};
\node[circle,draw=tapcoteal,fill=tapcotealfill,line width=0.85pt,minimum size=0.30cm,inner sep=0pt,font=\sffamily\bfseries\tiny,text=tapcoteal] (e2) at (5.58,0.92) {$e_2$};

\draw[line width=0.75pt,draw=tapconavy,->] (u1) -- (u3);
\draw[line width=0.75pt,draw=tapconavy,->] (u2) -- (u4);
\draw[line width=0.7pt,draw=tapconavy,->] (u1) -- (u4);
\draw[line width=0.9pt,draw=tapconavy,->] (u3) -- (Dr);
\draw[line width=0.9pt,draw=tapconavy,->] (u4) -- (Dr);

\draw[line width=0.9pt,draw=tapcoteal,->] (Qr) -- (w1);
\draw[line width=0.9pt,draw=tapcoteal,->] (Qr) -- (w2);
\draw[line width=0.85pt,draw=tapcoteal,->] (w1) -- (e1);
\draw[line width=0.8pt,draw=tapcoteal,->] (w1) -- (e2);
\draw[line width=0.85pt,draw=tapcoteal,->] (w2) -- (e2);

\node[text=tapconavy,font=\sffamily\bfseries\tiny] at (1.28,0.24) {Setup shift $\delta h_D(q)$};
\node[text=tapcoteal,font=\sffamily\bfseries\tiny] at (4.85,0.24) {Launch wave $\delta a_Q(q)$};
\end{tikzpicture}
\end{minipage}}
\endgroup
  \caption{Cross-stage sequential graph coupling across boundary vertex $v_{\mathrm{FF}} = (D_r, Q_r)$ on $G_d$. (a) Severing the timing graph at flip-flop boundaries isolates upstream cone $G_{\mathrm{up}}$ from downstream cone $G_{\mathrm{dn}}$ and misses capture setup degradation ($\delta h_D$). (b) \tapco{} couples capture setup shifts $\delta h_D$ on incoming edges to $D_r$ with launch Clock-to-Q arrival waves $\delta a_Q$ on outgoing edges from $Q_r$ across $v_{\mathrm{FF}}$.}
  \Description{Comparison between severed single-cone DAGs at flip-flop boundaries versus coupled bi-stage DAG cones propagating both capture setup shifts and launch arrival waves across sequential vertices.}
  \label{fig:mot-4-3}
\end{figure}

Propagating arrival shifts to path endpoints is still insufficient if all arrival reductions are valued equally or if sequential elements are treated as combinational gates. First, improving the arrival time of an endpoint that already has positive slack ($s_e^s > 0$) contributes zero improvement to setup TNS $T_d$, wasting power and area (\cref{fig:mot-4-3-a}). Second, replacing a flip-flop master $c \to c'$ simultaneously shifts its launch Clock-to-Q delay (which propagates forward into downstream data paths) and modifies its capture setup time requirement (which directly shifts the required arrival at its own data input). To price both effects accurately, our endpoint valuation model decouples launch and capture sequential timing obligations and evaluates candidates strictly through clipped endpoint slack differentials (\cref{fig:mot-4-3-b}).

Specifically, when candidate $q = (i, c, c')$ replaces a sequential flip-flop instance, \tapco{} decomposes its timing impact into two distinct boundary mechanisms:
\begin{enumerate}
  \item \emph{Launch Clock-to-Q Perturbation:} On the launch side, the Clock-to-Q arc delay change $\Delta D_{\mathrm{c2q}} = D_{c',\mathrm{c2q}}(\sigma_{\mathrm{clk}}^s, \widehat{C}_Q) - D_{c,\mathrm{c2q}}(\sigma_{\mathrm{clk}}^s, C_Q^s)$ enters \cref{eq:graphdelay} at the flip-flop output $Q$ and propagates forward through the downstream combinational cone via \cref{eq:grapharrival} to produce arrival shifts $\delta a_{e,\rho}(q)$ at downstream endpoints $e$.
  \item \emph{Capture Setup-Time Perturbation:} On the capture side, where instance $i$ serves as endpoint $e_i \in \mathcal{E}_d$, replacing $c \to c'$ alters the intrinsic setup-time constraint table $H_c(\sigma_D, \sigma_{\mathrm{clk}})$ for each analyzed data transition $\rho \in \{\mathrm{rise}, \mathrm{fall}\}$:
  \begin{equation}
    \delta h_{e_i,\rho}(q) = H_{c',\rho}\bigl(\sigma_{e_i,\rho}^s, \sigma_{\mathrm{clk}}^s\bigr) - H_{c,\rho}\bigl(\sigma_{e_i,\rho}^s, \sigma_{\mathrm{clk}}^s\bigr),
    \label{eq:graphsetup}
  \end{equation}
  with $\delta h_{e,\rho}(q) = 0$ for all endpoints $e \ne e_i$ whose capture cell is unchanged.
\end{enumerate}

Combining the propagated data-path arrival shift $\delta a_{e,\rho}(q)$ and the capture setup requirement shift $\delta h_{e,\rho}(q)$, the projected summary slack $\widehat{s}_e(q)$ of endpoint $e$ is anchored to the baseline endpoint slack $s_e^s$:
\begin{equation}
  \widehat{s}_e(q) = s_e^s + \min_{\rho \in \{\mathrm{rise},\mathrm{fall}\}} \bigl(s_{e,\rho}^s - \delta a_{e,\rho}(q) - \delta h_{e,\rho}(q)\bigr) - \min_{\rho \in \{\mathrm{rise},\mathrm{fall}\}} s_{e,\rho}^s.
  \label{eq:graphslack}
\end{equation}
Finally, we map the projected endpoint slacks to the normalized circuit-level timing objective gain $\widehat{g}_T(q)$ consistent with \cref{eq:score}:
\begin{equation}
  \widehat{g}_T(q) = \frac{w_T}{|T_d^0|} \sum_{e \in \mathcal{E}_d(q)} \Bigl[\min\bigl(0, \widehat{s}_e(q)\bigr) - \min\bigl(0, s_e^s\bigr)\Bigr],
  \label{eq:graphgain}
\end{equation}
where $\mathcal{E}_d(q) \subseteq \mathcal{E}_d$ is the sparse subset of endpoints reached by the propagation wave or capture setup update.

With \cref{eq:graphsetup,eq:graphslack,eq:graphgain}, \tapco{} aligns microscopic candidate valuation directly with signoff TNS: it prevents wasted power expenditure on already-closed paths ($s_e^s \ge 0$), penalizes moves that push non-violating paths into negative slack ($\widehat{s}_e(q) < 0$), and enables safe, high-gain multi-VT and drive-strength co-optimization across both sequential flip-flops and combinational logic.

\subsection{Coarse-to-Fine Instance-Aware Power Screening}
\label{sec:graph-power}

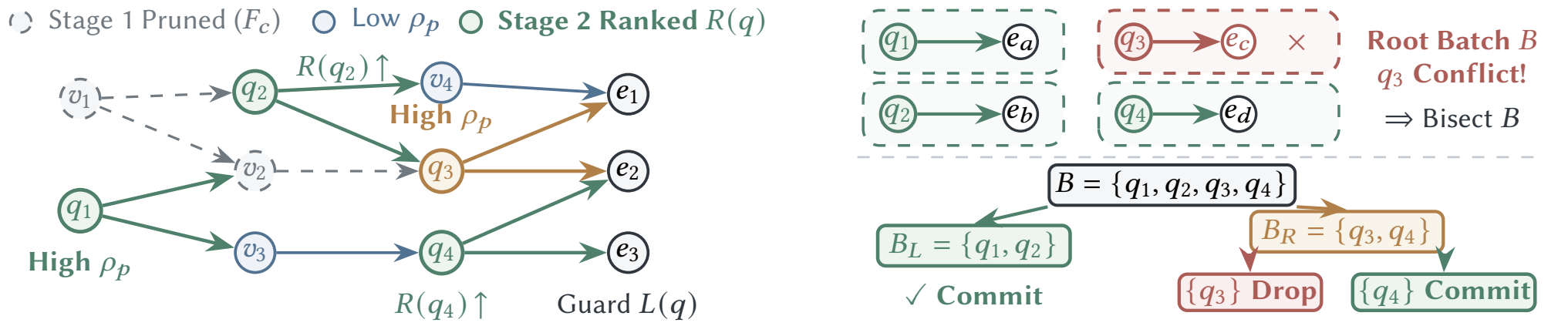
\begin{figure}[!t]
  \centering
\begingroup
\let\tiny\scriptsize
\captionsetup[subfloat]{font=small,labelfont=normalfont,justification=centering,singlelinecheck=false,captionskip=3pt}
\subfloat[Activity-weighted coarse-to-fine vertex screening on $G_d$\label{fig:mot-4-4-a}]{%
\begin{minipage}[b]{0.485\linewidth}\centering
\begin{tikzpicture}[x=1cm,y=1cm,font=\sffamily\scriptsize,>=Stealth]
\path[use as bounding box] (0,0.05) rectangle (6.2,2.75);

\node[circle,draw=tapcomuted,fill=tapcograyfill,dashed,line width=0.6pt,minimum size=0.18cm,inner sep=0pt] at (0.14,2.54) {};
\node[text=tapcomuted,font=\sffamily\tiny,anchor=west] at (0.25,2.54) {Stage 1 Pruned ($F_c$)};
\node[circle,draw=tapconavy,fill=tapconavyfill,line width=0.65pt,minimum size=0.18cm,inner sep=0pt] at (2.74,2.54) {};
\node[text=tapconavy,font=\sffamily\tiny,anchor=west] at (2.85,2.54) {Low $\rho_p$};
\node[circle,draw=tapcoteal,fill=tapcotealfill,line width=0.75pt,minimum size=0.18cm,inner sep=0pt] at (4.00,2.54) {};
\node[text=tapcoteal,font=\sffamily\bfseries\tiny,anchor=west] at (4.11,2.54) {Stage 2 Ranked $R(q)$};

\node[circle,draw=tapcomuted,fill=tapcograyfill,dashed,line width=0.65pt,minimum size=0.34cm,inner sep=0pt,font=\sffamily\tiny,text=tapcomuted] (v1) at (0.65,1.88) {$v_1$};
\node[circle,draw=tapcoteal,fill=tapcotealfill,line width=0.9pt,minimum size=0.36cm,inner sep=0pt,font=\sffamily\bfseries\tiny,text=tapcoteal] (q1) at (0.65,0.92) {$q_1$};
\node[text=tapcoteal,font=\sffamily\bfseries\tiny,anchor=north] at (0.65,0.70) {High $\rho_p$};

\node[circle,draw=tapcoteal,fill=tapcotealfill,line width=0.9pt,minimum size=0.36cm,inner sep=0pt,font=\sffamily\bfseries\tiny,text=tapcoteal] (q2) at (2.15,1.94) {$q_2$};
\node[text=tapcoteal,font=\sffamily\bfseries\tiny,anchor=west] at (2.38,2.14) {$R(q_2)\!\uparrow$};
\node[circle,draw=tapcomuted,fill=tapcograyfill,dashed,line width=0.65pt,minimum size=0.34cm,inner sep=0pt,font=\sffamily\tiny,text=tapcomuted] (v2) at (2.15,1.26) {$v_2$};
\node[circle,draw=tapconavy,fill=tapconavyfill,line width=0.7pt,minimum size=0.34cm,inner sep=0pt,font=\sffamily\tiny,text=tapconavy] (v3) at (2.15,0.56) {$v_3$};

\node[circle,draw=tapconavy,fill=tapconavyfill,line width=0.7pt,minimum size=0.34cm,inner sep=0pt,font=\sffamily\tiny,text=tapconavy] (v4) at (3.75,2.02) {$v_4$};
\node[circle,draw=tapcoamber,fill=tapcoamberfill,line width=0.9pt,minimum size=0.36cm,inner sep=0pt,font=\sffamily\bfseries\tiny,text=tapcoamber] (q3) at (3.75,1.26) {$q_3$};
\node[text=tapcoamber,font=\sffamily\bfseries\tiny,anchor=south] at (3.75,1.46) {High $\rho_p$};
\node[circle,draw=tapcoteal,fill=tapcotealfill,line width=0.9pt,minimum size=0.36cm,inner sep=0pt,font=\sffamily\bfseries\tiny,text=tapcoteal] (q4) at (3.75,0.56) {$q_4$};
\node[text=tapcoteal,font=\sffamily\bfseries\tiny,anchor=north] at (3.75,0.35) {$R(q_4)\!\uparrow$};

\node[circle,draw=tapcoink,fill=tapcograyfill,line width=0.7pt,minimum size=0.34cm,inner sep=0pt,font=\sffamily\tiny] (e1) at (5.35,1.92) {$e_1$};
\node[circle,draw=tapcoink,fill=tapcograyfill,line width=0.7pt,minimum size=0.34cm,inner sep=0pt,font=\sffamily\tiny] (e2) at (5.35,1.26) {$e_2$};
\node[circle,draw=tapcoink,fill=tapcograyfill,line width=0.7pt,minimum size=0.34cm,inner sep=0pt,font=\sffamily\tiny] (e3) at (5.35,0.56) {$e_3$};

\draw[line width=0.6pt,draw=tapcomuted,dashed,->] (v1) -- (q2);
\draw[line width=0.6pt,draw=tapcomuted,dashed,->] (v1) -- (v2);
\draw[line width=0.85pt,draw=tapcoteal,->] (q1) -- (v2);
\draw[line width=0.85pt,draw=tapcoteal,->] (q1) -- (v3);
\draw[line width=0.85pt,draw=tapcoteal,->] (q2) -- (v4);
\draw[line width=0.85pt,draw=tapcoteal,->] (q2) -- (q3);
\draw[line width=0.6pt,draw=tapcomuted,dashed,->] (v2) -- (q3);
\draw[line width=0.7pt,draw=tapconavy,->] (v3) -- (q4);
\draw[line width=0.7pt,draw=tapconavy,->] (v4) -- (e1);
\draw[line width=0.85pt,draw=tapcoamber,->] (q3) -- (e1);
\draw[line width=0.85pt,draw=tapcoamber,->] (q3) -- (e2);
\draw[line width=0.85pt,draw=tapcoteal,->] (q4) -- (e2);
\draw[line width=0.85pt,draw=tapcoteal,->] (q4) -- (e3);

\node[text=tapcoink,font=\sffamily\tiny,anchor=north] at (5.35,0.35) {Guard $L(q)$};
\end{tikzpicture}
\end{minipage}}\hfill
\subfloat[Disjoint-cone batching \& recursive bisection ($B \to B_L, B_R$)\label{fig:mot-4-4-b}]{%
\begin{minipage}[b]{0.485\linewidth}\centering
\begin{tikzpicture}[x=1cm,y=1cm,font=\sffamily\scriptsize,>=Stealth]
\path[use as bounding box] (0,0.05) rectangle (6.2,2.75);

\draw[rounded corners=4pt,draw=tapcoteal,dashed,line width=0.65pt,fill=tapcotealfill,fill opacity=0.35]
  (0.20,2.10) rectangle (1.92,2.64);
\node[circle,draw=tapcoteal,fill=tapcotealfill,line width=0.8pt,minimum size=0.30cm,inner sep=0pt,font=\sffamily\bfseries\tiny,text=tapcoteal] (bq1) at (0.50,2.37) {$q_1$};
\node[circle,draw=tapcoink,fill=white,line width=0.65pt,minimum size=0.28cm,inner sep=0pt,font=\sffamily\tiny] (s1) at (1.55,2.37) {$e_a$};
\draw[line width=0.8pt,draw=tapcoteal,->] (bq1) -- (s1);

\draw[rounded corners=4pt,draw=tapcoteal,dashed,line width=0.65pt,fill=tapcotealfill,fill opacity=0.35]
  (0.20,1.48) rectangle (1.92,2.02);
\node[circle,draw=tapcoteal,fill=tapcotealfill,line width=0.8pt,minimum size=0.30cm,inner sep=0pt,font=\sffamily\bfseries\tiny,text=tapcoteal] (bq2) at (0.50,1.75) {$q_2$};
\node[circle,draw=tapcoink,fill=white,line width=0.65pt,minimum size=0.28cm,inner sep=0pt,font=\sffamily\tiny] (s2) at (1.55,1.75) {$e_b$};
\draw[line width=0.8pt,draw=tapcoteal,->] (bq2) -- (s2);

\draw[rounded corners=4pt,draw=tapcored,dashed,line width=0.75pt,fill=tapcoredfill,fill opacity=0.45]
  (2.22,2.10) rectangle (4.28,2.64);
\node[circle,draw=tapcored,fill=tapcoredfill,line width=0.85pt,minimum size=0.30cm,inner sep=0pt,font=\sffamily\bfseries\tiny,text=tapcored] (bq3) at (2.52,2.37) {$q_3$};
\node[circle,draw=tapcored,fill=white,line width=0.75pt,minimum size=0.28cm,inner sep=0pt,font=\sffamily\tiny,text=tapcored] (s3) at (3.42,2.37) {$e_c$};
\draw[line width=0.85pt,draw=tapcored,->] (bq3) -- (s3);
\node[text=tapcored,font=\sffamily\bfseries\tiny] at (3.92,2.37) {$\times$};

\draw[rounded corners=4pt,draw=tapcoteal,dashed,line width=0.65pt,fill=tapcotealfill,fill opacity=0.35]
  (2.22,1.48) rectangle (4.28,2.02);
\node[circle,draw=tapcoteal,fill=tapcotealfill,line width=0.8pt,minimum size=0.30cm,inner sep=0pt,font=\sffamily\bfseries\tiny,text=tapcoteal] (bq4) at (2.52,1.75) {$q_4$};
\node[circle,draw=tapcoink,fill=white,line width=0.65pt,minimum size=0.28cm,inner sep=0pt,font=\sffamily\tiny] (s4) at (3.42,1.75) {$e_d$};
\draw[line width=0.8pt,draw=tapcoteal,->] (bq4) -- (s4);

\node[text=tapcored,font=\sffamily\bfseries\tiny,align=center] at (5.25,2.22) {Root Batch $B$\\$q_3$ Conflict!};
\node[text=tapcoink,font=\sffamily\tiny] at (5.25,1.72) {$\Rightarrow$ Bisect $B$};

\draw[line width=0.5pt,draw=tapcorule,dashed] (0.15,1.38) -- (6.05,1.38);

\node[rounded corners=2pt,draw=tapcoink,fill=tapcograyfill,line width=0.75pt,minimum width=1.95cm,minimum height=0.32cm,inner sep=1.5pt,font=\sffamily\bfseries\tiny] (root) at (2.85,1.14) {$B=\{q_1,q_2,q_3,q_4\}$};

\node[rounded corners=2pt,draw=tapcoteal,fill=tapcotealfill,line width=0.85pt,minimum width=1.65cm,minimum height=0.32cm,inner sep=1.5pt,font=\sffamily\bfseries\tiny,text=tapcoteal] (bl) at (1.15,0.62) {$B_L=\{q_1,q_2\}$};
\node[text=tapcoteal,font=\sffamily\bfseries\tiny,anchor=north] at (1.15,0.42) {$\checkmark$ Commit};

\node[rounded corners=2pt,draw=tapcoamber,fill=tapcoamberfill,line width=0.8pt,minimum width=1.65cm,minimum height=0.32cm,inner sep=1.5pt,font=\sffamily\bfseries\tiny,text=tapcoamber] (br) at (4.35,0.74) {$B_R=\{q_3,q_4\}$};

\node[rounded corners=2pt,draw=tapcored,fill=tapcoredfill,line width=0.8pt,minimum width=1.18cm,minimum height=0.30cm,inner sep=1pt,font=\sffamily\bfseries\tiny,text=tapcored] (l3) at (3.52,0.22) {$\{q_3\}$ Drop};
\node[rounded corners=2pt,draw=tapcoteal,fill=tapcotealfill,line width=0.85pt,minimum width=1.28cm,minimum height=0.30cm,inner sep=1pt,font=\sffamily\bfseries\tiny,text=tapcoteal] (l4) at (5.18,0.22) {$\{q_4\}$ Commit};

\draw[line width=0.8pt,draw=tapcoteal,->] (root.south west) -- (bl.north);
\draw[line width=0.8pt,draw=tapcoamber,->] (root.south east) -- (br.north);
\draw[line width=0.75pt,draw=tapcored,->] (br.south west) -- (l3.north);
\draw[line width=0.75pt,draw=tapcoteal,->] (br.south east) -- (l4.north);
\end{tikzpicture}
\end{minipage}}
\endgroup
  \caption{Graph-wide coarse-to-fine node power screening and conflict-free batch bisection on $G_d$. (a) \tapco{} prunes low-leverage vertex families via $F_c$ and refines active graph vertices using pin switching density $\rho_p$ and slack-risk priority $R(q)$. (b) Top-ranked vertices with disjoint fanout cones are grouped into conflict-free batches $B \subset V_d$, while any batch triggering nonlinear regression is recursively bisected ($B \to B_L, B_R$, \cref{sec:graph-trials}) to isolate conflicting vertices.}
  \Description{Two-panel graph view showing coarse-to-fine activity-weighted vertex screening on G_d on the left, and independent vertex batching with recursive bisection conflict isolation on the right.}
  \label{fig:mot-4-4}
\end{figure}

Accurate power co-optimization faces a fundamental scalability-accuracy dilemma: ranking cell replacements solely by static Liberty leakage or capacitance ratios ignores orders-of-magnitude variations in instance toggle rates and local slack headroom, whereas running detailed instance-level power and graph-timing queries across all $|\mathcal{I}_d| \times |\mathcal{V}(c)|$ cell-variant pairs incurs excessive runtime overhead (\cref{fig:mot-4-4-a}). We address this tradeoff through a two-stage coarse-to-fine power screening funnel on $G_d$: Stage~1 rapidly scores master families using lightweight sampled instance statistics to allocate the candidate pool, and Stage~2 refines selected vertices using pin-level switching-power densities and endpoint-pressure-aware priority ranking (\cref{fig:mot-4-4-a}).

In \emph{Stage~1 (Coarse Master-Family Screening)}, for each instantiated library master $c$, \tapco{} samples a small representative subset of instances to obtain mean empirical internal, switching, and leakage powers $(\overline{P}_{c,\mathrm{int}}, \overline{P}_{c,\mathrm{sw}}, \overline{P}_{c,\mathrm{leak}})$. Let $K_c$ and $\ell_c$ denote the nominal input capacitance and static leakage of master $c$, and let $\gamma_{c,c'}$ denote the library internal-energy scaling ratio between $c'$ and $c$. The family-level power reduction potential $F_c$ is evaluated as
\begin{equation}
  F_c = \max_{c' \in \mathcal{V}(c)} \left[
    \frac{w_L \overline{P}_{c,\mathrm{leak}}}{P_{d,\mathrm{leak}}^0} \left(1 - \frac{\ell_{c'}}{\ell_c}\right)
    + \frac{w_D}{P_{d,\mathrm{dyn}}^0} \left(
      \overline{P}_{c,\mathrm{int}}\bigl(1 - \gamma_{c,c'}\bigr)
      + \overline{P}_{c,\mathrm{sw}}\left(1 - \frac{K_{c'}}{K_c}\right)
    \right)
  \right]_+,
  \label{eq:familypower}
\end{equation}
where $[z]_+ = \max(0,z)$. The coarse screening pool allocates half of its capacity to critical timing-cone instances and populates the remainder from high-$F_c$ cell families, ensuring balanced coverage of timing recovery and power reduction.

In \emph{Stage~2 (Fine Instance-Level Refinement)}, for each candidate instance $i$ admitted from the coarse pool, \tapco{} extracts its exact snapshot internal power $P_{i,\mathrm{int}}^s$, leakage power $P_{i,\mathrm{leak}}^s$, and the pin-level switching-power density $\rho_p = P_{u(p),\mathrm{sw}}^s / C_{u(p)}^s$ on each input net driven by $u(p)$. The instance-specific dynamic and leakage power savings of replacement $q = (i, c, c')$ are formulated as
\begin{align}
  \widehat{\Delta P}_{i,\mathrm{dyn}}(q) &= P_{i,\mathrm{int}}^s \bigl(1 - \gamma_{c,c'}\bigr) + \sum_{p \in \operatorname{inputs}(i)} \rho_p \bigl(C_p(c) - C_p(c')\bigr), \label{eq:instancepower-dyn}\\
  \widehat{\Delta P}_{i,\mathrm{leak}}(q) &= P_{i,\mathrm{leak}}^s \left(1 - \frac{\ell_{c'}}{\ell_c}\right). \label{eq:instancepower}
\end{align}
Combining the normalized power gain $\widehat{g}_P(q)$ with the graph-propagated timing gain $\widehat{g}_T(q)$ from \cref{eq:graphgain} yields the composite predicted score gain $\widehat{g}(q)$:
\begin{equation}
  \widehat{g}_P(q) = w_D \frac{\widehat{\Delta P}_{i,\mathrm{dyn}}(q)}{P_{d,\mathrm{dyn}}^0} + w_L \frac{\widehat{\Delta P}_{i,\mathrm{leak}}(q)}{P_{d,\mathrm{leak}}^0}, \qquad
  \widehat{g}(q) = \widehat{g}_T(q) + \widehat{g}_P(q).
  \label{eq:graphselection}
\end{equation}
To prioritize candidates that deliver high net gain $\widehat{g}(q) > 0$ without endangering critical endpoints, \tapco{} ranks admissible moves by a slack-degradation-penalized priority score $R(q)$:
\begin{equation}
  R(q) = \frac{\widehat{g}(q)}{1 + \beta L(q)}, \qquad
  L(q) = \max_{e \in \mathcal{E}_d(q)} \bigl[s_e^s - \widehat{s}_e(q)\bigr]_+,
  \label{eq:graphrank}
\end{equation}
where $L(q)$ measures the worst-case endpoint slack degradation induced by $q$ and $\beta > 0$ controls endpoint risk aversion.

With \cref{eq:familypower,eq:instancepower-dyn,eq:instancepower,eq:graphselection,eq:graphrank}, \tapco{} concentrates fine-grained graph propagation and physical trial budgets onto the highest-leverage instances, accurately pricing pin-specific dynamic switching savings against multi-endpoint timing slack risk.

\subsection{Batched Physical Trials and Bisection Recovery}
\label{sec:graph-trials}

Executing physical STA and power updates one cell at a time is too slow for large netlists, so high-performance optimizers group top-ranked candidates into multi-cell batches. However, under conventional monolithic batch transactions, if a single candidate inside a batch of $B$ moves triggers an unexpected electrical slew violation or nonlinear path conflict, rolling back the entire batch discards the $B-1$ beneficial moves (\cref{fig:mot-4-4-b}). To retain high batch throughput without sacrificing valid moves, our batched trial engine pairs state-guarded batch execution with recursive bisection recovery ($B \to B_L, B_R$), automatically isolating conflicting replacements while preserving all valid co-batched improvements (\cref{fig:mot-4-4-b}).

Specifically, candidates with $R(q) > 0$ are dequeued in descending order of $R(q)$ and grouped into conflict-free root batches $B = \{q_1, \dots, q_b\}$ containing at most one replacement per instance. Before applying batch $B$, \tapco{} checkpoints the current cell masters and placement coordinates. After applying $B$ and refreshing incremental STA and power analysis on the working state $x$, the engine measures the exact internal PPA objective $J(x)$:
\begin{equation}
  J(x) = w_T \frac{T(x)}{|T_d^0|} - w_D \frac{P_{\mathrm{dyn}}(x)}{P_{d,\mathrm{dyn}}^0} - w_L \frac{P_{\mathrm{leak}}(x)}{P_{d,\mathrm{leak}}^0} - \lambda_{\mathrm{elec}} p_{\mathrm{elec}}(x).
  \label{eq:localobjective}
\end{equation}
A batch trial $x' = \operatorname{Apply}(x, B)$ is immediately committed if $J(x') - J(x) > \epsilon_J$ and all configured endpoint slack floor and electrical guard conditions hold. If a multi-cell batch ($|B| > 1$) fails this acceptance test, \tapco{} restores the checkpointed state $x$ and invokes recursive bisection recovery:
\begin{equation}
  \operatorname{Recover}(x, B) =
  \begin{cases}
    x', & \text{if } J\bigl(\operatorname{Apply}(x,B)\bigr) - J(x) > \epsilon_J \text{ and guards hold}, \\
    \operatorname{Recover}\bigl(\operatorname{Recover}(x, B_L),\, B_R\bigr), & \text{if } |B| > 1 \text{ (with } B = B_L \cup B_R,\; |B_L| = \lfloor|B|/2\rfloor\text{)}, \\
    x, & \text{if } |B| = 1 \text{ and trial rejected}.
  \end{cases}
  \label{eq:bisection}
\end{equation}
At periodic rebuild intervals, \tapco{} refreshes the full STA snapshot $s$ and reconstructs the graph priorities $R(q)$ to reflect the evolved circuit state before returning the final legalized netlist to the outer loop (\cref{sec:method}).

With \cref{eq:localobjective,eq:bisection}, \tapco{} combines the throughput of batched physical evaluation with the precision of singleton conflict isolation, guaranteeing that every retained inner-loop transformation strictly improves measured circuit PPA before outer-loop signoff verification.

\section{Experiments}
\label{sec:experiments}

We conduct a comprehensive experimental campaign to evaluate \tapco{}. First, we benchmark \tapco{} against the MLCAD 2026 Contest 1st-place winner \tool{CERBERUS}~\cite{zhou2026cerberus,mlcad26results} and standard \openroad{} optimization flows across all eight contest benchmarks (\cref{sec:exp-main}). Second, we perform a head-to-head comparison against our preliminary conference version (\tool{TAPCO})~\cite{tang2026tapco} to quantify the gains in convergence reliability, timing closure, and multi-VT allocation brought by our bi-level graph-guided architecture (\cref{sec:exp-conf}). Third, we validate the physical fidelity and individual contributions of the inner-loop graph priors through $77$ independent physical singleton probes and controlled component ablations (\cref{sec:exp-ablation}). Finally, we analyze outer-loop LLM policy evolution and design-specific specialization across the benchmark suite (\cref{sec:exp-policy}).

\subsection{Experimental Setup}
\label{sec:exp-setup}

\paragraph{Benchmarks and Evaluation Protocol.} We evaluate all eight industrial-scale benchmarks from the MLCAD 2026 Contest suite~\cite{mlcad26contest} mapped to the ASAP7 predictive 7-nm FinFET technology~\cite{clark2016asap7}. As summarized in \cref{tab:journal-benchmarks}, the benchmark suite spans $20{,}097$ to $312{,}696$ instances and over two orders of magnitude in initial setup TNS ($-12.44\text{ ns}$ to $-7568.43\text{ ns}$). Crucially, the suite encompasses two distinct physical regimes: five logic-dominated circuits (\texttt{AES}, \texttt{JPEG}, \texttt{NVDLA-P}, \texttt{NVDLA-A}, and \texttt{NVDLA-M}) where standard-cell sizing and multi-VT assignment control the majority of total power, and three macro-rich SoC blocks (\texttt{Ariane}, \texttt{NVDLA-C}, and \texttt{MemPool}) containing fixed SRAM macros that establish a high baseline power floor and severe routing/placement blockages.

\begin{table}[!t]
  \centering
  \caption{Initial benchmark characteristics of the eight-design ASAP7 suite under the reference contest constraints. Instance counts include all physically instantiated standard cells and hard macros.}
  \label{tab:journal-benchmarks}
  \small\setlength{\tabcolsep}{6pt}
  \begin{tabular}{lrrrrc}
\toprule
Design & Instances & TNS (ns) & Total (mW) & Leakage ($\mu$W) & SRAM \\
\midrule
AES & 20,097 & -12.44 & 432.0 & 96.3 & No \\
JPEG & 64,602 & -48.77 & 294.0 & 174.0 & No \\
NVDLA-P & 106,994 & -198.18 & 40.0 & 306.0 & No \\
Ariane & 210,455 & -7568.43 & 658.0 & 17900.0 & Yes \\
NVDLA-A & 75,686 & -203.23 & 165.0 & 321.0 & No \\
NVDLA-C & 312,696 & -55.58 & 601.0 & 17300.0 & Yes \\
MemPool & 205,063 & -3680.33 & 279.0 & 3070.0 & Yes \\
NVDLA-M & 20,346 & -13.99 & 44.6 & 61.5 & No \\
\bottomrule
\end{tabular}

\end{table}

All implementations are evaluated using the official contest signoff flow based on \openroad{} and \tool{OpenSTA}~\cite{ajayi2019openroad,openroadrsz}. Following \cref{eq:score}, the signoff quality score $S_Q$ uses PPA tradeoff weights $(w_T, w_D, w_L) = (30, 50, 50)$ together with the total penalty deduction $\mathrm{Pen.} = -\sum_j \lambda_j p_{d,j}$ combining normalized electrical slew/capacitance violations, average cell displacement $D_{\mathrm{avg}}$ (mean Manhattan displacement per standard cell in placement rows relative to the initial placement, weighted by $-0.5 D_{\mathrm{avg}}$), and routing congestion overflow. Every emitted netlist and placed layout must pass strict placement legality checks and independent formal LEC against the original benchmark netlist; any implementation that fails legality or LEC is marked invalid (\texttt{Fail}) and disqualified from incumbent selection.

To compare implementation quality across different execution environments, all quality scores in this section exclude the runtime term while retaining the original timing and power weights and other score adjustments. The conference \tool{TAPCO} scores are recomputed using this definition; for \tool{CERBERUS}, we use the authors' reported $S_{\mathrm{final}}^{\mathrm{noR}}$ values.

\paragraph{Compared Baselines.} We compare \tapco{} against four representative baselines:
\begin{enumerate}
  \item \textbf{\tool{OR-LR}}~\cite{ajayi2019openroad,openroadrsz}: The upstream \openroad{} Lagrangian-relaxation global gate sizer and timing-driven repair flow executed for up to $100$ iterations from the initial placed design.
  \item \textbf{\tool{OR-LR+PR}}~\cite{openroadrsz}: A two-stage sequential baseline that follows \tool{OR-LR} timing repair with native slack-driven power recovery ($100\%$ recovery effort). We report its direct post-recovery signoff quality in \cref{tab:journal-baselines} alongside best-of-stage checkpoint fallback to \tool{OR-LR}.
  \item \textbf{\tool{TAPCO}}~\cite{tang2026tapco}: Our preliminary MLCAD conference version, which applies outer-loop LLM adaptation over conventional resizer operators~\cite{liu2026hypas} without the inner-loop graph-guided physical priors, driver-load coupling, or coarse-to-fine instance power screening developed in \cref{sec:graph}.
  \item \textbf{\tool{CERBERUS}}~\cite{zhou2026cerberus,mlcad26results}: The 1st-place winning system of the MLCAD 2026 Contest, representing the state-of-the-art agentic optimization reference on this benchmark suite.
\end{enumerate}

\subsection{Dominance Over Contest Winner \tool{CERBERUS} and Conventional Flows}
\label{sec:exp-main}

\Cref{tab:journal-baselines} presents the overall signoff quality comparison across all eight benchmarks against conventional flows, \tool{TAPCO}, and the contest champion \tool{CERBERUS}, while \cref{tab:journal-ppa} reports \tapco{}'s worst negative slack (WNS), setup TNS, and power, together with relative PPA reductions over the initial placed designs (\cref{tab:journal-benchmarks}) and decomposed score credits.

\begin{table}[!t]
  \centering
  \caption{Main quality score ($S_Q$) comparison across all eight ASAP7 benchmarks. \tool{OR-LR+PR} reports the direct post-power-recovery score; because uncoupled power recovery degrades every design, a best-of-stage checkpoint selector between \tool{OR-LR} and \tool{OR-LR+PR} simply reverts to \tool{OR-LR} on all eight benchmarks (mean $21.164$). \texttt{Fail} indicates that the method failed to produce a placement-legal, LEC-verified final-phase implementation. Bold highlights the best result on each design.}
  \label{tab:journal-baselines}
  \footnotesize\setlength{\tabcolsep}{3.0pt}
  \begin{tabular}{lrrrrrrr}
\toprule
Design & \tool{OR-LR}~\shortcite{openroadrsz} & \tool{OR-LR+PR}~\shortcite{openroadrsz} & \tool{TAPCO}~\shortcite{tang2026tapco} & \tool{CERBERUS}~\shortcite{zhou2026cerberus} & \textbf{\tapco{}} & $\Delta$ \tool{TAPCO} & $\Delta$ \tool{CERB.} \\
\midrule
AES     &  1.354 &   $-18.360$ & 37.778 & 58.502 & \textbf{64.729} & $+26.951$ & $+6.227$ \\
JPEG    & 29.951 &   $-32.809$ & 29.801 & 42.991 & \textbf{56.928} & $+27.127$ & $+13.937$ \\
NVDLA-P & 51.261 &  $-121.743$ & Fail   & 59.341 & \textbf{63.325} & Valid ($\checkmark$) & $+3.984$ \\
Ariane  &  0.102 &    $-0.290$ & 17.074 & 27.549 & \textbf{27.870} & $+10.796$ & $+0.322$ \\
NVDLA-A & 42.620 &    $-7.892$ & Fail   & 58.779 & \textbf{71.028} & Valid ($\checkmark$) & $+12.249$ \\
NVDLA-C & 25.071 & $-1607.154$ & Fail   & 27.844 & \textbf{28.754} & Valid ($\checkmark$) & $+0.910$ \\
MemPool &  7.825 &     $1.008$ & Fail   & 12.429 & \textbf{14.060} & Valid ($\checkmark$) & $+1.630$ \\
NVDLA-M & 11.129 &   $-15.543$ & 24.737 & 41.586 & \textbf{46.455} & $+21.718$ & $+4.869$ \\
\midrule
Valid / Total & $8/8$ & $8/8$ & $4/8$ & $8/8$ & $\mathbf{8/8}$ & $+4$ Valid & $8/8$ Wins \\
Mean (4 Valid)& 10.634 & $-16.751$ & 27.348 & 42.657 & \textbf{48.996} & $\mathbf{+21.648}$ & $\mathbf{+6.339}$ \\
Mean (All 8)  & 21.164 & $-225.348$ & --     & 41.128 & \textbf{46.644} & --        & $\mathbf{+5.516}$ \\
\bottomrule
\end{tabular}

\end{table}

\begin{table}[!t]
  \centering
  \caption{Detailed physical signoff metrics and decomposed score components of \tapco{} across all eight benchmarks. Relative percentages in parentheses indicate reductions relative to the initial placed baselines in \cref{tab:journal-benchmarks}. Following \cref{eq:score}, $\text{Timing} + \text{Dyn.} + \text{Leak.} + \text{Pen.} = S_Q$, where $\text{Pen.} = -\sum_j \lambda_j p_{d,j}$ combines the displacement penalty $-0.5 D_{\mathrm{avg}}$ (mean $D_{\mathrm{avg}} = 0.044$ across the suite) and residual electrical slew/capacitance penalties (e.g., $-0.34$ electrical and $-0.01$ displacement points on \texttt{NVDLA-A}).}
  \label{tab:journal-ppa}
  \footnotesize\setlength{\tabcolsep}{2.2pt}
  \begin{tabular}{lrrrrrrrrr}
    \toprule
    & \multicolumn{4}{c}{Final Physical Signoff Metrics (\tapco{})} & \multicolumn{5}{c}{Score Decomposition (pts)} \\
    \cmidrule(lr){2-5}\cmidrule(lr){6-10}
    Design & WNS (ns) & Final TNS (ns) & Total Power (mW) & Leakage ($\mu$W) & Timing & Dyn. & Leak. & Pen. & $S_Q$ \\
    \midrule
    AES     & $-0.096$ & $-12.15$ ($-2.3\%$)   & $259.0$ ($-40.0\%$) & $11.5$ ($-88.1\%$)   & $+0.70$  & $+20.02$ & $+44.03$ & $-0.02$ & $64.73$ \\
    JPEG    & $-0.102$ & $-38.34$ ($-21.4\%$)  & $225.0$ ($-23.5\%$) & $38.9$ ($-77.6\%$)   & $+6.42$  & $+11.72$ & $+38.82$ & $-0.03$ & $56.93$ \\
    NVDLA-P & $-0.351$ & $-69.05$ ($-65.2\%$)  & $38.4$ ($-4.0\%$)   & $48.3$ ($-84.2\%$)   & $+19.55$ & $+1.69$  & $+42.11$ & $-0.02$ & $63.33$ \\
    Ariane  & $-0.256$ & $-602.92$ ($-92.0\%$) & $658.0$ ($0.0\%$)   & $17800.0$ ($-0.6\%$) & $+27.61$ & $-0.01$  & $+0.28$  & $-0.01$ & $27.87$ \\
    NVDLA-A & $-0.188$ & $-118.98$ ($-41.5\%$) & $111.0$ ($-32.7\%$) & $47.3$ ($-85.3\%$)   & $+12.44$ & $+16.31$ & $+42.63$ & $-0.35$ & $71.03$ \\
    NVDLA-C & $-0.193$ & $-8.46$ ($-84.8\%$)   & $585.0$ ($-2.7\%$)  & $16600.0$ ($-4.0\%$) & $+25.43$ & $+1.31$  & $+2.02$  & $-0.01$ & $28.75$ \\
    MemPool & $-0.555$ & $-2870.91$ ($-22.0\%$)& $268.0$ ($-3.9\%$)  & $2730.0$ ($-11.1\%$) & $+6.60$  & $+1.93$  & $+5.54$  & $-0.01$ & $14.06$ \\
    NVDLA-M & $-0.350$ & $-15.89$ ($+13.6\%$)  & $38.5$ ($-13.7\%$)  & $7.6$ ($-87.6\%$)    & $-4.07$  & $+6.79$  & $+43.80$ & $-0.06$ & $46.45$ \\
    \midrule
    Mean (8) & $-0.261$ & -- ($-39.4\%$)       & -- ($-15.1\%$)      & -- ($-54.8\%$)       & $+11.83$ & $+7.47$  & $+27.40$ & $-0.06$ & $46.64$ \\
    \bottomrule
  \end{tabular}
\end{table}

\paragraph{Clean Sweep Over Contest Champion \tool{CERBERUS}.} As shown in \cref{tab:journal-baselines,tab:journal-ppa}, \tapco{} achieves a \textbf{100\% ($8/8$) win rate} against the contest winner \tool{CERBERUS}, elevating the eight-design mean quality score from $41.128$ to \textbf{46.644}, an average gain of \textbf{$+5.516$ points ($+13.4\%$ relative improvement)}. On logic-dominated designs where discrete sizing and multi-VT substitutions offer rich coupling opportunities, \tapco{} outperforms \tool{CERBERUS} by wide margins: \textbf{$+13.937$ points on \texttt{JPEG}} ($56.928$ vs.\ $42.991$), \textbf{$+12.249$ points on \texttt{NVDLA-A}} ($71.028$ vs.\ $58.779$), \textbf{$+6.227$ points on \texttt{AES}} ($64.729$ vs.\ $58.502$), \textbf{$+4.869$ points on \texttt{NVDLA-M}} ($46.455$ vs.\ $41.586$), and \textbf{$+3.984$ points on \texttt{NVDLA-P}} ($63.325$ vs.\ $59.341$). Even on the three macro-dominated designs (\texttt{Ariane}, \texttt{NVDLA-C}, and \texttt{MemPool}), where uneditable SRAM macros dominate total chip power and restrict standard-cell placement whitespace, \tapco{} consistently surpasses \tool{CERBERUS} by $+0.322$, $+0.910$, and $+1.630$ points, respectively. While agentic frameworks that tune outer scripts alone hit a ceiling when the resizer misjudges the timing and power effects of local moves, \tapco{} breaks through this barrier by pairing outer policy adaptation with inner graph-guided candidate valuation.

\paragraph{Failure of Conventional Decoupled Sizing and Power Recovery.} Compared with standard \openroad{} Lagrangian relaxation (\tool{OR-LR}, mean score $21.164$), \tapco{} more than doubles overall quality ($+25.480$ mean gain), improving \texttt{AES} by $47.8\times$ ($1.354 \to 64.729$) and \texttt{Ariane} by $273\times$ ($0.102 \to 27.870$). Notably, the two-stage \tool{OR-LR+PR} baseline fails to improve upon pure \tool{OR-LR} on any of the eight designs: invoking native slack-based power recovery after timing repair degrades the raw score on every benchmark (collapsing \texttt{MemPool} from $7.825$ to $1.008$ and driving the remaining seven circuits into negative scores, such as $-18.360$ on \texttt{AES} and $-32.809$ on \texttt{JPEG}), so even a best-of-stage checkpoint selector merely reverts to \tool{OR-LR} ($21.164$). This confirms that sequential, uncoupled power recovery greedily consumes slack margins without accounting for shared reconvergent cones or upstream driver loading, whereas \tapco{}'s unified graph prior co-optimizes timing and power simultaneously.

\begin{figure}[!t]
  \centering
  \includegraphics[width=\linewidth]{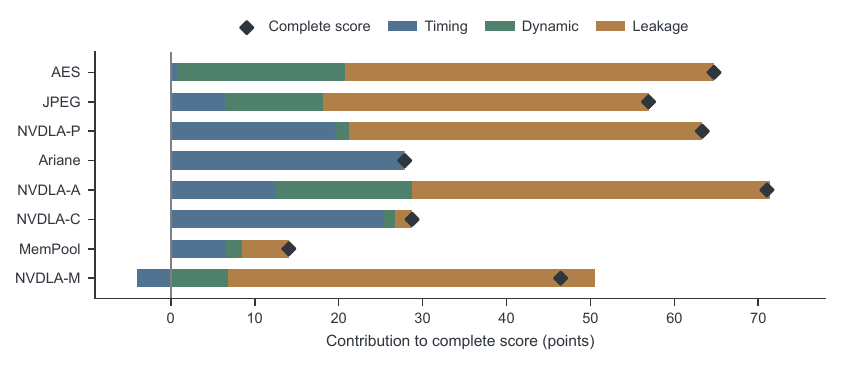}
  \caption{Decomposition of \tapco{} signoff quality scores into normalized setup timing ($\Delta T_d$), dynamic power ($\Delta P_{\mathrm{dyn}}$), and leakage power ($\Delta P_{\mathrm{leak}}$) contributions across all eight benchmarks. Diamonds mark the complete signoff score $S_Q$ after electrical, displacement, and routing adjustments.}
  \Description{Horizontal bar chart decomposing TAPCO-G scores across eight benchmarks into timing, dynamic power, and leakage gains.}
  \label{fig:journal-quality-composition}
\end{figure}

\paragraph{Adaptive Regime Specialization Across the Suite.} \Cref{fig:journal-quality-composition} reveals how \tapco{} automatically tailors its optimization trajectory to the physical regime of each circuit:
\begin{itemize}
  \item \emph{Balanced Timing and Power Co-Optimization Regime:} On \texttt{AES}, \texttt{JPEG}, \texttt{NVDLA-P}, and \texttt{NVDLA-A}, \tapco{} simultaneously achieves positive setup timing recovery and massive power reductions. On \texttt{AES}, leakage drops by $88.1\%$ ($96.3 \to 11.5\,\mu\text{W}$, contributing $+44.029$ points) and total power drops by $40.0\%$ ($432.0 \to 259.0\text{ mW}$, contributing $+20.018$ dynamic-power points) while TNS also improves from $-12.44$ to $-12.15\text{ ns}$. On \texttt{NVDLA-A} and \texttt{NVDLA-P}, \tapco{} recovers $41.5\%$ and $65.2\%$ of initial TNS alongside $85.3\%$ and $84.2\%$ leakage reductions.
  \item \emph{Macro-Dominated Timing-First Regime:} On \texttt{Ariane} and \texttt{NVDLA-C}, where fixed SRAM macros account for over $90\%$ of total leakage ($\sim 17.8\text{ mW}$), standard-cell power downsizing yields negligible chip-level return. \tapco{} detects this regime from evaluator diagnostics and concentrates effort on aggressive critical-path repair, slashing \texttt{Ariane}'s TNS by $92.0\%$ ($-7568.43 \to -602.92\text{ ns}$, $+27.610$ timing points) and \texttt{NVDLA-C}'s TNS by $84.8\%$ ($-55.58 \to -8.46\text{ ns}$, $+25.434$ timing points).
  \item \emph{Slack-for-Power Exchange Regime:} On \texttt{NVDLA-M}, with a mild $-13.99\text{ ns}$ initial TNS and $61.5\,\mu\text{W}$ cell leakage, \tapco{} executes a controlled $-4.074$-point timing trade ($-13.99 \to -15.89\text{ ns}$) to unlock $+43.805$ leakage points ($87.6\%$ leakage reduction to $7.6\,\mu\text{W}$) and $+6.788$ dynamic-power points.
\end{itemize}

\subsection{Comprehensive Superiority Over the Preliminary Conference Version}
\label{sec:exp-conf}

\begin{figure}[!t]
  \centering
  \includegraphics[width=\linewidth]{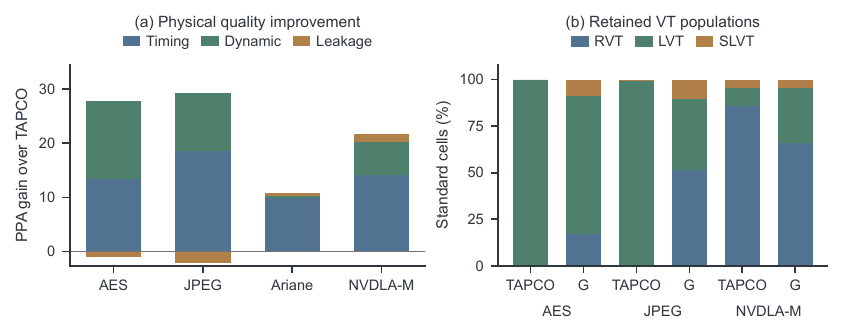}
  \caption{Physical comparison between \tapco{} and \tool{TAPCO}. (a) Component-wise timing ($\Delta T_d$), dynamic-power ($\Delta P_{\mathrm{dyn}}$), and leakage ($\Delta P_{\mathrm{leak}}$) score gains of \tapco{} over \tool{TAPCO} under identical $(30,50,50)$ weights. (b) Standard-cell VT (RVT, LVT, SLVT) distributions in the emitted implementations.}
  \Description{Left panel shows stacked bar improvements of TAPCO-G over conference TAPCO in timing and dynamic power; right panel shows multi-VT cell population breakdowns.}
  \label{fig:journal-conference}
\end{figure}

To isolate the advancement of \tapco{} over \tool{TAPCO}~\cite{tang2026tapco}, we compare their convergence reliability, physical PPA metrics (\cref{tab:conf-comparison}), and standard-cell VT distributions (\cref{fig:journal-conference}).

\begin{table}[!t]
  \centering
  \caption{Head-to-head physical PPA comparison between \tool{TAPCO}~\cite{tang2026tapco} and \tapco{} on the four designs where \tool{TAPCO} produced valid final-phase solutions (out of eight total benchmarks). Units: $\text{TNS}$ in $\text{ns}$, dynamic power $P_{\mathrm{dyn}}$ in $\text{mW}$, and leakage power $P_{\mathrm{leak}}$ in $\mu\text{W}$.}
  \label{tab:conf-comparison}
  \footnotesize\setlength{\tabcolsep}{1.8pt}
  \begin{tabular}{lrrrrrrrrrrr}
    \toprule
    & \multicolumn{3}{c}{Initial Baseline} & \multicolumn{4}{c}{\tool{TAPCO}~\shortcite{tang2026tapco}} & \multicolumn{4}{c}{\textbf{\tapco{} (Ours)}} \\
    \cmidrule(lr){2-4}\cmidrule(lr){5-8}\cmidrule(lr){9-12}
    Design & TNS & $P_{\mathrm{dyn}}$ & $P_{\mathrm{leak}}$ & TNS & $P_{\mathrm{dyn}}$ & $P_{\mathrm{leak}}$ & Score & TNS & $P_{\mathrm{dyn}}$ & $P_{\mathrm{leak}}$ & Score ($\Delta$) \\
    \midrule
    AES     & $-12.44$   & $431.9$ & $96.3$    & $-17.67$   & $385.0$ & $9.7$     & $37.778$ & $\mathbf{-12.15}$  & $\mathbf{259.0}$ & $11.5$            & $\mathbf{64.729}$ ($+26.951$) \\
    JPEG    & $-48.77$   & $293.8$ & $174.0$   & $-68.51$   & $288.0$ & $31.5$    & $29.801$ & $\mathbf{-38.34}$  & $\mathbf{225.0}$ & $38.9$            & $\mathbf{56.928}$ ($+27.127$) \\
    Ariane  & $-7568.43$ & $640.1$ & $17900.0$ & $-3092.71$ & $645.0$ & $18000.0$ & $17.074$ & $\mathbf{-602.92}$ & $\mathbf{640.2}$ & $\mathbf{17800.0}$& $\mathbf{27.870}$ ($+10.796$) \\
    NVDLA-M & $-13.99$   & $44.5$  & $61.5$    & $-22.49$   & $43.9$  & $9.5$     & $24.737$ & $\mathbf{-15.89}$  & $\mathbf{38.5}$  & $\mathbf{7.6}$    & $\mathbf{46.455}$ ($+21.718$) \\
    \bottomrule
  \end{tabular}
\end{table}

\paragraph{1. Breakthrough in Feasibility and Convergence ($4/8 \to 8/8$ Valid).} As shown in \cref{tab:journal-baselines}, \tool{TAPCO} failed to produce valid final-phase implementations on four out of the eight contest benchmarks (\texttt{NVDLA-P}, \texttt{NVDLA-A}, \texttt{NVDLA-C}, and \texttt{MemPool}), suffering a $50\%$ failure rate on large and macro-congested designs. Because \tool{TAPCO} relied exclusively on coarse-grained resizer sweeps without microscopic graph physical priors or state-guarded trial recovery, aggressive parameter settings on congested designs either triggered unrecoverable placement legalization failures or violated electrical/equivalence constraints. By introducing snapshot-anchored graph screening, bisection-recovered physical trials, and strict valid-incumbent gating, \tapco{} achieves \textbf{100\% ($8/8$) valid, LEC-verified convergence} across the entire benchmark suite.

\paragraph{2. Massive Timing and Dynamic-Power Dominance ($+21.648$ Mean Score Gain).} Even on the four benchmarks where \tool{TAPCO} succeeded (\cref{tab:conf-comparison} and \cref{fig:journal-conference}(a)), \tool{TAPCO} achieved its scores by severely sacrificing setup timing: relative to the initial designs, \tool{TAPCO} degraded TNS by $+42.0\%$ on \texttt{AES} ($-12.44 \to -17.67\text{ ns}$), $+40.5\%$ on \texttt{JPEG} ($-48.77 \to -68.51\text{ ns}$), and $+60.8\%$ on \texttt{NVDLA-M} ($-13.99 \to -22.49\text{ ns}$). In contrast, \textbf{relative to \tool{TAPCO}}, \tapco{} simultaneously improves setup timing and reduces dynamic power across all four circuits (while relative to the initial baseline, \tapco{} improves TNS on \texttt{AES}, \texttt{JPEG}, and \texttt{Ariane}, and on \texttt{NVDLA-M} contains the slack-for-leakage trade to $-15.89\text{ ns}$ vs.\ \tool{TAPCO}'s $-22.49\text{ ns}$):
\begin{itemize}
  \item \textbf{Setup TNS Reductions of $29.3\%$ to $80.5\%$ over \tool{TAPCO}:} Relative to \tool{TAPCO}, \tapco{} slashes TNS magnitude by \textbf{31.2\% on \texttt{AES}} ($-17.67 \to -12.15\text{ ns}$), \textbf{44.0\% on \texttt{JPEG}} ($-68.51 \to -38.34\text{ ns}$), \textbf{80.5\% on \texttt{Ariane}} ($-3092.71 \to -602.92\text{ ns}$), and \textbf{29.3\% on \texttt{NVDLA-M}} ($-22.49 \to -15.89\text{ ns}$).
  \item \textbf{Dynamic Power ($P_{\mathrm{dyn}}$) Reductions of up to $32.7\%$ over \tool{TAPCO}:} Relative to \tool{TAPCO}, \tapco{} directly reduces dynamic power $P_{\mathrm{dyn}}$ from $385.0$ to $259.0\text{ mW}$ ($-32.7\%$) on \texttt{AES}, from $288.0$ to $225.0\text{ mW}$ ($-21.9\%$) on \texttt{JPEG}, from $645.0$ to $640.2\text{ mW}$ ($-0.7\%$) on \texttt{Ariane}, and from $43.9$ to $38.5\text{ mW}$ ($-12.3\%$) on \texttt{NVDLA-M}, while also further reducing \texttt{NVDLA-M} leakage from $9.5$ to $7.6\,\mu\text{W}$ ($-20.0\%$).
  \item \textbf{Net Quality Surges of $+10.80$ to $+27.13$ Points:} Across the four designs, \tapco{} elevates mean quality from $27.348$ to \textbf{$48.996$ ($+21.648$ points, or $+79.2\%$ relative gain)}, nearly doubling the score on \texttt{AES} ($+26.951$), \texttt{JPEG} ($+27.127$), and \texttt{NVDLA-M} ($+21.718$).
\end{itemize}

\paragraph{3. Physical Mechanism: Surgical Multi-VT Allocation vs.\ Blind Global Swapping.} \Cref{fig:journal-conference}(b) explains the physical root cause of this performance gap by inspecting the multi-VT cell composition of the emitted netlists. Lacking endpoint-aware graph propagation, \tool{TAPCO} applied a blunt, circuit-wide cell preference: on \texttt{AES} and \texttt{JPEG}, it converted $99.6\%$ ($19{,}459 / 19{,}536$) and $99.9\%$ of matched cells into uniform LVT gates, leaving critical paths starved of fast SLVT drive while wasting dynamic and leakage power on non-critical cones. Conversely, \tapco{}'s graph-guided prior (\cref{sec:graph}) precisely quantifies each instance's clipped endpoint slack impact $\widehat{g}_T(q)$ and pin switching density $\rho_p$. On \texttt{AES}, \tapco{} surgically deploys $1{,}601$ ultra-fast SLVT cells on the most critical timing cones, retains $13{,}380$ LVT cells on moderate paths, and assigns $3{,}034$ low-leakage RVT cells to non-critical paths; on \texttt{NVDLA-M}, it selectively expands the LVT population from $1{,}946$ to $5{,}563$ on violating cones to recover $6.60\text{ ns}$ of TNS while preserving RVT across the rest of the netlist.

\subsection{Fidelity and Ablation of Graph-Guided Physical Priors}
\label{sec:exp-ablation}

We next look inside the inner loop to verify (i) how accurately our graph-guided physical prior predicts gains in the joint timing and power objective for isolated cell replacements, and (ii) how each inner-loop modeling component contributes to end-to-end optimization quality.

\paragraph{1. Physical Probe Verification of Predictor Fidelity.} To evaluate prediction accuracy free from trajectory drift, we conduct $77$ independent singleton physical probe experiments ($45$ candidate replacements on \texttt{AES} and $32$ on \texttt{NVDLA-M}, stratified across the top-ranked candidates and gain quartiles). For every single probe, we apply the replacement to the physical netlist, run full incremental STA and power measurement to obtain the ground-truth inner objective change $\Delta J$, and then restore all cell masters and placements to the exact baseline snapshot (verified by $77/77$ zero-deviation restoration checks).

\begin{figure}[!t]
  \centering
  \includegraphics[width=\linewidth]{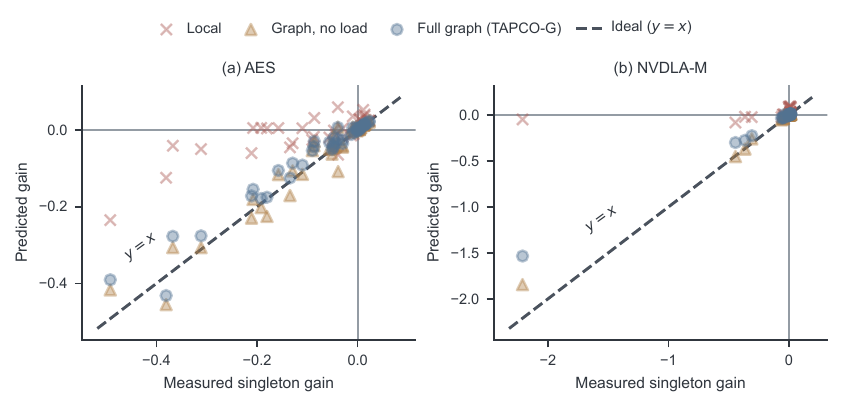}
  \caption{Predicted versus physically measured singleton timing--power gain across $77$ independent cell-replacement probes ($45$ on \texttt{AES} and $32$ on \texttt{NVDLA-M}). Each point is physically applied, evaluated via full signoff STA and power analysis, and exactly restored to the baseline state. The dashed diagonal ($y = x$) marks ideal zero-error prediction.}
  \Description{Scatter plots on AES and NVDLA-M showing that graph-guided predictions closely track measured physical gains along the y=x diagonal whereas local predictions exhibit large scatter and sign errors.}
  \label{fig:journal-prediction}
\end{figure}

\begin{table}[!t]
  \centering
  \caption{Singleton prediction fidelity across $77$ independent physical replacement probes on \texttt{AES} ($45$ probes) and \texttt{NVDLA-M} ($32$ probes), comparing myopic local estimation, uncoupled graph propagation, and the full \tapco{} graph model with driver-load coupling.}
  \label{tab:journal-prediction}
  \footnotesize\setlength{\tabcolsep}{4.5pt}
  \begin{tabular}{llccc}
    \toprule
    Design & Timing--Power Predictor & MAE (pts) $\downarrow$ & Sign Accuracy $\uparrow$ & Top-8 Hit Rate $\uparrow$ \\
    \midrule
    \texttt{AES}     & Myopic Local Estimator     & $0.0683$ & $34/45$ ($75.6\%$) & $5/8$ ($62.5\%$) \\
    \texttt{AES}     & Graph Propagation (No Load)& $0.0183$ & $40/45$ ($88.9\%$) & $7/8$ ($87.5\%$) \\
    \texttt{AES}     & \textbf{Full Graph Model (\tapco{})} & $\mathbf{0.0185}$ & $\mathbf{42/45}$ ($\mathbf{93.3\%}$) & $\mathbf{8/8}$ ($\mathbf{100.0\%}$) \\
    \midrule
    \texttt{NVDLA-M} & Myopic Local Estimator     & $0.1283$ & $26/32$ ($81.3\%$) & $6/8$ ($75.0\%$) \\
    \texttt{NVDLA-M} & Graph Propagation (No Load)& $0.0175$ & $29/32$ ($90.6\%$) & $7/8$ ($87.5\%$) \\
    \texttt{NVDLA-M} & \textbf{Full Graph Model (\tapco{})} & $\mathbf{0.0364}$ & $\mathbf{31/32}$ ($\mathbf{96.9\%}$) & $\mathbf{8/8}$ ($\mathbf{100.0\%}$) \\
    \bottomrule
  \end{tabular}
\end{table}

As demonstrated in \cref{fig:journal-prediction,tab:journal-prediction}, the myopic local estimator suffers from severe overestimation and frequent sign reversals (mean absolute error $\mathrm{MAE} = 0.0683$ on \texttt{AES} and $0.1283$ on \texttt{NVDLA-M}), predicting the wrong gain direction on nearly one-fourth of the candidates ($34/45$ and $26/32$ sign accuracy) because it ignores reconvergent path masking and positive-slack clipping. Introducing incremental DAG propagation (\cref{sec:graph-prop,sec:graph-seq}) slashes MAE by more than $3.7\times$, and incorporating upstream driver-load coupling (\cref{sec:graph-snapshot}) further elevates gain-direction sign accuracy to \textbf{$42/45$ (93.3\%) on \texttt{AES}} and \textbf{$31/32$ (96.9\%) on \texttt{NVDLA-M}} while reducing MAE by \textbf{72.9\%} ($0.0683 \to 0.0185$) and \textbf{71.6\%} ($0.1283 \to 0.0364$) relative to the local estimator. Most importantly for candidate selection, among the top $8$ candidates ranked by our full graph model, \textbf{100\% ($8/8$) deliver verified positive physical gains on both designs}, compared to only $5/8$ ($62.5\%$) and $6/8$ ($75.0\%$) under myopic local ranking.

\begin{table}[!t]
  \centering
  \caption{Matched component ablations of the inner-loop physical priors under identical initial checkpoints, binaries, and root-action budgets. Each row isolates the effect of enabling a single physical-guidance mechanism.}
  \label{tab:journal-components}
  \footnotesize\setlength{\tabcolsep}{3.6pt}
  \begin{tabular}{llrrrrr}
    \toprule
    Ablated Component & Design & Budget & Accepted (OFF $\to$ ON) & Score OFF & Score ON & Gain ($\Delta S_Q$) \\
    \midrule
    Graph Prop.\ (COMB)     & \texttt{AES}     & $256$ & $70 \to 185$ ($2.64\times$) & $60.201$ & $\mathbf{60.821}$ & $+0.620$ \\
    Graph Prop.\ (COMB)     & \texttt{NVDLA-M} & $256$ & $39 \to 188$ ($4.82\times$) & $42.284$ & $\mathbf{43.448}$ & $+1.164$ \\
    Graph Prop.\ (FF+COMB)  & \texttt{AES}     & $128$ & $64 \to 112$ ($1.75\times$) & $60.293$ & $\mathbf{60.635}$ & $+0.341$ \\
    Graph Prop.\ (FF+COMB)  & \texttt{NVDLA-M} & $128$ & $51 \to 104$ ($2.04\times$) & $42.236$ & $\mathbf{42.489}$ & $+0.253$ \\
    Graph Prop.\ (Late-Stage)& \texttt{AES}    & $256$ & $82 \to 164$ ($2.00\times$) & $61.848$ & $\mathbf{62.164}$ & $+0.316$ \\
    Graph Prop.\ (Late-Stage)& \texttt{NVDLA-M}& $256$ & $48 \to 142$ ($2.96\times$) & $45.038$ & $\mathbf{45.264}$ & $+0.226$ \\
    \midrule
    Driver-Load Coupling ($\Delta C_p$) & \texttt{AES}     & $256$ & $80 \to 185$ ($2.31\times$) & $60.360$ & $\mathbf{60.821}$ & $+0.461$ \\
    Driver-Load Coupling ($\Delta C_p$) & \texttt{NVDLA-M} & $256$ & $170 \to 188$ ($1.11\times$)& $43.268$ & $\mathbf{43.448}$ & $+0.181$ \\
    \midrule
    Coarse-to-Fine Power Funnel  & \texttt{AES}     & $128$ & $118 \to 124$ ($1.05\times$)& $64.362$ & $\mathbf{64.390}$ & $+0.028$ \\
    Coarse-to-Fine Power Funnel  & \texttt{NVDLA-A} & $128$ & $109 \to 121$ ($1.11\times$)& $70.210$ & $\mathbf{70.315}$ & $+0.104$ \\
    Coarse-to-Fine Power Funnel  & \texttt{NVDLA-P} & $128$ & $122 \to 128$ ($1.05\times$)& $63.097$ & $\mathbf{63.325}$ & $+0.229$ \\
    \midrule
    Expanded Candidate Horizon   & \texttt{AES}     & $128$ & $29 \to 116$ ($4.00\times$) & $59.926$ & $\mathbf{60.135}$ & $+0.209$ \\
    \bottomrule
  \end{tabular}
\end{table}

\begin{figure}[!t]
  \centering
  \includegraphics[width=\linewidth]{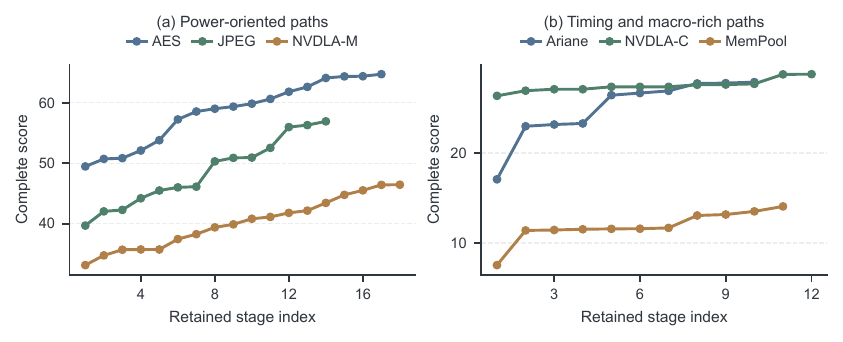}
  \caption{Monotonic signoff quality ($S_Q$) progression across retained adaptation stages on all eight ASAP7 benchmarks, illustrating how \tapco{} combines outer-loop policy/parameter specialization with inner-loop graph-guided continuations to reach design-specific optima.}
  \Description{Line plots showing monotonic score improvements across retained adaptation stages for all eight benchmarks.}
  \label{fig:journal-adaptation-paths}
\end{figure}

\paragraph{2. Controlled Component Ablations.} \Cref{tab:journal-components} reports strictly matched ON/OFF ablation pairs where every factor except the tested physical prior is held identical:
\begin{itemize}
  \item \emph{Impact of Incremental Graph Propagation:} Enabling graph-guided propagation and clipped endpoint valuation on combinational replacements (\texttt{Graph Prop.\ (COMB)}) improves signoff quality by $+0.620$ points on \texttt{AES} and $+1.164$ points on \texttt{NVDLA-M}. Crucially, because graph priorities eliminate false-positive moves, the number of physically accepted replacements within the same $256$-action budget surges by \textbf{$2.64\times$ ($70 \to 185$) on \texttt{AES}} and \textbf{$4.82\times$ ($39 \to 188$) on \texttt{NVDLA-M}}. Consistent gains hold when jointly optimizing flip-flops and combinational gates (\texttt{FF+COMB}, $+0.341$ and $+0.253$ points) and on heavily optimized late-stage checkpoints (\texttt{Late-Stage}, $+0.316$ and $+0.226$ points).
  \item \emph{Impact of Upstream Driver-Load Coupling:} Adding driver-capacitance perturbation modeling (\cref{eq:graphload}) on top of graph propagation yields an additional $+0.461$ points on \texttt{AES} and $+0.181$ points on \texttt{NVDLA-M}, more than doubling accepted moves on \texttt{AES} ($80 \to 185$) by filtering out oversized cells that overload weak fanin drivers.
  \item \emph{Impact of Coarse-to-Fine Power Screening and Horizon Expansion:} Activating instance-sampled coarse-to-fine power screening (\cref{sec:graph-power}) consistently boosts quality across \texttt{AES} ($+0.028$), \texttt{NVDLA-A} ($+0.104$), and \texttt{NVDLA-P} ($+0.229$), achieving a $100\%$ ($128/128$) trial acceptance rate on \texttt{NVDLA-P}. Furthermore, expanding the coarse/\allowbreak{}detailed candidate screening horizon from $512/128$ to $2048/512$ (\texttt{Expanded Candidate Horizon}) yields a $+0.209$-point improvement on \texttt{AES} and successfully recovers $8$ beneficial multi-endpoint replacements that had negative local delay estimates.
\end{itemize}

\subsection{Outer-Loop LLM Policy Evolution and Design-Specific Specialization}
\label{sec:exp-policy}

Finally, we examine how the outer-loop LLM adaptation engine (\cref{sec:method}) drives monotonic signoff quality progression (\cref{fig:journal-adaptation-paths}) by specializing both parameter envelopes ($\theta$, \cref{tab:parameter-policy}) and executable repair policies ($\mathcal{C}$, \cref{tab:source-policy}) across the eight circuits.

\begin{table}[!t]
  \centering
  \caption{Representative design-specialized parameter configurations ($\theta_d^\star$) discovered by \tapco{} across the eight benchmarks. \texttt{Default} denotes the inherited baseline setting of the underlying repair pipeline.}
  \label{tab:parameter-policy}
  \footnotesize\setlength{\tabcolsep}{3.5pt}
  \begin{tabular}{lccccc}
    \toprule
    Design & Wire ($\mu$m) & Slew/Cap.\ (\%) & Setup Margin (ns) & Repairs/Pass & Primary Active Move Classes \\
    \midrule
    \texttt{AES}     & $60$      & $10 / 10$ & $+0.500$ & $50$      & Selective Sizing + Graph Multi-VT \\
    \texttt{JPEG}    & \texttt{Default} & \texttt{Default} & $-0.105$ & \texttt{Default} & Graph Multi-VT + Load-Coupled Sizing \\
    \texttt{NVDLA-P} & $150$     & $19 / 19$ & \texttt{Default} & $2$ & Conservative Sizing + Power Funnel \\
    \texttt{Ariane}  & $70$      & \texttt{Default} & $+0.210$ & $200$     & Full Buffering, Cloning, Sizing \& VT \\
    \texttt{NVDLA-A} & $78$      & $9 / 9$   & $+0.050$ & $2$       & Tight-Envelope Sizing + Graph Multi-VT \\
    \texttt{NVDLA-C} & $30$      & $20 / 20$ & $+0.032$ & $25$      & Targeted Critical-Cone VT \& Pin Swap \\
    \texttt{MemPool} & \texttt{Default} & \texttt{Default} & $+0.040$ & \texttt{Default} & Macro-Aware Buffering + Graph Sizing \\
    \texttt{NVDLA-M} & \texttt{Default} & \texttt{Default} & $-0.230$ & \texttt{Default} & Slack-Trading Leakage + Graph Sizing \\
    \bottomrule
  \end{tabular}
\end{table}

\begin{table}[!t]
  \centering
  \caption{Source-level executable policy extensions ($\mathcal{C}_d^\star$) synthesized by \tapco{} beyond fixed parameter interfaces.}
  \label{tab:source-policy}
  \footnotesize\setlength{\tabcolsep}{4.5pt}
  \begin{tabular}{@{}lp{0.34\linewidth}p{0.36\linewidth}@{}}
    \toprule
    Policy Dimension & Expressivity Limit of Fixed Parameters $\theta$ & Synthesized Source-Level Policy $\mathcal{C}_d^\star$ \\
    \midrule
    Dynamic Frontier ($\mathcal{D}_{\mathrm{sched}}$)
      & Static endpoint traversal order ignores timing shifts induced by earlier moves.
      & Dynamically recomputes and re-sorts the violating endpoint frontier after intermediate repair passes. \\
    \addlinespace[2pt]
    Severity Effort ($\mathcal{D}_{\mathrm{effort}}$)
      & Uniform pass budgets waste whitespace on shallow paths or hammer blocked cones.
      & Allocates transformation effort proportionally to path violation depth and bypasses saturated cones. \\
    \addlinespace[2pt]
    Partial Commit ($\mathcal{D}_{\mathrm{commit}}$)
      & All-or-nothing pass commit rules discard valid sub-pass improvements.
      & Checkpoints intermediate netlist states and pairs inner batch trials with bisection recovery. \\
    \addlinespace[2pt]
    Graph Schedule ($\mathcal{D}_{\mathrm{cand}}$)
      & Binary move switches cannot express multi-stage timing--power candidate ranking.
      & Interleaves conventional topological repair with graph-guided sizing/VT passes and pin-swap sweeps. \\
    \bottomrule
  \end{tabular}
\end{table}

\paragraph{Design-Specialized Parameter and Source Policies.} Across the complete eight-design campaign, \tapco{} accumulates $106$ retained optimization stages that interleave outer-loop policy/parameter adaptations with inner-loop graph-guided continuations (\cref{fig:journal-adaptation-paths}). \Cref{tab:parameter-policy} demonstrates that \tapco{} never converges to a one-size-fits-all configuration: across the eight designs, adapted wire-length limits span a $5\times$ range ($30$ to $150\,\mu\text{m}$), repair-per-pass budgets span a $100\times$ range ($2$ to $200$), and setup slack margins span both aggressive positive over-design ($+0.500\text{ ns}$ on \texttt{AES} and $+0.210\text{ ns}$ on \texttt{Ariane} to force deeper timing repair) and negative slack relaxation ($-0.105\text{ ns}$ on \texttt{JPEG} and $-0.230\text{ ns}$ on \texttt{NVDLA-M} to protect power-saving cell assignments on near-zero-slack paths).

When parameter specialization reaches an expressivity limit, \tapco{} synthesizes the source-level policy extensions summarized in \cref{tab:source-policy}. Furthermore, in controlled multi-context LLM adaptation trials starting from a late-stage \texttt{AES} checkpoint ($S_Q = 59.898$), autonomous LLM proposals consistently achieve $100\%$ ($27/27$) placement-legal, LEC-verified executions across three independent runs, automatically discovering singleton batching, two-sample coarse power screening, and history-based variant filtering that cut wasted root trials by $47.3\%$ ($55 \to 29$) while monotonically elevating signoff quality to $59.950$.

\section{Conclusion}
\label{sec:conclusion}

We presented \tapco{}, a bi-level framework for post-placement timing and power co-optimization that unites outer-loop LLM-guided hierarchical policy adaptation with inner-loop graph-based physical priors. At the macroscopic level, \tapco{} leverages an LLM program prior conditioned on multi-channel physical diagnostics and evaluator-grounded history to jointly evolve exposed tool configurations and executable repair policies, overcoming the expressivity limits of static hyperparameter tuning. At the microscopic level, \tapco{} anchors differential NLDM delay lookups to live STA snapshots, couples local cell replacements with upstream driver capacitance perturbations, propagates arrival changes incrementally across reconvergent timing cones and sequential launch/capture boundaries, and filters candidates through a coarse-to-fine instance-aware power screening funnel with bisection-recovered physical trials. Evaluated across all eight ASAP7 7-nm benchmarks from the MLCAD 2026 Contest under strict placement legality and formal LEC verification, \tapco{} achieves a $100\%$ ($8/8$) valid convergence rate and a mean quality score of $46.644$ (reducing setup TNS by up to $92.0\%$ and standard-cell leakage by up to $88.1\%$ over the initial baselines), defeating the contest champion \tool{CERBERUS} on every benchmark ($+5.516$ mean gain, up to $+13.937$ points) and surpassing \tool{TAPCO} by $10.80$ to $27.13$ net PPA points ($+21.648$ mean gain) while slashing setup TNS by $29.3\%$ to $80.5\%$ across the four designs where \tool{TAPCO} converges.

\begin{acks}
\PaperAcknowledgment
\end{acks}

\bibliographystyle{ACM-Reference-Format}
\bibliography{references}

\end{document}